\documentclass[11pt]{article}

\usepackage{latexsym}

\usepackage{amsmath}
\usepackage{amssymb}
\usepackage{latexsym}
\usepackage{epic}
\usepackage{eepic}
\usepackage{epsfig}
\usepackage{url}
\usepackage{xspace}

\newcommand{\be}{\begin{itemize}}
\newcommand{\ee}{\end{itemize}}
\newcommand{\bn}{\begin{enumerate}}
\newcommand{\en}{\end{enumerate}}

\newcommand{\bp}{\begin{proposition}}
\newcommand{\ep}{\end{proposition}}
\newcommand{\bl}{\begin{lemma}}
\newcommand{\el}{\end{lemma}}
\newcommand{\bco}{\begin{corollary}}
\newcommand{\eco}{\end{corollary}}
\newcommand{\bt}{\begin{theorem}}
\newcommand{\et}{\end{theorem}}
\newcommand{\bpr}{\begin{proof}}
\newcommand{\epr}{\end{proof}}
\newcommand{\bd}{\begin{definition}}
\newcommand{\ed}{\end{definition}}

\newcommand{\beqn}{\begin{centeqn}}
\newcommand{\eeqn}{\end{centeqn}}

\newcommand{\bleqn}[1]{\begin{centlabeqn}{#1}}
\newcommand{\eleqn}{\end{centlabeqn}}

\newcommand{\bc}{\begin{center}}
\newcommand{\ec}{\end{center}}
\newcommand{\ul}{\underline}
\newcommand{\bs}{\bigskip}
\renewcommand{\ss}{\smallskip}

\newcommand{\bfg}{\begin{figure}}
\newcommand{\efg}{\end{figure}}

\newcommand{\PNAME}[1]{\mbox{\textsc{#1}}\xspace}       

\newcommand{\EXECUTE}{\PNAME{Execute}}
\newcommand{\POLL}{\PNAME{Poll}}
\newcommand{\MAIN}{\PNAME{Main}}
\newcommand{\CHOOSE}{\PNAME{Choose}}
\newcommand{\CREATE}{\PNAME{Create}}

\newenvironment{centeqn}	%
   {{\ss\\ \hspace*{\fill}}} 
   {\hspace*{\fill}\ss\\}

\newsavebox{\EqnLabel}
\newenvironment{centlabeqn}[1]		%
   {\sbox{\EqnLabel}{#1}
    {\bs\\ \hspace*{\fill}}
   } 
   {\hfill{\makebox[0in][r]{\usebox{\EqnLabel}}}\bs\\}

\newenvironment{centeqn-nbsp} %
   {{\ms\\ \hspace*{\fill}}}
   {\hspace*{\fill}}

\newcommand{\intrdef}[1]{\emph{#1\/}}   %
\newcommand{\intr}[1]{\emph{#1\/}}      %

\newcommand{\smpage}{\noindent \parbox{\textwidth}}

\newlength{\parboxwidth}
\newcommand{\AND}{\bigwedge}

\newcommand{\bigor}{\bigvee}	%
\newcommand{\OR}{\bigvee}
\newcommand{\UN}{\bigcup}

\newcommand{\ar}{\rightarrow}	%

\renewcommand{\d}{\, : \,}	%
\newcommand{\df}{\mbox{$\:\stackrel{\rm df}{=\!\!=}\:$}}

\newcommand{\ex}{\exists}
\newcommand{\fa}{\forall}
\newcommand{\halfind}{\hspace*{1.5em}}
\newcommand{\imp}{\Rightarrow}		%
\newcommand{\ind}{\hspace*{3.0em}}
\newcommand{\ints}{\cap}
\renewcommand{\l}{\ell}
\newcommand{\la}[1]{\mbox{$\, \stackrel{#1}{\rightarrow} \,$}}
\newcommand{\lra}{\mbox{$\longrightarrow$}}
\newcommand{\oneton}{[n]}
\newcommand{\pind}{\hspace*{3.0em}}
\newcommand{\pl}{\!\parallel\!}

\newcommand{\sat}{\models}

\newcommand{\sub}{\subseteq}
\newcommand{\un}{\cup}
\newcommand{\up}{\pj}

\newcommand{\struct}[2]{\raisebox{-0.1in}{$\stackrel { \displaystyle
#1} {\scriptstyle #2}\,$}}

\newcommand{\set}[1]{\{ #1 \}}
\newcommand{\tpl}[1]{\lpb #1 \rpb}
\newcommand{\lpb}{\langle \hspace{-0.34em} \langle \hspace{-0.34em} \langle}	
\newcommand{\rpb}{\rangle \hspace{-0.34em} \rangle \hspace{-0.34em} \rangle}
\newcommand{\pj}{\raisebox{0.2ex}{$\upharpoonright$}}

\newcommand{\gc}{\otimes}	%
\newcommand{\gd}{\oplus}	%

\newcommand{\gadd}{\mbox{${\oplus}_{\l \in [1:n]}$}} %

\newcommand{\gdi}[1]{{\oplus}_{#1}}

\newcommand{\gci}[1]{{\otimes}_{#1}}

\newcommand{\cmp}{\mbox{${\otimes}_{j \in I(i)}$}}   %

\newenvironment{lst}{\begin{list}	
                       {}
                       {\setlength{\topsep}{0em}
                        \setlength{\itemsep}{0em}
			\setlength{\leftmargin}{0.3in}
                        \setlength{\rightmargin}{0em}
		     }}
                    {\end{list}}

\newenvironment{blst}{\begin{list}		%
                       {$\bullet$}
                       {\setlength{\topsep}{0em}
                        \setlength{\itemsep}{0em}
			\setlength{\leftmargin}{0.25in}
		     }}
                    {\end{list}}

\newcounter{levelone}
\newenvironment{nlst1}{\begin{list}	%
                       {\arabic{levelone}.}
                       {\usecounter{levelone}
			\setlength{\topsep}{0.2ex}
                        \setlength{\itemsep}{0.1ex}
			\setlength{\leftmargin}{0.25in}
		     }}
                    {\end{list}}

\newcounter{leveltwo}
\newenvironment{nlst2}{\begin{list}	%
                       {(\alph{leveltwo})}
                       {\usecounter{leveltwo}
			\setlength{\topsep}{0em}
                        \setlength{\itemsep}{0em}
			\setlength{\leftmargin}{0.4in}
		     }}
                    {\end{list}}

\newcommand{\case}[2]{\vspace{1.5ex} \noindent \textit{Case} #1: #2.}

\newenvironment{proof}{\vspace{-1.0ex}\textit{Proof.} }	%
                      {\hfill{$\Box$}}

\newcommand{\IF}{\mbox{$\mathbf{if}$}\xspace}
\newcommand{\THEN}{\mbox{$\mathbf{then}$}\xspace}

\newcommand{\FOR}{\mbox{$\mathbf{for}$}\xspace}
\newcommand{\FORALL}{\mbox{$\mathbf{for all}$}\xspace}
\newcommand{\DO}{\mbox{$\mathbf{do}$}\xspace}

\newcommand{\REPEAT}{\mbox{$\mathbf{repeat}$}\xspace}
\newcommand{\UNTIL}{\mbox{$\mathbf{until}$}\xspace}

\newcommand{\RETURN}{\mbox{$\mathbf{return}$}\xspace}

\newcommand{\RW}{\mathit{RW}}	%
\newcommand{\WW}{\mathit{WW}}	%

\newcommand{\igcrw}[2]{\mbox{$\struct{\otimes}{#2 \in \RW(#1)}$}}

\newcommand{\igcww}[2]{\mbox{$\struct{\otimes}{#2 \in \WW(#1)}$}}

\newcommand{\andt}[2]
      {\textup{\mbox{${\bigwedge\hspace{-0.95em}\bigwedge\hspace{-0.95em}\bigwedge}_{#1 #2}\,$}}}

\newcommand{\Pids}{\mbox{\textsf{Pids}}}
\newcommand{\UPairs}{\mathcal{UI}}
\newcommand{\IPairs}{\mathcal{I}_0}
\newcommand{\Pairs}{\mathcal{I}}
\newcommand{\I}{\mathcal{I}}
\newcommand{\procs}[1]{\ms{procs}({#1})}
\newcommand{\pairs}[1]{\ms{pairs}({#1})}
\newcommand{\states}[1]{\ms{states}({#1})}

\newcommand{\pspec}[2]{\tpl{\{#1,#2\},spec_{#1#2}}}
\newcommand{\pspecij}{\pspec{i}{j}}

\newcommand{\aen}{\mathit{aen}}
\newcommand{\blk}{{\mit blk}}  %
\newcommand{\corr}{\sim}
\newcommand{\create}{\mathsf{create}}

\newcommand{\domI}{{\it dom}(I)}  %
\newcommand{\domJ}{{\it dom}(J)}  %

\newcommand{\edomI}{\{i_1,\ldots,i_K\}}  %
\newcommand{\f}{\mbox{$\mathrm{\Phi}$}}		%
\newcommand{\fin}{\scriptstyle \mathit{fin}}  %
\newcommand{\first}{\mathit{first}}

\newcommand{\guard}{\mathit{guard}}
\renewcommand{\inf}{\scriptstyle \mathit{inf}}  %
\newcommand{\normal}{\mathit{normal}}
\newcommand{\pnd}{\mathit{pnd}}  	%

\newcommand{\spec}{\mathit{spec}}

\newcommand{\start}{\mathit{start}}
\newcommand{\stof}[1]{\{\hspace*{-0.3em}|#1|\hspace*{-0.3em}\}}
\newcommand{\Wt}[2]{\mathit{Wt}_{#1}(#2)}

\newcommand{\CL}{\mathit{CL}}

\newcommand{\lP}[2]{\mathit{graph}({P}_{#1}^{#2})}

\newcommand{\graph}[1]{\mathit{graph}({#1})}

\newcommand{\B}[2]{\mbox{$B_{#1}^{#2}$}}  %
\newcommand{\CA}[3]{\mbox{$A_{#1,#3}^{#2}$}}  %
\newcommand{\CB}[3]{\mbox{$B_{#1,#3}^{#2}$}}  %
\newcommand{\MV}[2]{\mbox{$a_{#1}^{#2}$}}  %
\renewcommand{\P}[2]{\mbox{$P_{#1}^{#2}$}}  %
\newcommand{\PP}[2]{\mbox{$P_{#1}$}}   %
\newcommand{\SYij}{\mbox{$(S_{ij}^0, P_i^j \pl P_j^i)$}}  %
\newcommand{\SYik}{\mbox{$(S_{ik}^0, P_i^k \pl P_k^i)$}}  %
\newcommand{\SYjk}{\mbox{$(S_{jk}^0, P_j^k \pl P_k^j)$}}  %
\newcommand{\SY}{\mbox{$(S_I^0, P_{i_1}^I \pl \ldots \pl P_{i_K}^I)$}} %

\newcommand{\Pij}{P_{i}^{j}}  %
\newcommand{\Pji}{P_{j}^{i}}  %

\newcommand{\MP}{M_{\cal P}}
\newcommand{\SP}{{\cal P}}
\renewcommand{\Pr}{{\cal A}}
\newcommand{\Ps}{{\cal S}}

\newcommand{\satf}{\models_{\f}}	%
\newcommand{\AP}{\mbox{$\cal AP$}}  %
\newcommand{\ACTL}{\mbox{$\mathrm{ACTL}$}}   	%
\newcommand{\ACTLm}{\mbox{$\mathrm{ACTL}^-$}}  	%
\newcommand{\ACTLmij}{\mbox{$\mathrm{ACTL}^{-}_{ij}$}}  %
\newcommand{\ACTLS}{\mbox{$\mathrm{ACTL}^*$}}   %
\newcommand{\ACTLSm}{\mbox{$\mathrm{ACTL}^* - X$}}   %
\newcommand{\CTL}{\mbox{$\mathrm{CTL}$}}  	%
\newcommand{\CTLS}{\mbox{$\mathrm{CTL}^*$}}   	%

\newcommand{\SH}{\mathcal{SH}}
\newcommand{\SHij}{\mathcal{SH}_{i,j}}

\newcommand{\A}{\mathrm{\mathsf{A}}}
\newcommand{\E}{\mathrm{\mathsf{E}}}
\newcommand{\F}{\mathrm{\mathsf{F}}}
\newcommand{\G}{\mathrm{\mathsf{G}}}
\newcommand{\U}{\mathrm{\mathsf{U}}}
\newcommand{\Uw}{\mathrm{\mathsf{U_w}}}

\newcommand{\X}{\mathrm{\mathsf{X}}}

\newcommand{\lfalse}{\mathit{false}}
\newcommand{\ltrue}{\mathit{true}}

\newcommand{\AG}{\textup{\textsf{AG}}}

\newcommand{\AF}{\textup{\textsf{AF}}}
\newcommand{\EF}{\textup{\textsf{EF}}}

\newcommand{\AX}{\textup{\textsf{AX}}}
\newcommand{\EX}{\textup{\textsf{EX}}}

\newcommand{\ea}{\raisebox{0ex}{$\stackrel{\infty}{\textup{\textsf{G}}}$}}

\newcommand{\io}{\raisebox{0ex}{$\stackrel{\infty}{\textup{\textsf{F}}}$}}

\newcommand{\ab}{\mathit{ab}}
\newcommand{\cm}{\mathit{cm}}
\renewcommand{\sb}{\mathit{sb}}
\newcommand{\st}{\mathit{st}}
\newcommand{\term}{\mathit{term}}

\newtheorem{theorem}{Theorem}
\newtheorem{lemma}[theorem]{Lemma}
\newtheorem{proposition}[theorem]{Proposition}
\newtheorem{corollary}[theorem]{Corollary}

\newtheorem{definition}{Definition}

\newcommand{\ms}[1]{%
        \relax\ifmmode
                \mathord{\mathcode`\-="702D\it #1\mathcode`\-="2200}%
        \else
                {\it #1}%
        \fi
}

\newcommand{\replica}[1]{\ms{replica}({#1})}
\newcommand{\client}[1]{\ms{client}({#1})}
\newcommand{\Op}{{\cal O}}
\newcommand{\R}{{\cal R}}

\newcommand{\prev}{\ms{prev}}

\newcommand{\CSC}{\mathit{CSC}}

\newcommand{\citeN}{\cite}

\begin{document}

\bc
\textbf{\Large Synthesis of Large Dynamic Concurrent Programs
       from Dynamic Specifications}\\[0.2in]
{\large Paul C. Attie} \\[0.05in]
\vskip 0.05in
{\large Department of Computer Science}\\
{\large American University of Beirut}\\
{\large and}\\
{\large Center for Advanced Mathematical Sciences}\\
{\large American University of Beirut}\\
\texttt{paul.attie@aub.edu.lb}

\today
\ec

\begin{abstract}
We present a tractable method for synthesizing arbitrarily large
concurrent programs, for a shared memory model with 
common hardware-available primitives such as atomic registers,
compare-and-swap, load-linked/store conditional, etc.
The programs we synthesize are dynamic: new processes can be created
and added at run-time, and so
our programs are not finite-state, in general. Nevertheless, we
successfully exploit automatic synthesis and model-checking methods based on 
propositional temporal logic.
Our method is algorithmically efficient, with complexity polynomial in
the number of component processes (of the program) that are ``alive'' at any time.
Our method does not explicitly construct the automata-theoretic
product of all processes that are alive,
thereby avoiding \intr{state explosion}. Instead, 
for each pair of processes which interact, our method constructs an
automata-theoretic product (\intr{pair-machine}) which embodies all the
possible interactions of these two processes.
From each pair-machine, we can synthesize a correct \intr{pair-program} which
coordinates the two involved processes as needed.
We allow such pair-programs to
be added dynamically at run-time. They are then ``composed
conjunctively'' with the currently alive pair-programs to re-synthesize
the program as it results after addition of the new pair-program.
We are thus able to add new behaviors, which result in new properties
being satisfied, at run-time.
This ``incremental composition'' step has complexity independent of
the total number of processes, it only requires the mechanical
analysis of the two processes in the pair-program, and their immediate
neighbors, i.e., the other processes which they interact directly with.
We establish a ``large model'' theorem which shows
that the synthesized large program inherits correctness properties
from the pair-programs.

\end{abstract}

\section{Introduction}
\label{sec:intro}

We exhibit a method of mechanically synthesizing a concurrent program
consisting of a large, and dynamically varying, number of sequential
processes executing in parallel.  Our programs operate in shared
memory, commonly available hardware primitives, such as 
using read and write operations on atomic registers, compare-and-swap,
load-linked/store conditional.  Even
thought our synthesis method is largely mechanical, we only require
that each process have a finite number of actions, and that the data
referred to in action guards be finite. Underlying data that processes
operate on, and which does not affect action guards, can be
infinite. Also, since the number of processes can increase without
limit, the synthesized program as a whole is not finite-state.  In
addition, our method is computationally efficient, it does not
explicitly construct the automata-theoretic product of a large number
of processes (e.g., all processes that are ``alive'' at some point)
and is therefore not susceptible to the \emph{state-explosion
problem}, i.e., the exponential growth of the number of global states
with the number of processes, which is widely acknowledged to be the
primary impediment to large-scale application of mechanical
verification methods.

Rather than build a global product, our method constructs the product
of small numbers of sequential processes, and in particular, the
product of each pair of processes that interact, thereby avoiding the
exponential complexity in the number of processes that are ``alive''
at any time.
The product of each pair of interacting processes, or 
\intr{pair-machine}, is a
Kripke structure which embodies the interaction of the two processes.
The pair-machines can be constructed manually, and then efficiently
model-checked (since it is small) to verify \intr{pair-properties}:
behavioral properties of the interaction of the two processes, when
viewed in isolation from the remaining processes. Alternatively, the
pair-properties can be specified first, and the pair-machine
automatically synthesized from the pair-properties by the use of
mechanical synthesis methods such as \cite{EC82,MW84,KV97}. Again this
is efficient since the pair-machines are small.

Corresponding to each pair-machine is a \intr{pair-program}, a
syntactic realization of the pair-machine, which generates the
pair-machine as its global-state transition diagram.  Finally, we
syntactically compose all of the pair-programs. This composition has a
conjunctive nature: a process $P_i$ can make a transition iff that
transition is permitted by \emph{all} of the pair-programs in which
$P_i$ participates.
We allow such ``pair-programs'' to be added dynamically at run-time. They are then
composed with the currently alive pair-programs to
re-synthesize the program as it results after addition of the new
pair-program.  We are thus able to add new behaviors, which result in
new properties being satisfied, at run-time.  The use of pairwise
composition greatly facilitates this, since the addition of a new
pair-program does not disturb the correctness properties which are
satisfied by the currently present pair-programs.  We establish a
``large model'' theorem which shows that the synthesized large program
inherits correctness properties from the pair-programs.
 
Since the pair-machines are small, and since the composition step
operates on syntax, i.e., the pair-programs themselves, and not their
state-transition diagrams, our method is computationally efficient.
In particular, the dynamic addition of a single pair-program requires a 
mechanical synthesis or model checking step whose complexity is
independent of the total number of alive processes at the time, but
which depends only on the checking products of the two processes
involved in the pair-program, together with some of their neighbors,
i.e., the processes which they immediately interact with.
Our method thus overcomes the severe limitations previously
imposed by state-explosion on the applicability of
automatic synthesis methods, and extends these methods to the new
domain of dynamic programs.

Our method can generate systems under arbitrary {\em process
interconnection} schemes, e.g., fully connected, ring, star.  In our
model of parallel computation, two processes are interconnected if and
only if either
\label{MP-def-intercon}
(1) one process can inspect the local state of the other process or
(2) both processes read and/or write a common variable, or both.

The method requires the pair-programs to satisfy certain technical
assumptions, thus it is not completely
general. Nevertheless, it is applicable in many interesting cases. We
illustrate our method by synthesizing a ring-based two phase commit
protocol. Using the large model theorem, we show that
correctness properties that two processes of the ring satisfy when
interacting in isolation carry over when those processes are part of
the ring. We then easily construct a correctness proof for the ring
using these properties. We note that the ring can contain an
arbitrarily large number of processes, i.e., we really synthesize a
\intr{family} of rings, one for each natural number.

A crucial aspect of our method is its soundness: which correctness
properties can be established for our synthesized programs?  We
establish a ``large model'' theorem which shows that the synthesized
program inherits all of the correctness properties of the
pair-programs, i.e., the pair-properties.  We express our
pair-properties in the branching time temporal logic $\ACTL$
\cite{GL94} minus the nexttime operator.  In particular, propositional
invariants and some temporal leads-to properties of any pair-program
also hold of the synthesized program.  (A temporal leads-to property
has the following form: if condition 1 holds now, then condition 2
eventually holds. $\ACTL$ can express temporal leads-to if condition 1
is purely propositional.)  In addition, we can use a suitable
deductive system to combine the pair-properties to deduce correctness
properties of the large program which are not directly expressible in
pairwise fashion.

This paper extends our previous work \cite{AE98} on the synthesis
of large concurrent programs in four important directions:

\bn

\item It eliminates the requirement that all pair-programs be isomorphic
    to each other, which in effect constrains the synthesized program
    to contain only one type of interaction amongst its component
    processes.  In our method, every process can be nonisomorphic with
    every other process, and our method would still be computationally
    efficient.

\item It extends the set of correctness properties that are preserved
    from propositional invariants and propositional temporal leads-to
    properties (i,e., leads-to properties where the conditions are
    purely propositional) to formulae that can contain arbitrary
    nesting of temporal modalities.

\item It eliminates the requirement that the number of processes of the
    synthesized program be fixed: our previous work synthesized an
    infinite family of programs, each of which contains a large, but
    fixed, number of processes. By contrast, the current method
    produces a single program, in which the number of processes can
    dynamically increase at run-time.

\item It produces programs that do not require a large grain of
    atomicity: in \cite{Att99a,AE98}, each process needed to
    atomically inspect the state of all of its neighbors (i.e., all
    processes with which it is composed in some pair-program) in a
    single transition. By contrast, the current method produces
    programs that operate using only hardware-available primitives for
    interprocess communication and synchronization.

\en
To demonstrate the utility of our method, we apply it to synthesize a
two-phase commit protocol, and a replicated data service.

\paragraph{Related work.} Previous synthesis methods
\cite{AM94,DW90,EC82,KMTV00,KV97,MW84,PR89,PR89b} all rely on some
form of exhaustive state space search, and thus suffer from the
\emph{state-explosion problem}: synthesizing a concurrent program
consisting of $K$ sequential processes, each with $O(N)$ local states,
requires building the global state transition diagram of size
$O(N^K)$. There are a number of methods proposed for verifying
correctness properties of an infinite family of finite-state processes
\cite{APRXZ01,CGB86,EK00,EN96,PRZ01,SG92}. All of these deal with an
infinite family of concurrent programs, where each program consists of
a possibly large, but \emph{fixed} set of processes.  No method to
date can verify or synthesize a \emph{single} concurrent program in
which processes can be dynamically created \emph{at run time}.
Furthermore, all methods to date that deal with large concurrent
programs, apart from our own previous work \cite{Att99a,AE98} make the
``parametrized system'' assumption: the processes can be partitioned
into a small number of ``equivalence classes,'' within each of which
all processes are isomorphic.
Hence, in eliminating these two significant restrictions, our 
method is a significant improvement over the previous literature, and
moves automated synthesis methods close to the realm of practical
distributed algorithms. We illustrate this point by using our
method to synthesize a replicated data service based on the algorithms
of \cite{FGLLS99,LLSG92}. Our algorithm is actually more flexible,
since it permits the dynamic addition of more replicas at run time.
Some synthesis method in the literature synthesize ``open systems,''
or ``reactive modules,'' which interact with an environment, and are
required to satisfy a specification regardless of the environment's behavior.
The main argument for open systems synthesis is that open systems can
deal with any ``input'' which the environment presents.
We can achieve this effect by using the ``exists nexttime'' $(\EX)$ modality of 
the temporal logic CTL \cite{EC82,Em90}.
We illustrate this in our replicated data service example, where we
specify that a client can submit operations at any time.

The rest of the paper is as follows. 
Section~\ref{sec:model} presents our model of concurrent computation.
Section~\ref{sec:ACTLS} discusses temporal logic and fairness.
Section~\ref{sec:static:method} presents a restricted version of the
method, which is only applicable to static concurrent programs: those
with a fixed set of processes.
This approach simplifies the development and exposition of our method,
Section~\ref{sec:static:soundness} establishes the soundness of the
synthesis method for static programs.
Section~\ref{sec:example-twophase} presents the two phase commit
example, which can be treated with the restricted method.
Section~\ref{sec:dynamic:method} presents the general synthesis
method, which can produce dynamic concurrent programs.
Section~\ref{sec:dynamic:soundness} shows that the general method is sound.
Section~\ref{sec:atomic-implementation} outlines how the synthesized
programs can be implemented using atomic registers.
In Section~\ref{sec:example:esds}
we use our method to synthesize an eventually-serializable replicated
data service.
Section~\ref{sec:conclusions} discusses
further work and concludes.

\section{Model of Concurrent Computation}
\label{sec:model}

We assume the existence of a possibly infinite, universal set $\Pids$ of unique
process indices.
A concurrent program $P$ consists of a finite, unbounded, and possibly
varying number of sequential processes $P_i, i \in \Pids$ running in
parallel, i.e., $P = P_1 \| \cdots \| P_K$ where $P_1, \ldots, P_K$
execute in parallel and are the processes that have been ``created''
so far.  For technical convenience, we do not allow processes to be
``destroyed'' in our model. Process destruction can be easily emulated
by having a process enter a ``sink'' state, from which it has no
enabled actions.

With every process $P_i$, we
associate a single, unique index, namely $i$. Two
processes are {\em similar} if and only if one can be obtained from
the other by swapping their indices. Intuitively, this corresponds to
concurrent algorithms where a single ``generic'' indexed piece of code
gives the code body for all processes.

As stated above, we compose a dynamically varying number of
pair-programs to synthesize the overall program. To define the
syntax and semantics of the pair-programs, 
we use the \intr{synchronization skeleton} model of \cite{EC82}.
The synchronization skeleton of a process
$P_i$ is a state-machine where each state represents a region of code
that performs some sequential computation and each arc
represents a conditional transition (between different regions of
sequential code) used to enforce synchronization constraints. For
example, a node labeled $C_i$ may represent the critical
section of $P_i$. While in $C_i$, $P_i$ may
increment a single variable, or it may perform an extensive
series of updates on a large database.  In general, the internal
structure and intended application of the regions of sequential code
are unspecified in the synchronization
skeleton. The abstraction to synchronization skeletons thus
eliminates all steps of the sequential computation from consideration.

Formally, the synchronization skeleton of each process $P_i$ is a
directed graph where each node $s_i$ is a unique {\em local state} of
$P_i$, and each arc has a label of the form $\gdi{\l \in \oneton} B_\l
\ar A_\l$,\footnote{$\oneton$ denotes the integers from $1$ to $n$ inclusive.}
where each $B_\l \ar A_\l$ is a guarded command
\cite{Dij76}, and $\gd$ is guarded command ``disjunction,'' i.e., the
arc is equivalent to $n$ arcs, between the same pair of nodes, each
labeled with one of the  $B_\l \ar A_\l$.
Let $\hat{P}_i$ denote the synchronization skeleton of process $i$ with all the
arc labels removed.

Roughly, the operational semantics of 
	$\gdi{\l \in \oneton} B_\l \ar A_\l$
is that if one of the $B_\l$ evaluates to true, then the corresponding body
$A_\l$ can be executed. If none of the $B_\l$ evaluates to
true, then the command ``blocks,'' i.e., waits until one of
the $B_\l$ holds.\footnote{This interpretation
was proposed by \cite{Dij82}.}
Each node
must have at least one outgoing arc, i.e., a skeleton contains no
``dead ends,'' and two nodes are connected by at most one arc in each direction.
A {\em (global) state} is a
tuple of the form $(s_1, \ldots , s_K, v_1,\ldots ,v_m)$ where each
$s_i$ is the current local state of $P_i$, and $v_1,\ldots,v_m$ is a list
giving the current values of all the shared variables, $x_1,\ldots,x_m$ (we
assume these are ordered in a fixed way, so that $v_1,\ldots ,v_m$
specifies a unique value for each shared variable).
A guard $B$
is a predicate on states, and a body $A$ is a parallel assignment
statement that updates
the values of the shared variables. If $B$ is omitted from
a command, it is interpreted as $\ltrue$, and we write the
command as $A$. If $A$ is omitted, the shared variables are
unaltered, and we write the command as $B$.

We model parallelism in the usual way by the nondeterministic
interleaving of the ``atomic" transitions of the individual
synchronization skeletons of the processes $P_i$. Hence, at each step
of the computation, some process with an ``enabled'' arc is
nondeterministically selected to be executed next. Assume that the
current state is $s = (s_1, \ldots , s_i, \ldots , s_K, v_1,\ldots
,v_m)$ and that $P_i$ contains an arc from $s_i$ to
$s'_i$ labeled by the command $B \rightarrow A$. If $B$ is true in $s$,
then a permissible next state is $( s_1, \ldots , s'_i,
\ldots , s_K, v'_1,\ldots ,v'_m )$ where $v'_1,\ldots ,v'_m$ is the
list of updated values for the shared variables produced by executing
$A$ in state $s$. The arc from $s_i$ to $s'_i$ is said to be {\em enabled} in
state $s$. An arc that is not enabled is {\em disabled}, or
{\em blocked}. A \intr{(computation) path} is any sequence of states
where each successive pair of states is related by the above next-state relation.
If the number of processes is fixed, then the concurrent program can
be written as $P_1 \| \cdots \| P_K$, where $K$ is fixed.
In this case, we 
also specify a
a set $S_0$ of global states
in which execution is permitted to start. 
These are the \intr{initial states}. The program is then written as
$(S_0, P_1 \| \cdots \| P_K)$. An initialized (computation) path is a
computation path whose first state is an initial state.
A state is \intr{reachable} iff it lies along some initialized path.

\section{Temporal Logic and Fairness}
\label{sec:ACTLS}

$\CTLS$ is a propositional branching time temporal logic \cite{Em90}
whose formulae are built up from atomic propositions, propositional
connectives, the universal $(\A)$ and existential $(\E)$ path
quantifiers, and the linear-time modalities nexttime (by process $j$)
$\X_j$, and strong until $\U$. 
The sublogic $\ACTLS$ \cite{GL94} is the ``universal fragment'' of
$\CTLS$: it results
from $\CTL$ by restricting negation to propositions, and eliminating the
existential path quantifier $\E$.
The sublogic $\CTL$ \cite{EC82} results from
restricting $\CTLS$ so that every linear-time modality is paired with a
path quantifier, and vice-versa. The sublogic $\ACTL$ \cite{GL94} 
results from restricting $\ACTLS$ in the same way.
The linear-time temporal logic PTL
\cite{MW84} results from removing the path quantifiers from $\CTLS$.

We have the following syntax for 
$\CTLS$. We inductively define a class of state formulae
(true or false of states) using rules (S1)--(S3) below and a class of
path formulae (true or false of paths) using rules (P1)--(P3) below:

\begin{lst}
   \item[(S1)] The constants $\ltrue$ and $\lfalse$ are state formulae.
               $p$ is a state formulae for any atomic
               proposition $p$.

   \item[(S2)] If $f,g$ are state formulae, then so are $f \land g$,
   $\neg f$.

   \item[(S3)] If $f$ is a path formula, then $\A f$ is a state formula.

   \item[(P1)] Each state formula is also a path formula;

   \item[(P2)] If $f,g$ are path formulae, then so are $f \land g$, $\neg f$.

   \item[(P3)] If $f,g$ are path formulae, then so are $\X_j f$, $f \U g$.
\end{lst}
The linear-time temporal logic PTL \cite{MW84} consists of the
set of path formulae generated by rules (S1) and (P1)--(P3).
We also introduce some additional modalities as abbreviations:
$\F f$ (eventually) for $[\ltrue \U f]$, 
$\G f$ (always) for $\neg \F \neg f$, 
$[f \Uw g]$ (weak until) for $[f \U g] \lor \G f$,
$\io f$ (infinitely often) for $\G\F f$, and
$\ea f$ (eventually always) for $\F\G f$.

Likewise, we have the following syntax for $\ACTLS$. 
\begin{lst}
   \item[(S1)] The constants $\ltrue$ and $\lfalse$ are state formulae.
               $p$ and $\neg p$ are state formulae for any atomic
               proposition $p$.

   \item[(S2)] If $f,g$ are state formulae, then so are $f \land g$, $f \lor g$.

   \item[(S3)] If $f$ is a path formula, then $\A f$ is a state formula.

   \item[(P1)] Each state formula is also a path formula;

   \item[(P2)] If $f,g$ are path formulae, then so are $f \land g$, $f \lor g$.

   \item[(P3)] If $f,g$ are path formulae, then so are $\X_j f$, $f \U g$, and
               $f \Uw g$.
\end{lst}
The logic $\ACTL$ \cite{GL94} is obtained by replacing rules (S3),(P1)--(P3) by (S3'):

\begin{lst}
   \item[(S3')] If $f, g$ are state formulae, then so are $\AX_j f$, $\A[f \U g]$, and
               $\A[f \Uw g]$.
\end{lst}

\noindent

The set of state formulae generated by rules (S1)--(S3) and (P0) forms $\ACTL$.
The logic $\ACTLm$ is the logic $\ACTL$ without the $\AX_j$ modality.
We define the logic $\ACTLSm$ to be the logic $\ACTLS$ without the $X_j$
modality, and
the logic $\ACTLm$ to be $\ACTL$ without the $\AX_j$
modality, and the logic $\ACTLmij$ to be $\ACTLm$ where
the atomic propositions are drawn only from $\AP_i \un \AP_j$.

Formally, we define the semantics of $\CTLS$ formulae with respect to a
structure $M ~=~(S, R)$ consisting of
\begin{blst}

   \item $S$, a countable set of states. Each state is a mapping from
the set $\AP$ of atomic propositions into $\{\ltrue, \lfalse\}$, and

   \item $R = \UN_{i \in \Pids} R_i$, where 
         $R_i \subseteq S \times \{i\} \times S$ is a binary relation on $S$ giving the
         transitions of process $i$.

\end{blst}
Here $\AP = \UN_{i \in \Pids} \AP_{i}$, where $\AP_i$ is the
set of atomic propositions that ``belong'' to process $i$. Other
processes can read propositions in $\AP_i$, but only process $i$ can
modify these propositions (which collectively define the local state
of process $i$).

A {\em path} is a sequence of states $(s_1,s_2 \ldots)$ such that
$\forall i, (s_i,s_{i+1}) \in R$, and a \emph{fullpath} is a maximal 
path. A fullpath $(s_1, s_2, \ldots)$ is infinite unless for some
$s_k$ there is no $s_{k+1}$ such that ($s_k, s_{k+1}) \in R$.
We use the convention (1) that $\pi = (s_1,s_2,\ldots)$ denotes a
fullpath and (2) that $\pi^i$ denotes the suffix $(s_i,s_{i+1},s_{i+2},\ldots)$
of $\pi$, provided $i \leq |\pi|$, where $|\pi|$, the length of $\pi$, is
$\omega$ when $\pi$ is infinite and $k$ when $\pi$ is finite and of the form
$(s_1, \ldots, s_k)$; otherwise $\pi^i$ is undefined.
We also use the usual notation to indicate truth in a structure:
$M,s_1 \sat f$ (respectively $M, \pi \sat f$) means that $f$ is true in
structure $M$ at state $s_1$ (respectively of fullpath $\pi$).
In addition, we use $M, S \sat f$ to mean $\forall s \in S \d (M,s \sat f$),
where $S$ is a set of states. We define $\sat$ inductively:

\label{MP-def-sat}
\begin{lst}

\item[(S1)] $M, s_1 \sat \ltrue$ and $M,s_1 \not\sat \lfalse$.
            $M, s_1 \sat p$ ~iff~ $s_1(p) = \ltrue$.
            $M, s_1 \sat \neg p$ ~iff~ $s_1(p) = \lfalse$.

\item[(S2)]
$M, s_1 \sat f \land g$ ~iff~ $M, s_1 \sat f$ and $M, s_1 \sat g$\\
$M, s_1 \sat f \lor g$  ~iff~ $M, s_1 \sat f$ or  $M, s_1 \sat g$

\item[(S3)]
$M, s_1 \sat \A f$ ~iff~ for every fullpath $\pi = (s_1,s_2,\ldots)$
                       in $M$: \  $M, \pi \sat f$

\item[(P1)] $M, \pi \sat f$ ~iff~ $M, s_1 \sat f$
               
\item[(P2)]
$M, \pi \sat f \land g$ ~iff~ $M, \pi \sat f$ and $M, \pi \sat g$\\
$M, \pi \sat f \lor  g$ ~iff~ $M, \pi \sat f$ or  $M, \pi \sat g$

\item[(P3)]
$M, \pi \sat \X_j f$ ~iff~ $\pi^2$ is defined and $(s_1, s_2) \in R_j$
and $M, \pi^2 \sat f$\\
$M, \pi \sat f \U g$ ~iff~ 
             there exists $i \in [1:|\pi|]$ such that\\
\hspace*{1.25in}   $M, \pi^i \sat g$ and 
                   for all $j \in [1:(i-1)]$: \ $M, \pi^j \sat f$\\
$M, \pi \sat f \Uw g$ ~iff~ 
             for all $i \in [1:|\pi|]$\\
\hspace*{1.25in}   if $M, \pi^j \not\sat g$ for all $j \in [1:i]$, then
                    \ $M, \pi^i \sat f$

\end{lst}
When the structure $M$ is understood from context, it may be omitted (e.g., $M,
s_1 \sat p$ is written as $s_1 \sat p$).
Since the other logics are all sublogics of $\CTLS$, the above definition
provides semantics for them as well.
We refer the reader to \cite{Em90} for details in general, and to
\cite{GL94} for details of \ACTL.

\subsection{Fairness}
\label{sec:fairness}

To guarantee liveness properties of the synthesized program, we use
a form of weak fairness. Fairness is usually specified as a linear-time logic
(i.e., PTL) formula $\f$, and a fullpath is fair iff it satisfies $\f$.
 To state correctness properties under the assumption of
fairness, we relativize satisfaction ($\sat$)
so that only fair fullpaths are considered. The resulting notion of
satisfaction, $\satf$, is defined by \cite{EL87} as follows:
\begin{lst}
\item[(S3-fair)]
$M, s_1 \satf \A f$ ~iff~ for every $\f$-fair fullpath $\pi = (s_1,s_2,\ldots)$
                     in $M$: \  $M, \pi \sat f$
\end{lst}
Effectively, path quantification is only over the paths that satisfy $\f$.

\section{Synthesis of Static Concurrent Programs}
\label{sec:static:method}

To simplify the development and exposition of our method, we first present a
restricted case, where we synthesize \intr{static} concurrent programs, i.e.,
those with a fixed set of processes. 
We extend the method to dynamic concurrent programs in
Section~\ref{sec:dynamic:method} below.

As stated earlier, our aim is to synthesize a large concurrent program
$P =  P_{i_1} \pl \ldots \pl P_{i_K}$ without
explicitly generating its global state transition diagram, and thereby incurring
time and space complexity exponential in the number of component processes of $P$. We
achieve this by breaking the synthesis problem down into two steps:
\bn

\item For every pair of processes in $P$ that interact directly, synthesize a
\intr{pair-program} that describes their interaction.

\item Combine all the pair-programs to produce $P$.
\en
When we say $P_i$ and $P_j$ interact directly, we mean that each
process can read the other processe's atomic propositions (which,
recall, encode the processe's local state), and that they have
a set $\SH_{ij}$ of shared variables that they both read and write.
We define the \intr{interconnection relation} $I \sub \edomI \times
\edomI \times \ACTLm$ as follows: $(i,j,f_{ij}) \in I$ iff $P_i$ and
$P_j$ interact directly, and $f_{ij}$ is an $\ACTLm$ formula specifying
this interaction. In the sequel we let $\spec_{ij}$ denote the
specification associated with $i,j$, and we say that $\edomI$ is the
domain of $I$. We introduce the ``spatial
modality'' $\andt{i}{j}$ which quantifies over all pairs $(i,j)$ such
that $i$ and $j$ are related by $I$. Thus, $\andt{i}{j} \spec_{ij}$ is
equivalent to $\fa (i,j,\spec_{ij}) \in I : \spec_{ij}$.
We stipulate that $I$ is ``irreflexive,'' that is, $(i,i,f_{ij})
\not\in I$ for all $i, f_{ij}$, and that every process
interacts directly with at least one other process:
$\fa i \in \{i_1,\ldots, i_K\} :
	(\ex j, f_{ij} : (i,j,f_{ij}) \in I \lor (j,i,f_{ij}) \in I)$.
Furthermore, for any pair of process indices $i,j$, $I$ contains at
most one pair $(k,\l,f_{k\l})$ such that $k \in \{i,j\}$ and $\l \in \{i,j\}$.
In the sequel, we say that $i$ and $j$ are \emph{neighbors} when
$(i,j,f_{ij}) \in I$ or $(j,i,f_{ij}) \in I$, for some $f_{ij}$.
We shall sometimes abuse notation and write $(i,j) \in I$ (or $i\,I\,j$) for
$\ex f_{ij} : ((i,j,f_{ij}) \in I \lor (j,i,f_{ij}) \in I)$.
We also introduce the following abbreviations: $I(i)$ denotes the set $\{j ~|~
i\,I\,j\}$; and $\hat{I}(i)$ denotes the set $\{i\} \un \{j ~|~ i\,I\,j\}$.
Since the interconnection relation $I$ embodies a complete
specification, we shall refer to a program that has been synthesized
from $I$ as an \emph{$I$-program}, and to its component processes as
\emph{$I$-processes}.

Since our focus in this article is on avoiding state-explosion, we
shall not explicitly address step 1 of the synthesis method outlined
above. Any method for deriving concurrent programs from temporal logic
specifications can be used to generate the required pair-programs,
e.g., the synthesis method of \cite{EC82}.
Since a pair-program has only $O(N^2)$ states (where $N$ is the
size of each sequential process), the problem of deriving a
pair-program from a specification is considerably easier than
that of deriving an $I$-program from the specification. Hence,
the contribution of this article, namely the second step above, is to
reduce the more difficult problem (deriving the $I$-program) to
the easier problem (deriving the pair-programs). We proceed as
follows.

For sake of argument, let us first assume that all the pair-programs
are actually isomorphic to each other.
Let $i\,I\,j$. We denote the pair-program for processes $i$ and $j$ by
$\SYij$, where $S^0_{ij}$ is the set of initial states, $\P{i}{j}$ is
the synchronization skeleton for process $i$ in this pair-program, and
$\P{j}{i}$ is the synchronization skeleton for process $j$. 
We take $\SYij$
and generalize it in a natural way to an $I$-program.
We show that our generalization preserves a large class of correctness
properties.  Roughly the idea is as follows. Consider first the
generalization to three pairwise interconnected processes $i,j,k$,
i.e., $I = \{ (i,j), (j,k), (k,i) \}$\footnote{Note the abuse of
notation: we have omitted the $\ACTLm$ formulae.}.
With respect to process $i$, the
proper interaction (i.e., the interaction required to satisfy the
specification) between process $i$ and process $j$ is captured by the
synchronization commands that label the arcs of $\P{i}{j}$.
Likewise, the proper interaction between process $i$ and process $k$
is captured by the arc labels of $\P{i}{k}$. Therefore, in the
three-process program consisting of processes $i, j, k$ executing
concurrently, (and where process $i$ is interconnected to both process
$j$ and process $k$), the proper interaction for process $i$ with
processes $j$ and $k$ is captured as follows: when process $i$
traverses an arc, the synchronization command which labels that arc
in $\P{i}{j}$ is executed ``simultaneously'' with the synchronization
command which labels the corresponding arc in $\P{i}{k}$. For
example, taking as our specification the mutual exclusion problem, if
$P_i$ executes the mutual exclusion protocol with respect to both
$P_j$ and $P_k$, then, when $P_i$ enters its critical section, both
$P_j$ and $P_k$ must be outside their own critical sections.\\
\indent Based on the above reasoning, we determine that the synchronization
skeleton for process $i$ in the aforementioned three-process program
(call it $\P{i}{jk}$) has the same basic graph structure as $\P{i}{j}$
and $\P{i}{k}$, and an arc label in $\P{i}{jk}$ is a ``composition'' of
the labels of the corresponding arcs in $\P{i}{j}$ and $\P{i}{k}$.
In addition, the initial states $S_{ijk}^0$ of the three-process
program are exactly those states that ``project'' onto initial states
of all three pair-programs ($\SYij$, $\SYik$, and $\SYjk$).

Generalizing the above to the case of an arbitrary interconnection relation
$I$, we see that the skeleton for process $i$ in the $I$-program
(call it $\PP{i}{I}$) has the same basic graph structure as $\P{i}{j}$, and a
transition label in $\PP{i}{I}$ is a ``composition'' of the labels of
the corresponding transitions in $\P{i}{j_1}, \ldots, \P{i}{j_n}$,
where $\{j_1, \ldots, j_n\} = I(i)$, i.e., processes $j_1,\ldots,j_n$
are all the $I$-neighbors of process $i$.
Likewise the set $S_I^0$ of initial states of the $I$-program is
exactly those states all of whose ``projections'' onto all the pairs
in $I$ give initial states of the corresponding pair-program.

We now note that the above discussion does not use in any essential
way the assumption that pair-programs are isomorphic to each other. In
fact, the above argument can still be made if pair-programs are not
isomorphic, provided that they induce the same \emph{local structure} on all
common processes. That is, for pair-programs $\SYij$ and $\SYik$, we
require that $\lP{i}{j} = \lP{i}{k}$, where  $\lP{i}{j}, \lP{i}{k}$
result from removing all arc labels from $\P{i}{j}, \P{i}{k}$ respectively.
Also, the initial state sets of all the pair-programs must be so that
there is at least one $I$-state that projects onto some initial state
of every pair-program (and hence the initial state set of the
$I$-program will be nonempty). We assume, in the sequel, that these
conditions hold. Also, all quoted results from \cite{AE98} have been
reverified to hold in our setting, i.e., when the similarity
assumptions of \cite{AE98} are dropped.

Before formally defining our synthesis method, we need some technical
definitions. 

Since $\P{i}{j}$ and $\PP{i}{I}$ have the same local structure, they
have the same nodes (remember that  $\P{i}{j}$ and $\PP{i}{I}$ are
synchronization skeletons). A node of $\P{i}{j}$, $\PP{i}{I}$ is
a mapping of $\AP_i$ to $\{\ltrue, \lfalse\}$. We will refer to such
nodes as $i$-states.
A state of the pair-program $\SYij$ is a
tuple $(s_i, s_j, v_{ij}^1,\ldots,v_{ij}^m)$ where $s_i, s_j$ are
$i$-states, $j$-states, respectively, and
$v_{ij}^1,\ldots,v_{ij}^m$ give the values of all the variables in
$\SH_{ij}$.
We refer to states of $\P{i}{j} \pl \P{j}{i}$ as
$ij$-states. An $ij$-state inherits the assignments
defined by its component $i$- and $j$-states:
   $s_{ij} (p_i) = s_i(p_i)$, 
   $s_{ij} (p_j) = s_j(p_j)$, 
where $s_{ij} = (s_i, s_j, v_{ij}^1,\ldots,v_{ij}^m)$, and $p_i, p_j$
are arbitrary atomic propositions in $\AP_i$, $\AP_j$, respectively.

We now turn to $I$-programs.  If interconnection relation $I$ has
domain $\edomI$, then we denote an $I$-program by $\SY$. $S_I^0$ is
the set of initial states, and $\PP{i}{I}$ is the synchronization
skeleton for process $i$ ($i \in \edomI$) in this $I$-program.
A state of $\SY$ is a
tuple $(s_{i_1},\ldots,s_{i_K}, v^1,\ldots,v^n)$, where $s_i$, ($i \in
\{i_1,\ldots,i_K\}$) is an $i$-state and $v^1,\ldots,v^n$ give the
values of all the shared variables of the $I$-program (we
assume some fixed ordering of these variables, so that the values
assigned to them are uniquely determined by the list
$v^1,\ldots,v^n$).  We refer to states of an $I$-program as
$I$-states. An
$I$-state inherits the assignments defined by its component $i$-states
$(i \in \{i_1,\ldots,i_K\})$: $s_{ij} (p_i) = s_i(p_i)$, where $s =
(s_{i_1},\ldots,s_{i_K}, v^1,\ldots,v^n)$, and $p_i$ is an arbitrary
atomic proposition in $\AP_i$ $(i \in \{i_1,\ldots,i_K\})$.
We shall usually use $s, t, u$ to denote $I$-states.
If $J \sub I$, then we define a $J$-program exactly like an
$I$-program, but using interconnection relation $J$ instead of
$I$. $J$-state is similarly defined.

Let $s_i$ be an $i$-state. We define a state-to-formula operator
$\stof{s_i}$ that takes an $i$-state $s_i$ as an argument and returns a
propositional formula that characterizes $s_i$ in that $s_i \sat
\stof{s_i}$, and $s'_i \not\sat \stof{s_i}$ for all $i$-states $s'_i$
such that $s'_i \neq s_i$:
\label{MP-def-F}
$\stof{s_i} = ({\bigwedge}_{s_i(p_i) = true} p_i) ~\land~
            ({\bigwedge}_{s_i(p_i) = false} \neg p_i)$,
where $p_i$ ranges over the members of $\AP_i$.
$\stof{s_{ij}}$ is defined similarly.
We define the {\bf state projection operator} $\up $. This operator has
several variants.
First of all, we define
projection onto a single process from both $I$-states and $ij$-states:
if $s = (s_{i_1},\ldots,s_{i_K}, v^1,\ldots,v^n)$, then
   $s \up i = s_i$, and
if $s_{ij} = (s_i, s_j, v_{ij}^1,\ldots,v_{ij}^m)$, then
   $s_{ij} \up i = s_i$.
This gives the $i$-state corresponding to the $I$-state $s$, $ij$-state
$s_{ij}$, respectively.
Next we define projection of an $I$-state onto a pair-program:
if $s = (s_{i_1},\ldots,s_{i_K}, v^1,\ldots,v^n)$, then $s \up ij =
(s_i, s_j, v_{ij}^1,\ldots,v_{ij}^m)$, where
$v_{ij}^1,\ldots,v_{ij}^m$ are those values from $v^1,\ldots,v^n$ that
denote values of variables in $\SH_{ij}$.  This gives the $ij$-state
corresponding to the $I$-state $s$, and is well defined only when $i
\, I \, j$.
We also define projection onto the shared variables in $\SH_{ij}$ from
both $ij$-states and $I$-states: 
if $s_{ij} = (s_i, s_j, v_{ij}^1,\ldots,v_{ij}^m)$, then
   $s_{ij} \up \SH_{ij} = (v_{ij}^1,\ldots,v_{ij}^m)$, and
if $s = (s_{i_1},\ldots,s_{i_K}, v^1,\ldots,v^n)$, then
   $s \up \SH_{ij} = (v_{ij}^1,\ldots,v_{ij}^m)$, where
$v_{ij}^1,\ldots,v_{ij}^m$ are those values from $v^1,\ldots,v^n$ that
denote values of variables in $\SH_{ij}$.
Finally, we define projection of an $I$-state onto a $J$-program. 
If $s = (s_{i_1},\ldots,s_{i_K}, v^1,\ldots,v^n)$, then $s \up J =
(s_{j_1}, \ldots, s_{j_L}, v_J^1,\ldots,v_J^m)$, where
$\{j_1, \ldots, j_L\}$ is the domain of $J$, and 
$v_J^1,\ldots,v_J^m$ are those values from $v^1,\ldots,v^n$ that
denote values of variables in $\UN_{(i,j) \in J} \SH_{ij}$.
This gives the $J$-state (defined analogously to an $I$-state)
corresponding to the $I$-state $s$ and is well defined only when
$J \subseteq I$.

To define projection for paths, we first extend the definition of path (and
fullpath) to include the index of the process making the transition,
e.g., each transition is labeled by an index denoting this process.
For example, a path in $M_I$ would be represented as $s^1 \la{d_1}
s^{2} \cdots s^n \la{d_n} s^{n+1} \la{d_{n+1}} s^{n+2} \cdots$, where
$\fa m \geq 1 \d (d_m \in \domI)$.
Let $\pi$ be an arbitrary path in $M_I$. For any $J$ such that $J
\subseteq I$, define a {\em $J$-block} (cf.  \citeN{CGB86} and \citeN{BCG88}) of
$\pi$ to be a maximal subsequence of $\pi$ that starts and ends in a
state and does not contain a transition by any $P_i$ such that $i \in
dom(J)$.  Thus we can consider $\pi$ to be a sequence of $J$-blocks
with successive $J$-blocks linked by a single $P_i$-transition such that
$i \in dom(J)$ (note that a $J$-block can consist of a single state).
It also follows that $s \up J = t \up J$ for any pair of states $s, t$
in the same $J$-block. This is because a transition that is not by
some $P_i$ such that $i \in dom(J)$ cannot affect any atomic
proposition in $\UN_{i \in dom(J)} \AP_i$, nor can it change the value
of a variable in $\UN_{(i,j) \in J} \SH_{ij}$; and a $J$-block
contains no such $P_i$ transition.  Thus, if $B$ is a $J$-block, we
define $B \up J$ to be $s \up J$ for some state $s$ in $B$.
We now give the formal definition of path projection. 
We use the same notation $(\up)$ as for state projection.
Let $B^n$ denote the $n$th $J$-block of $\pi$.

\bd[Path projection]
\label{def:pproj}
Let $\pi$ be $B^1 \la{d_1} \cdots B^n \la{d_n} B^{n+1} \cdots$
where $B^m$ is a $J$-block for all $m \geq 1$.
Then the {\em Path Projection Operator $\up J$} is given by:
$\pi \up J = 
B^1 \up J \la{d_1} \cdots B^n \up J \la{d_n} B^{n+1} \up J \cdots$
\ed

Thus there is a one-to-one correspondence between $J$-blocks of $\pi$
and states of $\pi \up J$, with the $n$th $J$-block of $\pi$
corresponding to the $n$th state of $\pi \up J$ (note that path
projection is well defined when $\pi$ is finite).

The above discussion leads to the following definition of the synthesis
method, which shows how an $I$-process $\PP{i}{I}$ of the
$I$-program $\SY$ is derived from the pair-processes
$\{\P{i}{j} ~|~ j \in I(i)\}$ of the 
the pair-programs
$\{\SYij ~|~ j \in I(i)\}$:

\bd[Pairwise synthesis]
An {\em $I$-process} $\PP{i}{I}$ is derived from the
pair-processes $\P{i}{j}$, for all $j \in I(i)$ as follows: \ss\\
\halfind $\PP{i}{I}$ contains a move from $s_i$ to $t_i$ with label
           $\cmp \gadd \CB{i}{j}{\l} \ar \CA{i}{j}{\l}$ \\
iff \\
\halfind for every $j$ in $I(i)$$:$ $\P{i}{j}$ contains a move from $s_i$ to $t_i$
     with label
      $\gadd \CB{i}{j}{\l} \ar \CA{i}{j}{\l}$. \\[\medskipamount]
The {\em initial state set} $S_I^0$ of the $I$-program is derived from
the initial state $S_{ij}^0$ of the pair-program as follows: \ss\\
\hspace*{\fill}
$S_I^0 = \{ s ~|~ \fa (i,j) \in I \d (s \up ij \in S_{ij}^0) \}$.
\hspace*{\fill}
\label{def:comp-pair-syn}
\ed
Here $\gd$ and $\gc$ are guarded command ``disjunction'' and
``conjunction,'' respectively.
Roughly, the operational semantics of 
   $\CB{i}{j}{1} \ar \CA{i}{j}{1} \gd \CB{i}{j}{2} \ar \CA{i}{j}{2}$
is that if one of the guards
$\CB{i}{j}{1}, \CB{i}{j}{2}$ evaluates to true, then the corresponding body
$\CA{i}{j}{1}, \CA{i}{j}{2}$ respectively,
can be executed. If neither $\CB{i}{j}{1}$ nor $\CB{i}{j}{2}$ evaluates to
true, then the command ``blocks,'' i.e., waits until one of
$\CB{i}{j}{1}, \CB{i}{j}{2}$ evaluates to true.\footnote{This interpretation
was proposed by \citeN{Dij82}.}
We call an arc whose label has the form $\gadd \CB{i}{j}{\l} \ar
\CA{i}{j}{\l}$ a \emph{pair-move}.  In compact notation, a
pair-process has at most one move between any pair of local states.

The operational semantics of 
   $\CB{i}{j}{1} \ar \CA{i}{j}{1} \gc \CB{i}{j}{2} \ar \CA{i}{j}{2}$
is that if both of the guards
$\CB{i}{j}{1}, \CB{i}{j}{2}$ evaluate to true, then the bodies
$\CA{i}{j}{1}, \CA{i}{j}{2}$
can be executed in parallel. If at least one of $\CB{i}{j}{1}$, $\CB{i}{j}{2}$
evaluates to false, then the command ``blocks,'' i.e., waits until
both of $\CB{i}{j}{1}, \CB{i}{j}{2}$ evaluate to true.
We call an arc whose label has the form $\cmp \gadd \CB{i}{j}{\l} \ar
\CA{i}{j}{\l}$ an \emph{$I$-move}.  In compact notation, an
$I$-process has at most one move between any pair of local states.

The above definition is, in effect, a {\em syntactic
transformation} that can be carried out in linear time and space (in
both $\SYij$ and $I$).  In particular, we avoid explicitly
constructing the global state transition diagram of $\SY$, which is of
size exponential in $K=|\edomI|$.

Let $M_{ij}, M_I$ be the global state transition diagrams of $\SYij,
\SY$, respectively. The technical definitions are given below,
and follow the operational semantics given in 
Section~\ref{sec:model}.

\bd[Pair-structure]
\label{def:pair-structure}
Let $i\,I\,j$. The semantics of 
$\SYij$ is given by the \intrdef{pair-structure}
$M_{ij} = ( S^0_{ij}, S_{ij}, R_{ij} )$ where
\bn

\item
$S_{ij}$ is a set of $ij$-states,

\item
$S^0_{ij} \sub S_{ij}$ gives the initial states of $\SYij$, and

\item
$R_{ij} \subseteq S_{ij} \times \{i,j\} \times S_{ij}$ is a
transition relation giving the transitions of $\SYij$.
A transition $(s_{ij}, h, t_{ij})$ by $\P{h}{\bar{h}}$ is in $R_{ij}$
if and only if all of the following hold:
   \bn
   \item $h \in \{i,j\}$,

   \item $s_{ij}$ and $t_{ij}$ are $ij$-states, and

   \item there exists a move 
           $(s_{ij} \up h,
             {\oplus}_{\l \in [1:n]}
                  \CB{h}{\bar{h}}{\l} \ar \CA{h}{\bar{h}}{\l},
             t_{ij} \up h)$
         in $\P{h}{\bar{h}}$ such that there exists $m \in [1:n]$:
      \bn
      \item[(i)] $s_{ij}(\CB{h}{\bar{h}}{m}) = true$,

      \item[(ii)] $< s_{ij} \up \SH_{ij} >
                  \CA{h}{\bar{h}}{m}
                  < t_{ij} \up \SH_{ij} >$, and

      \item[(iii)] $s_{ij} \up \bar{h} = t_{ij} \up \bar{h}$.
      \en
   \en
Here $\bar{h} = i$ if $h = j$ and $\bar{h} = j$ if $h = i$.
\en
\ed

In a transition $(s_{ij}, h, t_{ij})$, we say that $s_{ij}$ is the
{\em start} state and that $t_{ij}$ is the {\em finish} state.
The transition $(s_{ij}, h, t_{ij})$ is called a $\P{h}{\bar{h}}$-transition.
In the sequel, we use $s_{ij} \la{h} t_{ij}$ as
an alternative notation for the transition $(s_{ij}, h, t_{ij})$.
$<\ s_{ij} \up \SH_{ij} > A < t_{ij} \up \SH_{ij} >$
is Hoare triple notation \cite{Ho69} for total correctness,
which in this case means that execution of $A$ always
terminates,\footnote{Termination is obvious, since the right-hand side
of $A$ is a list of constants.}
and, when the shared variables in $\SH_{ij}$ have the values
assigned by $s_{ij}$, leaves these variables with the values assigned by $t_{ij}$.
$s_{ij}(\B{h}{\bar{h}}) = true$ states that the value of guard $\B{h}{\bar{h}}$ in
state $s_{ij}$ is $true$.\footnote{$s_{ij}(\B{h}{\bar{h}})$ is defined
by the usual inductive scheme:
$s_{ij}($``$x_{ij} = h_{ij}$"$) = true$ iff $s_{ij}(x_{ij}) = h_{ij}$,
$s_{ij}(B1_{h}^{\bar{h}} \land B2_{h}^{\bar{h}}) = true$ iff
        $s_{ij}(B1_{h}^{\bar{h}}) = true$ and $s_{ij}(B2_{h}^{\bar{h}}) = true$,
$s_{ij}(\neg B1_{h}^{\bar{h}}) = true$ iff $s_{ij}(B1_{h}^{\bar{h}}) = false$.}
We consider that $\SYij$ possesses a
correctness property expressed by an $\CTLS$ formula 
$f_{ij}$ if and only if $M_{ij}, S_{ij}^0 \sat f_{ij}$.

The semantics of $\SY$ is given by the global state transition diagram
$M_I$ generated by its execution.  We call the global state transition
diagram of an $I$-system an {\em $I$-structure}.

\bd[$I$-structure]
\label{def:static:I-structure}
The semantics of $\SY$ is given by the
\intrdef{$I$-structure} $M_I = (S_I^0, S_I, R_I)$ where
\bn

\item $S_I$ is a set of $I$-states,

\item $S_I^0 \sub S_I$ gives the initial states of $\SY$, and

\item $R_I \subseteq S_I \times \domI \times S_I$ is a
transition relation giving the transitions of $\SY$.
A transition $(s, i, t)$ by $\PP{i}{I}$ is in $R_I$ if and only if
   \bn
   \item $i \in \domI$,

   \item $s$ and $t$ are $I$-states, and

   \item there exists a move
           $(s \up i,
             \cmp {\oplus}_{\l \in [1:n]}
                    \CB{i}{j}{\l} \ar \CA{i}{j}{\l}, 
             t \up i)$
         in $\PP{i}{I}$ such that all of the following hold:
      \bn
      \item[(i)] for all $j$ in $I(i)$, there exists $m \in [1:n]$:\\
         \ind $s \up ij(\CB{i}{j}{m}) = true$ and
              $< s \up \SH_{ij} > \CA{i}{j}{m} < t \up \SH_{ij} >$,

      \item[(ii)] for all $j$ in $\domI - \{i\}$: $s \up j = t \up j$, and

      \item[(iii)] for all $j,k$ in $\domI - \{i\}$, $j\,I\,k$:
                      $s \up \SH_{jk} = t \up \SH_{jk}$.
      \en
   \en
\en
\ed

In a transition $(s, i, t)$, we say that $s$ is the {\em start} state,
and $t$ is the {\em finish} state.  The transition $(s, i, t)$ is
called a $\PP{i}{I}$-transition.  In the sequel, we use $s \la{i} t$
as alternative notation for the transition $(s, i, t)$.  Also, if $I$
is set to $\{ \{i,j\} \}$ in Definition~\ref{def:static:I-structure}, then the
result is, as expected, the pair-structure
definition~(\ref{def:pair-structure}). In other words, the two definitions
are consistent. Furthermore, the semantics of a $J$-system,$J
\subseteq I$  is
given by the $J$-structure $M_J =(S^0_J, S_J, R_J)$, which is obtained
by using $J$ for $I$ in
Definition~\ref{def:static:I-structure}.

As $M_I$ gives the semantics of $\SY$, 
we consider that $\SY$ possesses a
correctness property expressed by a  formula $\andt{k}{\l}
f_{k\l}$ if and only if $M_I, S_I^0 \sat \andt{k}{\l} f_{k\l}$, 
i.e., $M_I, S_I^0 \sat  \fa (i,j) \in I \d (f_{ij})$.

$M_{ij}$ and $M_I$ can be interpreted
as $\CTLS$ structures. We call $M_{ij}$ a \intr{pair-structure}, since
it gives the semantics of a pair-program, ad $M_I$ an
\intr{$I$-structure}, since it gives the semantics of an $I$-program.
We state our main soundness result below by
relating the $\ACTL$ formulae that hold in $M_I$ to those that hold in $M_{ij}$.

This characterization of transitions in the $I$-program as compositions
of transitions in all the relevant pair-programs is formalized in
the transition mapping lemma:

\bl[Transition mapping \textup{\cite{AE98}}]
\label{lem:static:trans-map}
For all $I$-states $s, t \in S_I$ and $i \in \domI$,
$s \la{i} t \in R_I ~\mathrm{iff}:$\\
\ind \ind $\fa j \in I(i) \d
                  ( s \up ij \la{i} t \up ij \in R_{ij} )$  and\\
\ind \ind $\fa j \in \edomI - \hat{I}(i) \d
                  ( s \up j = t \up j )$  and\\
\ind \ind $\fa j,k \in \edomI - \{i\}, j\,I\,k \d
                  ( s \up \SH_{jk} = t \up \SH_{jk} )$.
\el
\bpr
This was established in \cite{AE98} as Lemma 6.4.1. The proof there
did not assume that the $M_{ij}$ are isomorphic. Hence, it carries
over to the setting of this paper.
\epr

In similar manner, we establish:

\bco[Transition mapping \textup{\cite{AE98}}]
\label{cor:static:trans-map}
Let $J \subseteq I$ and $i \in \domJ$.
If  $s \la{i} t \in R_I$, then $s \up J \la{i} t \up J \in R_J$.
\eco

By applying the transition-mapping corollary to every transition along
a path $\pi$ in $M_I$, we show that $\pi \up J$ is a path in $M_J$.
Again, the proof carries over from \cite{AE98}.

\bl[Path mapping \textup{\cite{AE98}}]
\label{lem:static:gen-pmap}
Let $J \subseteq I$. If $\pi$ is a path in $M_I$,
   then $\pi \up J$ is a path in $M_J$.
\el

In particular, when $J = \{ (i,j,\spec_{ij})\}$, Lemma~\ref{lem:static:gen-pmap} forms
the basis for our soundness proof, since it relates computations of
the synthesized program to computations of the pair-programs.

Since every reachable state lies at the end of some initialized path,
we can use the path-mapping corollary to relate reachable states in
$M_I$ to their projections in $M_J$:

\bco[State mapping \textup{\cite{AE98}}]
\label{cor:static:reach}
Let $J \subseteq I$. If $t$ is a reachable state in $M_I$,
   then $t \up J$ is a reachable state in $M_J$.
\eco

\section{Soundness of the Method for Static Programs}
\label{sec:static:soundness}

\subsection{Deadlock-freedom}
\label{sec:static:deadlock}

As we showed in \cite{AE98}, it is possible for the synthesized
program $P$ to be deadlock-prone even though all the pair-programs are
deadlock-free.
To ensure deadlock-freedom of $P$, we imposed a condition on the
``blocking behavior'' of processes: after a process executes a move, it
must either have another move enabled, or it must not be blocking any
other process. 
In general, any behavioral condition which prevents the occurrence of
certain patterns of blocking (``supercycles'') is sufficient.

We formalize our notion of blocking behavior by the notion of
\intr{wait-for-graph}.
The wait-for-graph in a particular $I$-state $s$ contains as nodes all
the processes, and all the moves whose start state is a component of
$s$. These moves have an outgoing edge to every process which blocks
them.

\bd[Wait-for-graph  $W_I(s)$]
\label{def:static:wait-for-graph}
Let $s$ be an arbitrary $I$-state. The {\em wait-for-graph} $W_I(s)$ of $s$
is a directed bipartite graph, where
\begin{nlst1}

\item the nodes of $W_I(s)$ are
   \begin{nlst2}
   \item the $I$-processes $\{ \PP{i}{I} ~|~ i \in \domI \}$, and
   \item the moves
             $\{ a_i^I ~|~ i \in \domI \mbox{~and~} a_i^I \in \PP{i}{I}
                           \mbox{~and~} s \up i = a_i^I.start \}$
   \end{nlst2}

\item there is an edge from $\PP{i}{I}$ to every node of the form
      $a_i^I$ in $W_I(s)$, and

\item there is an edge from $a_i^I$ to $\PP{j}{I}$ in $W_I(s)$ if and
                  only if
                  $i\,I\,j$ and $a_i^I \in W_I(s)$ and
                  $s \up ij(a_i^I.guard_j) = false$.
\end{nlst1}
\ed

Here $a_i^I.guard_j$ is the conjunct of the guard of move $a_i^I$
which is evaluated over the (pairwise) shared state with $\PP{j}{I}$.
We characterize a deadlock as the occurrence in the wait-for-graph of
a graph-theoretic construct that we call a {\em supercycle}:

\bd[Supercycle]
\label{def:supercycle}
$SC$ is a supercycle in $W_I(s)$ if and only if all of the following hold:
\begin{nlst1}
   \item $SC$ is nonempty,
   \item if $\PP{i}{I} \in SC$ then for all $a_i^I$ such that
$a_i^I \in W_I(s)$, $\PP{i}{I} \lra a_i^I \in SC$, and
   \item if $a_i^I \in  SC$ then there exists $\PP{j}{I}$ such that
$a_i^I \lra \PP{j}{I} \in W_I(s)$ and $a_i^I \lra \PP{j}{I} \in SC$.
\end{nlst1}
\ed
Note that this definition implies that $SC$ is a subgraph of $W_I(s)$.

Our conditions will be stated over ``small'' programs, i.e,. programs that result
from compositing a small number of processes together. To then infer that the
large program $P$ has similar behavior, we use the following proposition.

\bp[Wait-for-graph projection]
\label{prop:static:waitfg-proj}
Let $J \subseteq I$ and $i\,J\,j$. Furthermore, let $s_I$ be an arbitrary $I$-state. Then
\bn

\item
$\PP{i}{I} \lra a_i^I \in W_I(s_I)$ iff
$\PP{i}{J} \lra a_i^J \in W_J(s_I \up J)$, and

\item
$a_i^I \lra \PP{j}{I} \in W_I(s_I)$ iff
$a_i^J \lra \PP{j}{J} \in W_J(s_I \up J)$.

\en
\ep
\begin{proof}
By assumption, $i\,J\,j$ and $J \subseteq I$. Hence $i\,I\,j$.

\ss

Proof of clause (1).
By the wait-for-graph definition~(\ref{def:static:wait-for-graph}),
       $\PP{i}{I} \lra a_i^I \in W_I(s_I)$ iff
       $s_I \up i = a_i^I.start$.
Since $i \in \domJ$, we have $(s_I \up J) \up i = s_I \up i$ by
definition of $\up J$. Thus 
       $s_I \up i  = a_i^I.start$ iff
       $(s_I \up J) \up i  = a_i^J.start$
(since $a_i^I.start = a_i^J.start = s_i$).
Finally, by the wait-for-graph definition~(\ref{def:static:wait-for-graph}) and $i\,J\,j$,
       $(s_I \up J) \up i  = a_i^J.start$ iff
       $\PP{i}{J} \lra a_i^J \in W_J(s_I \up J)$.
These three equivalences together yield clause (1) (using
transitivity of equivalence).

\ss

Proof of clause (2).
By the wait-for-graph definition~(\ref{def:static:wait-for-graph}),
       $a_i^I \lra \PP{j}{I} \!\in\! W_I(s_I)$ iff
       $s \up ij \!\not\sat\! a_i^I.guard_j$.
Since $i\,J\,j$, we have $(s_I \up J) \up ij = s_I \up ij$ by
definition of $\up J$.
Also, $a_i^I.guard_j = a_i^J.guard_j = \bigor_{\l \in [1:n]} \CB{i}{j}{\l}$.
Thus 
       $s_I \up ij \not\sat a_i^I.guard_j$ iff
       $(s_I \up J) \up ij \not\sat  a_i^J.guard_j$
Finally, by the wait-for-graph definition~(\ref{def:static:wait-for-graph}) and $i\,J\,j$,
       $(s_I \up J) \up ij \!\not\sat\! a_i^J.guard_j$ iff
       $a_i^J \lra \PP{j}{J} \in W_J(s_I \up J)$.
These three equivalences together yield clause (2), (using
transitivity of equivalence, and noting that $s \not\sat B$ and $s(B)
= false$ have identical meaning).
\end{proof}

\subsubsection{The Wait-for-graph Condition}
\label{sec:static:wait-for-cond}

In \cite{AE98}, we give a criterion, the wait-for-graph assumption,
which can be evaluated over the product of a small number of
processes, thereby avoiding state-explosion. We show there that if the
wait-for-graph assumption holds, then $W_I(s)$ cannot contain a
supercycle for any reachable state $s$ of $M_I$.  The wait-for-graph
condition embodies the requirement that, after a process executes a
move, it must either have another move enabled, or it must not be
blocking any other process.

\bd[Static wait-for-graph condition]
\label{def:static:wait-for-cond}
Let $t_k$ be an arbitrary reachable local state of $\P{k}{\l}$ in $M_{k\l}$
for all $\l \in I(k)$, and let $n = |t_k.moves|$.
Also let $J$ be an arbitrary interconnection relation 
such that $J \sub I$ and $J$ has the form
$\{ (j,k,\spec_{jk}), 
    (k,\l_1,\spec_{k\l_1}),\ldots, (k,\l_n,\spec_{k\l_n}) \}$,
where 
$k \not\in \{j, \l_1,\ldots,\l_n\}$.
Then, for every reachable $J$-state $t_J$ in $M_J$ such that
                 $t_J \up k = t_k$
             and $s_J \la{k} t_J \in R_J$ for some reachable $J$-state $s_J$,
      we have \ms\\
\hspace*{\fill}
	$\fa a^J_j \d (a_j^J \lra \PP{k}{J} \not\in W_J(t_J))$
\hspace*{\fill} \\
or \\
\hspace*{\fill}
        $\ex a_k^J \in W_J(t_J) \d
               (\fa \l \in \{\l_1,\ldots,\l_n\} \d
                      a_k^J \lra \PP{\l}{J} \not\in W_J(t_J))$.
\hspace*{\fill}
\ed

\bt[Static supercycle-free wait-for-graph]
\label{thm:static:supercycle-free-waitfor}
If the wait-for-graph condition holds, and $W_I(s_I^0)$ is
supercycle-free for every initial state $s_I^0 \in S_I^0$, then for
every reachable state $t$ of $M_I$, $W_I(t)$ is supercycle-free.
\et
\begin{proof}
Let $t$ be an arbitrary reachable state of $M_I$, and let $s$ be an
arbitrary reachable state of $M_I$ such that $s \la{k} t$ for some $k \in
\domI$. We shall establish that\ms\\
\hspace*{\fill}
	if $W_I(t)$ is supercyclic, then $W_I(s)$ is supercyclic.
\hfill{\makebox[0in][r]{(P1)}}\ms\\ 
The contrapositive of P1 together with the assumption that
$W_I(s_I^0)$ is supercycle-free for all $s_I^0 \in S_I^0$ is
sufficient to establish the conclusion of the theorem (by induction on
the length of a path from some $s^0_I \in S_I^0$ to $t$).

We say that an edge is {\em $k$-incident\/} iff at least one of its vertices
is $\PP{k}{I}$ or $a_k^I$. The following (P2) will be useful in
proving P1 \ms\\
\hspace*{\fill}
	if edge $e$ is not $k$-incident, then $e \in W_I(t)$ iff  $e \in W_I(s)$.
\hfill{\makebox[0in][r]{(P2)}}\ms

Proof of P2. 
If $e$ is not $k$-incident, then, by the wait-for-graph
definition~(\ref{def:static:wait-for-graph}),  either $e = \PP{h}{I} \lra a_h^I$,
or $e = a_h^I \lra \PP{\l}{I}$, for some $h,\l$ such that $h \neq k, \l \neq k$.
From $h \neq k, \l \neq k$ and $s \la{k} t \in R_I$, we have
      $s \up h = t \up h$ and
      $s \up h\l = t \up h\l$
by the wait-for-graph definition~(\ref{def:static:wait-for-graph}).
Since $e \in W_I(t), e \in W_I(s)$ are determined solely by $t \up h\l,
s \up h\l$ respectively, (see the wait-for-graph
definition~(\ref{def:static:wait-for-graph}), P2 follows.
(End proof of P2.) \ms

Let $v$ be a vertex in a supercycle $SC$.  We define $depth_{SC}(v)$
to be the length of the longest backward path in $SC$ which starts in
$v$.  If there exists an infinite backward path (i.e., one that
traverses a cycle) in $SC$ starting in $v$, then $depth_{SC}(v) =
\omega$ ($\omega$ for ``infinity''). We now establish that\ms\\
\hspace*{\fill}
	every supercycle $SC$ contains at least one cycle.
\hfill{\makebox[0in][r]{(P3)}}\ms

Proof of P3.
Suppose P3 does not hold, and $SC$ is a supercycle containing no
cycles.
Therefore, all backward paths in $SC$ are finite, and so
by definition of $depth_{SC}$ all vertices of $SC$ have
finite depth. Thus, there is at least one vertex $v$ in $SC$ with
maximal depth. But, by definition of $depth_{SC}$, $v$ has no
successors in $SC$, which, by the supercycle definition~(\ref{def:supercycle}),
contradicts the assumption that $SC$ is a supercycle.
(End proof of P3.) \ms

Our final prerequisite for the proof of P1 is \ms\\
\hspace*{\fill}
\parbox[b]{\parboxwidth}{
if $SC$ is a supercycle in $W_I(s)$,
then the graph $SC'$ obtained from $SC$ by removing all vertices of finite
depth from $SC$ (along with incident edges) is also a supercycle
in $W_I(s)$.}
\hfill{\makebox[0in][r]{(P4)}}\ms

Proof of P4.
By P3, $SC' \neq \emptyset$. Thus $SC'$ satisfies clause (1) of the supercycle
definition~(\ref{def:supercycle}).
Let $v$ be an arbitrary vertex of $SC'$.
Thus $v \in SC$ and $depth_{SC}(v) = \omega$ by definition of
$SC'$. Let $w$ be an arbitrary successor of $v$ in $SC$. $depth_{SC}(w)
= \omega$ by definition of $depth$. Hence $w \in SC'$.
Furthermore, $w$ is a successor of $v$ in $SC'$, by definition of $SC'$.
Thus every vertex $v$ of $SC'$
is also a vertex of $SC$, and the successors of $v$ in $SC'$ are
the same as the successors of $v$ in $SC$
Now since $SC$ is a supercycle, every vertex $v$ in $SC$ has enough successors
in $SC$ to satisfy  clauses (2) and (3) of the supercycle
definition~(\ref{def:supercycle}). It follows that
every vertex $v$ in $SC'$ has enough successors
in $SC'$ to satisfy  clauses (2) and (3) of the supercycle
definition~(\ref{def:supercycle}).
(End proof of P4.) \ms

We now present the proof of (P1). We assume the antecedent of P1 and
establish the consequent. Let $SC$ be some supercycle in $W_I(t)$.
Let $SC'$ be the graph obtained from $SC$ by removing all
vertices of finite depth from $SC$ (along with incident edges).
We now show that $P_k^I \not\in SC'$ and that $SC'$ contains
no move vertex of the form $a_k^I$.
There are two cases.

\ms

\textit{Case 1:} $\PP{k}{I} \not\in SC$.
Then obviously $\PP{k}{I} \not\in SC'$.  Now suppose some node of the
form $a_k^I$ is in $SC'$. By definition of $SC'$, we have $a_k^I \in
SC$ and $depth_{SC}(a_k^I) = \omega$.  Hence, by definition of
$depth$, there exists an infinite backward path in $SC$ starting in
$a_k^I$. Thus $a_k^I$ must have a predecessor in $SC$. By the
supercycle definition~(\ref{def:supercycle}), $\PP{k}{I}$ is the
only possible predecessor of $a_k^I$ in $SC$, and hence $\PP{k}{I} \in
SC$, contrary to the case assumption.  We therefore conclude that
$SC'$ contains no vertices of the form $a_k^I$. (End of case 1.)

\ms

\textit{Case 2:} $\PP{k}{I} \in SC$.
By the supercycle definition~(\ref{def:supercycle}), \ms\\
\pind  $\fa a_k^I \in W_I(t) \d ( \ex \l \d (
               a_k^I \lra \PP{\l}{I} \in W_I(t) ) )$. \hfill{(a)}\ms\\
Since there are exactly $n$ moves $a_k^I$ of process $P_k^I$ in $W_I(t)$
($n = |t_k.moves|$), we can select $\l_1,\ldots,\l_n$ (where $\l_1, \ldots,
\l_n$ are not necessarily pairwise distinct) such that \ms\\
\pind  $\fa a_k^I \in W_I(t) \d  ( \ex \l \in \{\l_1,\ldots,\l_n\} \d (
               a_k^I \lra \PP{\l}{I} \in W_I(t) ) )$. \hfill{(b)}\ms\\
Now let $J = \{ \{j,k\}, \{k,\l_1\},\ldots,\{k,\l_n\} \}$ where
$j$ is an arbitrary element of $I(k)$.
Applying the wait-for-graph projection proposition~(\ref{prop:static:waitfg-proj})
to (b) gives us \ms\\
\pind  $\fa a_k^J \in W_J(t \up J) \d  ( \ex \l \in \{\l_1,\ldots,\l_n\} \d (
                     a_k^J \lra \PP{\l}{J} \in W_J(t \up J) ) )$.
                                                     \hfill{(c)}\ms\\
Now $s \la{k} t \in R_I$ by assumption. Hence $s \up J \la{k} t \up J
\in R_J$ by the transition-mapping
corollary~(\ref{cor:static:trans-map}). Also, by the state-mapping
corollary~(\ref{cor:static:reach}) $s \up J$ is reachable in $M_J$, since
$s$ is reachable in $M_I$. Thus we can apply the wait-for-graph
assumption to $t \up J$ to get \ms\\
\pind $\fa a_j^J \d ( a_j^J \lra \PP{k}{J} \not\in W_J(t \up J) )$ \\
or\\
\pind $\ex a_k^J \in W_J(t \up J) \d ( \fa \l \in \{\l_1,\ldots,\l_n\}\d (
             a_k^J \lra \PP{\l}{J} \not\in W_J(t \up J) ) )$.
                                                      \hfill{(d)}\ms\\
Now (c) contradicts the second disjunct of (d). Hence \ms\\
\pind  $\fa a_j^J \d ( a_j^J \lra \PP{k}{J} \not\in W_J(t \up J) )$, \ms\\
and applying the wait-for-graph projection
proposition~(\ref{prop:static:waitfg-proj}) to this gives us \ms\\
\pind   $\fa a_j^J \d ( a_j^I \lra \PP{k}{I} \not\in W_I(t) )$. \ms\\
Since $j$ is an arbitrary element of $I(k)$, we conclude that
$\PP{k}{I}$ has no incoming edges in $W_I(t)$.
Thus, by definition of $depth$, $depth_{SC}(\PP{k}{I}) = 0$, and so
$\PP{k}{I} \not\in SC'$.

Now suppose some node of the form $a_k^I$ is in $SC'$. By definition
of $SC'$, we have $a_k^I \in SC$ and $depth_{SC}(a_k^I) = \omega$.
Hence, by definition of $depth$, there exists an infinite backward
path in $SC$ starting in $a_k^I$.  Thus $a_k^I$ must have a
predecessor in $SC$. By the supercycle
definition~(\ref{def:supercycle}), $\PP{k}{I}$ is the only possible
predecessor of $a_k^I$ in $SC$, and hence there exists an infinite
backward path in $SC$ starting in $P_k^I$. Thus $depth_{SC}(P_k^I) =
\omega$ by definition of $depth$. But we have established
$depth_{SC}(\PP{k}{I}) = 0$, so we conclude that $SC'$ contains no
vertices of the form $a_k^I$. (End of case 2.)

\ms

In both cases, $P_k^I \not\in SC'$, and $SC'$ contains no move vertex
of the form $a_k^I$.  Thus every edge of $SC'$ is not
$k$-incident. Hence, by P2, every edge of $SC'$ is an edge of $W_I(s)$
(since $SC' \sub W_I(t)$).  By P4, $SC'$ is a supercycle, so
$W_I(s)$ is supercyclic. Thus P1 is established, which establishes the
theorem.
\end{proof}

\subsubsection{Establishing Deadlock-freedom}

We show that the absence of supercycles in the wait-for-graph of a
state implies that there is at least one enabled move in that state.

\bp[Supercycle \cite{AE98}]
\label{prop:static:supercycle}
If $W_I(s)$ is supercycle-free, then
     some move $a_i^I$ has no outgoing edges in $W_I(s)$.
\ep
\begin{proof}
We establish the contrapositive. Since every local
state of a process has at least one outgoing arc
(Section~\ref{sec:model}), there exists at least one move of
the form $a_i^I$ for every $i \in \domI$ in $W_I(s)$. Suppose that
every such move has at least one outgoing edge in $W_I(s)$. Consider
the subgraph $SC$ of $W_I(s)$ consisting of these edges together with
all edges of the form $\PP{i}{I} \lra a_i^I$ in $W_I(s)$. By the
wait-for-graph definition~(\ref{def:static:wait-for-graph}), and the supercycle
definition~(\ref{def:supercycle}), it is clear that $SC$ is a
supercycle in $W_I(s)$. Thus $W_I(s)$ is not supercycle-free.
\end{proof}

\bp[Move enablement]
\label{prop:static:move-enablement}
Let $s$ be an arbitrary $I$-state such that $s \up i = a_i^I.start$.
If $a_i^I$ has no outgoing edges in $W_I(s)$, then $a_i^I$ can be executed in state
$s$.
\ep
\bpr
If $a_i^I$ has no outgoing edges in $W_I(s)$, then by
the wait-for-graph definition~(\ref{def:static:wait-for-graph}),
$s \up ij(a_i^I.guard_j) = true$
for all $j \in I(i)$. Hence, by the $I$-structure
definition~(\ref{def:static:I-structure}), $a_i^I$ can be executed in state
$s$.
\epr

\bt[Deadlock freedom \cite{AE98}]
\label{thm:static:dead-free}
If, for every reachable state $s$ of $M_I$, $W_I(s)$ is supercycle-free, then
$M_I, S_I^0 \sat \AG \EX \ltrue$.
\et
\begin{proof}
Let $s$ be an arbitrary reachable state of $M_I$. By the antecedent,
$W_I(s)$ is supercycle-free. Hence, by
the supercycle proposition~(\ref{prop:static:supercycle}), some move $a_i^I$ has no
outgoing edges in $W_I(s)$. By Proposition~\ref{prop:static:move-enablement}, 
$a_i^I$ can be executed in state $s$.
Since $s$ is an arbitrary reachable state of $M_I$, we
conclude that every reachable state of $M_I$ has at least one enabled
move $a_i^I$, (where, in general, $a_i^I$ depends on
$s$). Hence $M_I, S_I^0 \sat \AG \EX true$.
\end{proof}

\subsection{Liveness}
\label{sec:static:liveness}

To assure liveness properties of the synthesized programs, we need to
assume a form of weak fairness. Let $\CL(f)$ be the set of all
subformulae of $f$, including $f$ itself. Let $ex_i$ be an assertion
that is true along a transition in a structure iff that transition
results from executing process $i$. We give our fairness criterion as
a formula of the linear time temporal logic PTL \cite{MW84}.

\bd[Sometimes-blocking, $\blk_i^j, \blk_i$]
\label{def:sometimes-blocking}
An $i$-state $s_i$ is \emph{sometimes-blocking in $M_{ij}$} if and
only if:\ss\\
\hspace*{\fill}
$\ex s_{ij}^0 \in S_{ij}^0 \d
  (M_{ij}, s_{ij}^0 \sat
    \EF( \ \stof{s_i} \land 
        (\ex a_j^i \in \P{j}{i} \d (\stof{a_j^i.start} \land \neg a_j^i.guard))
    \ )        
  )$.
\hspace*{\fill} \ms\\
Also,
$\blk_i \df (\OR ~ \stof{s_i} : s_i \mbox{ is sometimes-blocking in } M_{ij})$,
and
$\blk_i \df \OR_{j \in I(i)} \blk_i^j$.
\ed
Note that $a_j^i.start$ is the start state of the two-process move
$a_j^i$, and $a_j^i.guard$ is its guard.

\bd[Weak blocking fairness $\f_b$]
   \mbox{$~~{\f_b} ~\df~  \AND_{i \in \domI} \ea(\blk_i \land en_i) \imp \io ex_i$.}
\label{def:static:weak-blocking-fairness}
\ed

\bd[Pending eventuality, $\pnd_i$]
\label{def:pending-eventuality}
An $ij$-state $s_{ij}$ has a \emph{pending eventuality} if and
only if: \ss\\
\hspace*{\fill}
$\ex f_{ij} \in \CL(\spec_{ij}) :
  (M_{ij}, s_{ij} \sat \neg f_{ij} \land \AF f_{ij})
$.
\hspace*{\fill}\\
Also,
   $\pnd_{ij} \df (\OR ~ \stof{s_{ij}} : s_{ij} \mbox{ has a pending eventuality})$.
\ed
In other words, $s_{ij}$ has a pending eventuality if there is a
subformula of the pair-specification $\spec_{ij}$ which does not hold
in $s_{ij}$, but is guaranteed to eventually hold along every fullpath
of $M_{ij}$ that starts in $s_{ij}$.

\bd $($\textup{\textbf{Weak eventuality fairness,}} $\f_\l$$)$\\
\label{def:static:weak-eventuality-fairness}
\hspace*{\fill}
\mbox{${\f_\l} ~\df~ \AND_{(i,j) \in I}
          (\ea en_i \lor \ea en_j) \land \ea \pnd_{ij} \imp \io (ex_i \lor ex_j)$.}
\hspace*{\fill}
\ed

\vspace{-1ex}

Our overall fairness notion $\f$ is then the conjunction of weak
blocking and weak eventuality fairness: $\f \df \f_b \land \f_\l$.

\bd[Liveness condition for static programs]
\label{def:static:liveness-cond}
For every reachable state $s_{ij}$ in $M_{ij}$,
	$M_{ij}, s_{ij} \sat \A(\G ex_i \imp \ea \aen_j)$,\\
where
$\aen_j \df
	\fa a_j^i \in \P{j}{i} \d (\stof{a_j^i.start} \imp  a_j^i.guard))$.
\ed
$\aen_j$ means that every move of $\P{j}{i}$  whose start state is a
component of the current global state is also enabled in the current
global state. The liveness condition requires, in every pair-program
$\SYij$, that if $\Pij$ can execute continuously along some path, then
there exists a suffix of that path along which $\Pij$ does not block
any move of $\Pji$.

\bl[Progress for static programs]
\label{lem:static:progress}
If
\bn
   \item \label{ass:static:liveness} the liveness condition holds, and
   \item \label{ass:static:supercycle} for every reachable $I$-state $u$, $W_I(u)$ is supercycle-free, and
   \item \label{ass:static:pending} $M_{ij}, s \up ij \sat \neg h_{ij} \land \AF h_{ij}$
		             for some $h_{ij} \in \CL(\spec_{ij})$, then
\en
\hspace*{\fill}
	$M_I, s \sat_{\f} \AF(ex_i \lor ex_j)$
\hspace*{\fill}
\el
\bpr
By assumption~\ref{ass:static:supercycle} and Theorem~\ref{thm:static:dead-free}, 
$M_I, S_I^0 \sat \AG \EX \ltrue$. Hence every fullpath in $M_I$ is infinite.
Let $\pi$ be an arbitrary $\f$-fair fullpath starting in $s$. If $M_I,
\pi \sat \F(ex_i \lor ex_j)$, then we are done. Hence we assume
\bleqn{(*)}
	$\pi \sat \G(\neg ex_i \land \neg ex_j)$
\eleqn
in the remainder of the proof. Now define
	$\psi_{\inf} \df \{k ~|~ \pi \sat \io ex_k\}$
and
	$\psi_{\fin} \df \{k ~|~ \pi \sat \ea \neg ex_k\}$.

Let $\rho$ be a suffix of $\pi$ such that no process in $\psi_{\fin}$
executes along $\rho$, and let $t$ be the first state of $\rho$. Note
that, by (*), $i \in \psi_{\fin}$, $j \in \psi_{\fin}$.

Let $W$ be the portion of $W_I(t)$ induced by starting in $\PP{i}{I},
\PP{j}{I}$ and following wait-for edges that enter processes in
$\psi_{\fin}$ or their moves. By assumption~\ref{ass:static:supercycle}, $W$
is supercycle-free. Hence, there exists a process $\PP{k}{I}$ in $W$
such that $\PP{k}{I}$ has some move $\MV{k}{I}$ with no wait-for edges
to any process in $W$, by Proposition~\ref{prop:static:supercycle}. Hence,
in state $t \up k\l$, $\MV{k}{\l}$ is enabled in all pair-machines $M_{k\l}$
such that $\l \in \psi_{\fin}$, i.e.,
$t \up k\l \sat \stof{\MV{k}{\l}.\start} \land en(\MV{k}{\l})$.
Also, $k \in \psi_{\fin}$, by definition of $W$.
Since $t$ is the first state of $\rho$ and no process in $\psi_{\fin}$
executes along $\rho$, we have from above, that
  $\AND \l \in \psi_{\fin} \ints I(k) : \rho \up k\l \sat \G en(\MV{k}{\l})$.

Now consider a pair-machine $M_{k\l}$ such that $\l \in \psi_{\inf}$
(if any). Hence $\rho \sat \io ex_{\l} \land \G \neg ex_k$, since $k
\in \psi_{\fin}$. Hence $\rho \up k\l \sat \G ex_\l \land \G \neg
ex_k$. By Lemma~\ref{lem:static:gen-pmap}, $\rho \up k\l$ is a path in
$M_{k\l}$. Since $\rho$ is an infinite path and $\rho \sat \io
ex_{\l}$,  $\rho \up k\l$ is an infinite path. Hence 
$\rho \up k\l$ is a fullpath in $M_{k\l}$.
By the liveness condition for static programs
(Definition~\ref{def:static:liveness-cond}), $\rho \up k\l \sat \ea \aen_k$. Now
$t \up k\l \sat \stof{\MV{k}{\l}.\start}$. Since $\rho \up k\l \sat \G \neg
ex_k$, $\PP{k}{\l}$'s local state does not change along $\rho \up
k\l$. Hence $\rho \up k\l \sat \G \stof{\MV{k}{\l}.\start}$.
Hence, by definition of $\aen_k$, $\rho \up k\l \sat \ea en(\MV{k}{\l})$.
Since $\l$ is an arbitrary element of $\psi_{\inf} \ints I(k)$, we have
$\AND \l \in \psi_{\inf} \ints I(k) : \rho \up k\l \sat \ea en(\MV{k}{\l})$.
Since $(\psi_{\inf} \ints I(k)) \un (\psi_{\fin} \ints I(k)) = I(k)$,
we conclude
	$\AND \l \in I(k) : \rho \up k\l \sat \ea en(\MV{k}{\l})$.
By Definitions~\ref{def:pproj} and \ref{def:comp-pair-syn}, we have
$\rho \sat \ea en(\MV{k}{I})$. Hence, we conclude
\bleqn{(a)}
	$\rho \sat \ea en_k$.
\eleqn
Assume $k \not\in \{i,j\}$. Then, by definition of $W$, in state
$t$ $\PP{k}{I}$ blocks some move $\MV{\l}{k}$ of some process
$\PP{\l}{I}$, i.e.,
	$t \sat \stof{\MV{\l}{k}.\start} \land \neg \MV{\l}{k}.\guard$.
By Definition~\ref{def:sometimes-blocking}, $t \up k$ is
sometimes-blocking in $M_{k\l}$ (since $t$ is reachable, so is $t \up k$, by
\cite[Corollary 6.4.5]{AE98}). Hence $t \up k \sat \blk_k^\l$, and so $t
\sat \blk_k^\l$. Now $\rho \sat \G \neg ex_k$. Since $t$ is the first
state of $\rho$, this means that $t \up k = u \up k$ for any state $u$
of $\rho$, i.e., the local state of $\PP{k}{I}$ does not change along
$\rho$. Thus, $\rho \sat \G \blk_k^\l$, since $t \sat \blk_k^\l$. Thus
$\rho \sat \G \blk_k$, by definition of $\blk_k$.
From this and (a), we have $\rho \sat \ea(\blk_k \land en_k)$.
Hence, by weak blocking fairness,
(Definition~\ref{def:static:weak-blocking-fairness}), $\rho \sat \io ex_k$,
which contradicts $\rho \sat \G \neg ex_k$. Hence the assumption $k
\not\in \{i,j\}$ does not hold, and so $k \in \{i,j\}$.

Since $\pi \sat \G(\neg ex_i \land \neg ex_j)$, by assumption (*), and
$s = \first(\pi)$, we have $u \up ij = s \up ij$ for every state $u$
along $\pi$. Now $M_{ij}, s \up ij \sat \neg h_{ij} \land \AF h_{ij}$
for some $h_{ij} \in \CL(\spec_{ij})$ by
assumption~\ref{ass:static:pending}. Hence 
$M_{ij}, u \up ij \sat \neg h_{ij} \land \AF h_{ij}$ for all $u$ along
$\pi$.
Hence $M_{ij}, u \up ij \sat \pnd_{ij}$ for all $u$ along $\pi$ by
Definition~\ref{def:pending-eventuality}.
Hence, $M_I, u \sat \pnd_{ij}$  for all $u$ along $\pi$, since
$\pnd_{ij}$ is purely propositional, and so
$M_I, \pi \sat \G \pnd_{ij}$.
Since $\rho$ is a suffix of $\pi$ and $k \in \{i,j\}$, we conclude from
(a) that $\pi \sat \ea en_i \lor \ea en_j$.
Hence 
$M_I, \pi \sat (\ea en_i \lor \ea en_j) \land \ea \pnd_{ij}$.
By weak eventuality fairness
(Definition~\ref{def:static:weak-eventuality-fairness}), $\pi \sat \io(ex_i
\lor ex_j)$. This contradicts the assumption (*), which is therefore
false. Hence $\pi \sat \F(ex_i \lor ex_j)$. Since $\pi$ is an
arbitrary $\f$-fair fullpath starting in $s$, the lemma follows.  \epr

\subsection{The Large Model Theorem for Static Programs}
\label{sec:static:large-model}

\bt[Large model]
\label{thm:static:large-model}
Let $(i,j,\spec_{ij}) \in I$, where $\spec_{ij} \in \ACTLmij$,
and let $s$ be an arbitrary reachable $I$-state.
If
\bn
\item the liveness condition for static programs holds,

\item $W_I(u)$ is supercycle-free for every reachable $I$-state $u$, and

\item  $M_{ij}, s \up {ij} \sat f_{ij}$ for some $f_{ij} \in \CL(\spec_{ij})$, 
\en
then \ms\\
\hspace*{\fill}
     $M_I, s \satf f_{ij}$.
\hspace*{\fill}
\et
\bpr
The proof is by induction on the structure of $f_{ij}$.  
Throughout, let $s_{ij} = s \up ij$.

\bigskip

$f_{ij} = p_i$, or $f_{ij} = \neg p_i$, where $p_i \in \AP_i$, i.e.,
$p_i$ is an atomic proposition.\\
By definition of $\up ij$, $s$ and $s \up ij$ agree on all atomic
propositions in $\AP_i \un \AP_j$. The result follows.

\bigskip

{$f_{ij} = g_{ij} \land h_{ij}$}.
The antecedent is
$M_{ij}, s_{ij} \sat  g_{ij} \land h_{ij}$.
So, by $\CTLS$ semantics,
     $M_{ij}, s_{ij} \sat  g_{ij}$ and
     $M_{ij}, s_{ij} \sat  h_{ij}$.
Since $f_{ij} \in \CL(\spec_{ij})$, we have $g_{ij} \in
\CL(\spec_{ij})$ and $h_{ij} \in \CL(\spec_{ij})$. Hence, applying
the induction hypothesis, we get
     $M_I, s \satf  g_{ij}$ and
     $M_I, s \satf  h_{ij}$.
So by $\CTLS$ semantics we get
$M_I, s \satf (g_{ij} \land h_{ij})$.

\bigskip

{$f_{ij} = g_{ij} \lor h_{ij}$}.
The antecedent is
$M_{ij}, s_{ij} \sat  g_{ij} \lor h_{ij}$.
So, by $\CTLS$ semantics,
     $M_{ij}, s_{ij} \sat  g_{ij}$ or
     $M_{ij}, s_{ij} \sat  h_{ij}$.
Since $f_{ij} \in \CL(\spec_{ij})$, we have $g_{ij} \in
\CL(\spec_{ij})$ and $h_{ij} \in \CL(\spec_{ij})$. Hence, applying
the induction hypothesis, we get
     $M_I, s \satf  g_{ij}$ or
     $M_I, s \satf  h_{ij}$.
So by $\CTLS$ semantics we get
$M_I, s \satf (g_{ij} \lor h_{ij})$.

\bigskip

{$f_{ij} = \A[g_{ij} \Uw h_{ij}]$}.
Let $\pi$ be an arbitrary $\f$-fair fullpath starting in $s$. We
establish $\pi \sat [g_{ij} \Uw h_{ij}]$. 
By Definition~\ref{def:pproj}, $\pi \up ij$ starts in $s \up ij =
s_{ij}$. Hence, by CTL semantics, $\pi \up ij \sat [g_{ij} \Uw
h_{ij}]$ (note that this holds even if $\pi \up ij$ is not a fullpath,
i.e., is a finite path). We have two cases. 

Case 1:  $\pi \up ij \sat \G g_{ij}$. Let $t$ be an arbitrary state
along $\pi$. By Definition~\ref{def:pproj}, $t \up ij$ lies along
$\pi \up ij$. Hence $t \up ij \sat g_{ij}$. By the induction
hypothesis, $t \sat g_{ij}$. Hence $\pi \sat \G g_{ij}$, since $t$ was
arbitrarily chosen. Hence $\pi \sat [g_{ij} \Uw h_{ij}]$ by \CTLS semantics.

Case 2: $\pi \up ij \sat [g_{ij} \U h_{ij}]$. Let $s^{m'}_{ij}$ be
the first state along $\pi \up ij$ that satisfies $h_{ij}$\footnote{We
use $s^n_{ij}$ to denote the $n'$th state along $\pi \up ij$, i.e.,
$\pi \up ij = s^0_{ij}, s^1_{ij}, \ldots$, and we let $s_{ij} =
s^0_{ij}$.}. By Definition~\ref{def:pproj}, there exists at least one
state $t$ along $\pi$ such that $t \up ij = s^{m'}_{ij}$. Let
$s^{n'}$ be the first such state. By the induction hypothesis,
$s^{n'} \sat h_{ij}$. Let $s^n$ be any state along $\pi$ up to but
not including $s^{n'}$ (i.e., $0 \leq n < n'$). Then, by
Definition~\ref{def:pproj}, $s^n \up ij$ lies along the portion of
$\pi \up ij$ up to, and possibly including, $s^{m'}_{ij}$. That is,
$s^n \up ij = s^m_{ij}$, where $0 \leq m \leq m'$. Now suppose $s^n
\up ij = s^{m'}_{ij}$ (i.e., $m = m'$). Then, by $s^{m'}_{ij} \sat
h_{ij}$ and the induction hypothesis, $s^n \sat h_{ij}$, contradicting
the fact that $s^{n'}$ is the first state along $\pi$ that satisfies
$h_{ij}$. Hence, $m \neq m'$, and so $0 \leq m < m'$. Since
$s^{m'}_{ij}$ is the first state along $\pi \up ij$ that satisfies
$h_{ij}$, and $\pi \up ij \sat [g_{ij} \U h_{ij}]$, we have $s^m_{ij}
\sat g_{ij}$ by \CTLS semantics. From $s^n \up ij = s^m_{ij}$ and the
induction hypothesis, we get $s^n \sat g_{ij}$. Since $s^n$ is any
state along $\pi$ up to but not including $s^{n'}$, and 
$s^{n'} \sat h_{ij}$, we have $\pi \sat [g_{ij} \U h_{ij}]$ by 
\CTLS semantics. Hence $\pi \sat [g_{ij} \Uw h_{ij}]$ by \CTLS semantics.

In both cases, we showed $\pi \sat [g_{ij} \Uw h_{ij}]$. 
Since $\pi$ is an arbitrary $\f$-fair fullpath starting in $s$, we
conclude $M_I, s \satf \A[g_{ij} \Uw h_{ij}]$.

\bigskip

{$f_{ij} = \A[g_{ij} \U h_{ij}]$}.
Since $f_{ij} \in \CL(\spec_{ij})$, we have $g_{ij} \in
\CL(\spec_{ij})$ and $h_{ij} \in \CL(\spec_{ij})$. Suppose $s_{ij}
\sat h_{ij}$. Hence $s \sat h_{ij}$ by the induction hypothesis, and
so $s \sat \A[g_{ij} \U h_{ij}]$ and we are done. Hence we assume 
$s_{ij} \sat \neg h_{ij}$ in the remainder of the proof. Since $s_{ij} \sat
\A[g_{ij} \U h_{ij}]$ by assumption, we have 
$s_{ij} \sat \neg h_{ij} \land \AF h_{ij}$. 
Let $\pi$ be an arbitrary $\f$-fair fullpath starting in $s$.
By Theorem~\ref{thm:static:dead-free}, $\pi$ is an infinite path.
We now establish $\pi \satf \F h_{ij}$.\ss

\textit{Proof of $\pi \satf \F h_{ij}$}. Assume $\pi \satf \neg \F
h_{ij}$, i.e., $\pi \satf \G \neg h_{ij}$.  Let $t$ be an arbitrary
state along $\pi$. Let $\rho$ be the segment of $\pi$ from $s$ to
$t$. By Definition~\ref{def:pproj}, $\rho \up ij$ is a path from
$s_{ij}$ to $t \up ij$. By Lemma~\ref{lem:static:gen-pmap}, $\rho \up ij$ is
a path in $M_{ij}$. Suppose $\rho \up ij$ contains a state
$u_{ij}$ such that $u_{ij} \sat h_{ij}$. By
Definition~\ref{def:pproj}, there exists a state $u$ along $\rho$ such
that $u \up ij = u_{ij}$. By the induction hypothesis, we have $u
\satf h_{ij}$, contradicting the assumption $\pi \satf \G \neg
h_{ij}$. Hence $\rho \up ij$ contains no state that satisfies
$h_{ij}$. Since $s_{ij} \sat \AF h_{ij}$ and $\rho \up ij$ is a path
from $s_{ij}$ to $t \up ij$ (inclusive) which contains no state
satisfying $h_{ij}$, we must have $t \up ij \sat \neg h_{ij} \land \AF
h_{ij}$ by CTL semantics. Let $\pi'$ be the suffix of $\pi$ starting
in $t$. Since $t \up ij \sat \neg h_{ij} \land \AF h_{ij}$ and $h_{ij}
\in \CL(\spec_{ij})$, we can apply the Progress Lemma to conclude
$M_I, t \satf \AF(ex_i \lor ex_j)$. Since $t$ is an arbitrary state
along $\pi$, we conclude $M_I, \pi \sat \io(ex_i \lor ex_j)$. Hence,
by Definition~\ref{def:pproj}, $\pi \up ij$ is a fullpath. By
Lemma~\ref{lem:static:gen-pmap}, $\pi \up ij$ is a fullpath in $M_{ij}$.
Since $\pi \up ij$ starts in $s_{ij} = s \up ij$, and $s_{ij} \sat \AF h_{ij}$,
$\pi \up ij$ must contain a state $v_{ij}$ such that $v_{ij} \sat
h_{ij}$. By Definition~\ref{def:pproj}, $\pi$ contains a state $v$
such that $v \up ij = v_{ij}$. By the induction hypothesis and $v_{ij}
\sat h_{ij}$, we have $v \satf h_{ij}$. Hence $\pi \satf \F h_{ij}$,
contrary to assumption, and we are done. (End of proof of $\pi \satf
\F h_{ij}$).\ss

By assumption, $s_{ij} \sat \A[g_{ij} \U h_{ij}]$. Hence
$s_{ij} \sat \A[g_{ij} \Uw h_{ij}]$.
From the above proof case for $\A[g_{ij} \Uw h_{ij}]$, we have 
$s \satf \A[g_{ij} \Uw h_{ij}]$. Hence $\pi \satf [g_{ij} \Uw h_{ij}]$, since
$\pi$ is a $\f$-fair fullpath starting in $s$. From this and 
$\pi \satf \F h_{ij}$, we have $\pi \satf [g_{ij} \U h_{ij}]$ by \CTLS
semantics. Since $\pi$ is an arbitrary $\f$-fair fullpath starting in
$s$, we have $s \satf \A[g_{ij} \U h_{ij}]$.
\epr

\bco[Large model]
If the liveness condition for static programs holds, and 
   $W_I(u)$ is supercycle-free for every reachable $I$-state $u$,
then \ms\\
\hspace*{\fill}
	$(\fa (i,j) \in I \d M_{ij}, S^0_{ij} \sat \spec_{ij})
 ~\mathit{implies}~
     M_I, S^0 \satf \AND_{(i,j) \in I} \spec_{ij}.$
\hspace*{\fill}
\label{cor:large-model}
\eco

Unlike \cite{AE98}, $\spec_{ij}$ and $\spec_{k\l}$, where $\{k,\l\}
\neq \{i,j\}$, can be completely different formulae, whereas in
\cite{AE98} these formulae had to be ``similar,'' i.e., one was
obtained from the other by substituting process indices.

\section{Example---A Two Phase Commit Protocol}
\label{sec:example-twophase}

\begin{figure}
\rule{\textwidth}{1pt}
\bc
\scalebox{0.8}{\input{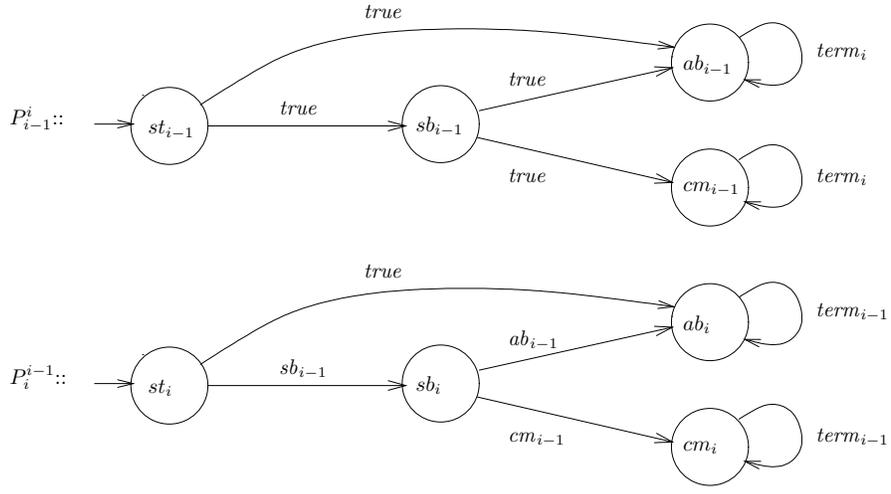}}
\ec
\rule{\textwidth}{1pt}
\caption{Pair program $\P{i-1}{i} \pl \P{i}{i-1}$.}
\label{fig:twophase-i}
\end{figure}

\begin{figure}
\hspace*{0.5in}
\scalebox{0.7}{\input{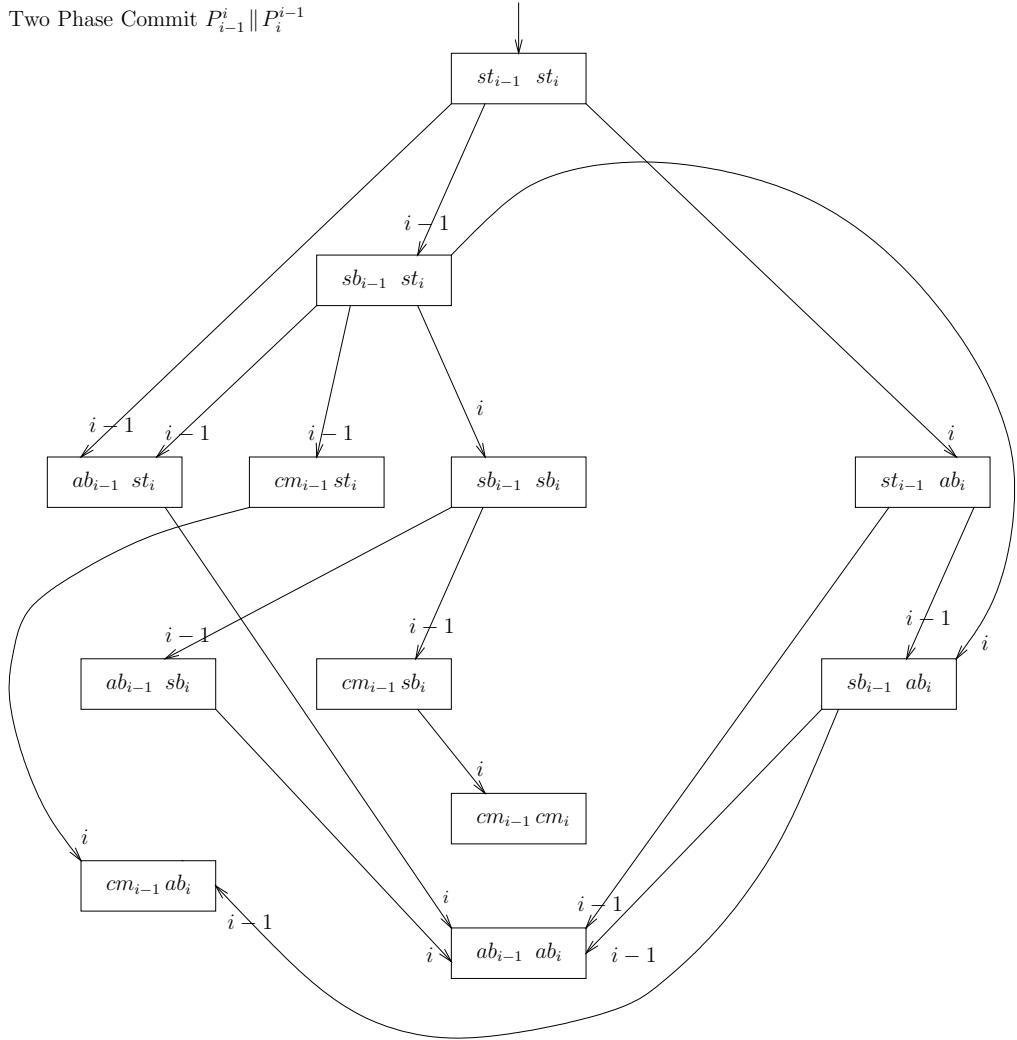}}
\caption{Global state transition diagram of the pair-program
         $\P{i-1}{i} \pl \P{i}{i-1}$.}
\label{fig:twophase-model-i}
\end{figure}

We illustrate our method by synthesizing a ring-based (non fault
tolerant) two-phase commit
protocol $P^I = \PP{0}{I} \pl \PP{1}{I} \pl \cdots \pl \PP{n-1}{I}$,
where $I$ specifies a ring.
$\PP{0}{I}$ is the
\intr{coordinator}, and $\PP{i}{I}, 1 \leq i < n$ are the participants: each
participant represents a transaction. The protocol proceeds in two
cycles around the ring. The coordinator initiates the first cycle, in
which each participant decides to either submit its transaction or
unilaterally abort. $\PP{i}{I}$ can submit only after it observes that
$\PP{i}{I}$ has submitted. After the first cycle, the coordinator
observes the state of $\PP{n-1}{I}$. If $\PP{n-1}{I}$ has submitted its
transaction, that means that all participants have submitted their
transactions, and so the coordinator decides commit. If $\PP{n-1}{I}$  has
aborted, that means that some participant $\PP{i}{I}$ unilaterally aborted
thereby causing all participants $\PP{j}{I}, i < j \leq n-1$ to abort. In
that case, the coordinator decides abort. The second cycle then relays
the coordinators decision around the ring. The participant processes
are all similar to each other, but the coordinator is not similar to
the participants. Hence, there are three pair-programs to consider: 
$\P{n-1}{0} \pl \P{0}{n-1}$,
$\P{0}{1} \pl \P{1}{0}$, and
$\P{i-1}{i} \pl \P{i}{i-1}$.
These are given in Figures~\ref{fig:twophase-0}, \ref{fig:twophase-1},
and \ref{fig:twophase-i}, respectively, where $term_i \df cm_i \lor ab_i$,
and an incoming arrow with no source indicates an initial local state.
Figures~\ref{fig:twophase-model-0}, \ref{fig:twophase-model-1}, and
\ref{fig:twophase-model-i} give the respective global state transition
diagrams (i.e., pair-structures).
The synthesized two phase commit
protocol $P^I$ is given in Figure~\ref{fig:twophase-ring}. 
We establish the correctness of $P^I$ as follows:

\begin{tabular}{llr}

1. &  $cm_0 \ar sb_{n-1}$	&  LMT\\
2. &  $\AND_{2 \leq i < n} (sb_i \ar sb_{i-1})$	&  LMT\\
3. &  $cm_0 \ar \AND_{1 \leq i < n} sb_i$	&  1, 2\\
4. &  $\AND_{1 \leq i < n} (cm_i \ar cm_{i-1})$   &  LMT\\
5. &  $\AND_{0 \leq i < n} (cm_i \ar (\AND_{1 \leq j < n} sb_j))$  & 3, 4\\
6. &  $\AND_{1 \leq i < n} ((cm_{i-1} \land sb_i) \leadsto cm_i)$ & LMT\\
7. &  $\AND_{0 \leq i < n} \AG(\neg cm_i \lor \neg ab_i) \land
       \AG(cm_i \imp \AG cm_i) $                &  LMT\\
8. &  $\AND_{1 \leq i < n}
\AG[sb_i \imp \A[sb_i \U (sb_i \land (cm_{i-1} \lor ab_{i-1}))]]$ &  LMT\\
9. &  $\AND_{1 \leq i < n} (cm_i \ar \A[sb_{i+1} \U (sb_{i+1} \land cm_i)])$ 
                                                & 5, 7, 8\\
10. & $\AND_{1 \leq i < n} ( (cm_{i-1} \land sb_i) \leadsto (cm_i \land sb_{i+1}))$
                                                & 6, 9 \\
11. &  $cm_0 \ar \AND_{1 \leq i < n} cm_{i}$   & 3, 10 \\
\label{proof-one}
\end{tabular}

Here the formula $f \ar g$ abbreviates $\A[(f \imp \AF g) \Uw g]$,
which intuitively means that if $f$ holds at some point, then $g$
holds at some (possibly different) point. There is no ordering on the
times at which $f$ and $g$ hold. $f \leadsto g$ abbreviates
$\AG[f \imp \AF g]$. The above formula hold in all initial states of
$M_I$, the global state transition diagram of $P^I$. The notation LMT
means that the formula was established first in the
relevant pair structure, and then we used the large model theorem to
deduce that the formula also hols in $M^I$. A notation of some formula
numbers means that the formula was deduced using the preceding
formulae, and using an appropriate CTL deductive system \cite{Em90}.
Formula 11 gives us a correctness property of two phase commit: if the
coordinator commits, then so does every participant. Using the large
model theorem, we deduce $\AND_{1 \leq i < n} (ab_{i-1} \ar ab_{i})$,
from which $ab_0 \ar \AND_{1 \leq i < n} ab_{i}$ follows, namely if
the coordinator aborts, then so does every participant. Likewise, we
establish $\AF(cm_0 \lor ab_0)$ (the coordinator eventually decides),
and $\AND_{1 \leq i < n} \AG(st_i \imp \EX_i ab_{i})$ (every
participant can abort unilaterally). This last formula is not in
$\ACTLmij$, but it was shown to be preserved in \cite{AE98}, and we
have extended the proof there to the setting of this paper.

\bfg
\rule{\textwidth}{1pt}
\bc
\scalebox{0.8}{\setlength{\unitlength}{0.00083333in}
\begingroup\makeatletter\ifx\SetFigFont\undefined%
\gdef\SetFigFont#1#2#3#4#5{%
  \reset@font\fontsize{#1}{#2pt}%
  \fontfamily{#3}\fontseries{#4}\fontshape{#5}%
  \selectfont}%
\fi\endgroup%
{\renewcommand{\dashlinestretch}{30}
\begin{picture}(6946,3652)(0,-10)
\put(1245,2662){\ellipse{600}{600}}
\drawline(1035,2902)(1035,2902)
\put(1080,2602){\makebox(0,0)[lb]{\smash{{{\SetFigFont{10}{12.0}{\rmdefault}{\mddefault}{\updefault}$st_{n-1}$}}}}}
\put(3360,2677){\ellipse{600}{600}}
\put(3165,2617){\makebox(0,0)[lb]{\smash{{{\SetFigFont{10}{12.0}{\rmdefault}{\mddefault}{\updefault}$sb_{n-1}$}}}}}
\put(5460,3157){\ellipse{600}{600}}
\put(3660,1282){\ellipse{600}{600}}
\put(1560,802){\ellipse{600}{600}}
\put(3660,307){\ellipse{600}{600}}
\put(5460,2182){\ellipse{600}{600}}
\path(1545,2662)(3085,2662)
\path(2950.000,2624.500)(3085.000,2662.000)(2950.000,2699.500)
\path(3660,2782)(5165,3117)
\path(5041.373,3051.064)(5165.000,3117.000)(5025.077,3124.272)
\path(3645,2572)(5165,2222)
\path(5025.028,2215.749)(5165.000,2222.000)(5041.857,2288.837)
\path(840.000,2647.000)(960.000,2677.000)(840.000,2707.000)
\path(960,2677)(660,2677)
\drawline(5160,3202)(5160,3202)
\drawline(1035,1027)(1035,1027)
\path(1860,907)(3365,1242)
\path(3241.373,1176.064)(3365.000,1242.000)(3225.077,1249.272)
\path(1845,697)(3365,347)
\path(3225.028,340.749)(3365.000,347.000)(3241.857,413.837)
\drawline(3360,1327)(3360,1327)
\path(1140.000,772.000)(1260.000,802.000)(1140.000,832.000)
\path(1260,802)(960,802)
\path(1485,2827)(1487,2828)(1491,2831)
	(1499,2836)(1511,2844)(1528,2855)
	(1550,2869)(1576,2885)(1605,2904)
	(1638,2925)(1673,2947)(1710,2969)
	(1747,2992)(1783,3015)(1820,3036)
	(1855,3057)(1888,3077)(1920,3096)
	(1951,3113)(1980,3129)(2009,3145)
	(2035,3159)(2061,3172)(2087,3184)
	(2112,3196)(2136,3207)(2161,3217)
	(2185,3227)(2210,3237)(2235,3246)
	(2260,3255)(2286,3264)(2313,3273)
	(2340,3281)(2368,3290)(2397,3298)
	(2426,3306)(2457,3313)(2487,3321)
	(2519,3328)(2551,3335)(2583,3342)
	(2615,3348)(2648,3354)(2681,3360)
	(2714,3365)(2747,3370)(2780,3375)
	(2812,3379)(2844,3383)(2876,3387)
	(2908,3391)(2940,3394)(2972,3397)
	(3003,3399)(3035,3402)(3063,3404)
	(3092,3406)(3120,3408)(3150,3410)
	(3180,3411)(3211,3413)(3242,3414)
	(3275,3415)(3308,3416)(3341,3417)
	(3375,3418)(3410,3419)(3445,3419)
	(3481,3420)(3517,3420)(3553,3420)
	(3589,3420)(3626,3420)(3662,3419)
	(3698,3419)(3734,3418)(3769,3417)
	(3804,3416)(3839,3415)(3873,3414)
	(3907,3413)(3940,3411)(3973,3410)
	(4005,3408)(4037,3406)(4069,3404)
	(4100,3402)(4131,3400)(4163,3397)
	(4194,3395)(4226,3392)(4259,3389)
	(4293,3386)(4327,3383)(4363,3379)
	(4400,3375)(4439,3371)(4480,3366)
	(4522,3361)(4566,3356)(4612,3350)
	(4660,3344)(4709,3338)(4759,3332)
	(4809,3325)(4859,3319)(4908,3312)
	(4955,3306)(4999,3301)(5038,3295)
	(5074,3291)(5103,3287)(5128,3283)
	(5146,3281)(5175,3277)
\path(5052.026,3263.678)(5175.000,3277.000)(5060.225,3323.115)
\path(5685,3352)(5686,3352)(5690,3355)
	(5699,3361)(5715,3371)(5736,3383)
	(5759,3397)(5783,3411)(5806,3424)
	(5827,3435)(5847,3444)(5864,3452)
	(5880,3457)(5895,3461)(5910,3464)
	(5923,3466)(5936,3467)(5949,3467)
	(5963,3467)(5977,3465)(5991,3462)
	(6005,3459)(6019,3454)(6033,3449)
	(6046,3442)(6058,3435)(6070,3427)
	(6081,3419)(6092,3410)(6101,3400)
	(6110,3389)(6120,3377)(6128,3363)
	(6137,3347)(6145,3331)(6152,3313)
	(6158,3294)(6164,3275)(6168,3256)
	(6172,3237)(6174,3218)(6175,3200)
	(6175,3183)(6174,3167)(6173,3152)
	(6170,3138)(6165,3124)(6160,3110)
	(6154,3096)(6147,3083)(6139,3071)
	(6129,3059)(6120,3048)(6110,3038)
	(6099,3029)(6089,3020)(6078,3013)
	(6068,3007)(6058,3002)(6045,2997)
	(6033,2992)(6020,2988)(6006,2985)
	(5992,2983)(5977,2982)(5963,2981)
	(5948,2980)(5934,2981)(5921,2982)
	(5908,2983)(5895,2984)(5882,2986)
	(5869,2989)(5855,2991)(5839,2995)
	(5822,2999)(5802,3004)(5781,3009)
	(5762,3014)(5730,3022)
\path(5853.693,3022.000)(5730.000,3022.000)(5839.141,2963.791)
\path(3885,1477)(3886,1477)(3890,1480)
	(3899,1486)(3915,1496)(3936,1508)
	(3959,1522)(3983,1536)(4006,1549)
	(4027,1560)(4047,1569)(4064,1577)
	(4080,1582)(4095,1586)(4110,1589)
	(4123,1591)(4136,1592)(4149,1592)
	(4163,1592)(4177,1590)(4191,1587)
	(4205,1584)(4219,1579)(4233,1574)
	(4246,1567)(4258,1560)(4270,1552)
	(4281,1544)(4292,1535)(4301,1525)
	(4310,1514)(4320,1502)(4328,1488)
	(4337,1472)(4345,1456)(4352,1438)
	(4358,1419)(4364,1400)(4368,1381)
	(4372,1362)(4374,1343)(4375,1325)
	(4375,1308)(4374,1292)(4373,1277)
	(4370,1263)(4365,1249)(4360,1235)
	(4354,1221)(4347,1208)(4339,1196)
	(4329,1184)(4320,1173)(4310,1163)
	(4299,1154)(4289,1145)(4278,1138)
	(4268,1132)(4258,1127)(4245,1122)
	(4233,1117)(4220,1113)(4206,1110)
	(4192,1108)(4177,1107)(4163,1106)
	(4148,1105)(4134,1106)(4121,1107)
	(4108,1108)(4095,1109)(4082,1111)
	(4069,1114)(4055,1116)(4039,1120)
	(4022,1124)(4002,1129)(3981,1134)
	(3962,1139)(3930,1147)
\path(4053.693,1147.000)(3930.000,1147.000)(4039.141,1088.791)
\path(3885,525)(3886,525)(3890,528)
	(3899,534)(3915,544)(3936,556)
	(3959,570)(3983,584)(4006,597)
	(4027,608)(4047,617)(4064,625)
	(4080,630)(4095,634)(4110,637)
	(4123,639)(4136,640)(4149,640)
	(4163,640)(4177,638)(4191,635)
	(4205,632)(4219,627)(4233,622)
	(4246,615)(4258,608)(4270,600)
	(4281,592)(4292,583)(4301,573)
	(4310,562)(4320,550)(4328,536)
	(4337,520)(4345,504)(4352,486)
	(4358,467)(4364,448)(4368,429)
	(4372,410)(4374,391)(4375,373)
	(4375,356)(4374,340)(4373,325)
	(4370,311)(4365,297)(4360,283)
	(4354,269)(4347,256)(4339,244)
	(4329,232)(4320,221)(4310,211)
	(4299,202)(4289,193)(4278,186)
	(4268,180)(4258,175)(4245,170)
	(4233,165)(4220,161)(4206,158)
	(4192,156)(4177,155)(4163,154)
	(4148,153)(4134,154)(4121,155)
	(4108,156)(4095,157)(4082,159)
	(4069,162)(4055,164)(4039,168)
	(4022,172)(4002,177)(3981,182)
	(3962,187)(3930,195)
\path(4053.693,195.000)(3930.000,195.000)(4039.141,136.791)
\path(5685,2400)(5686,2400)(5690,2403)
	(5699,2409)(5715,2419)(5736,2431)
	(5759,2445)(5783,2459)(5806,2472)
	(5827,2483)(5847,2492)(5864,2500)
	(5880,2505)(5895,2509)(5910,2512)
	(5923,2514)(5936,2515)(5949,2515)
	(5963,2515)(5977,2513)(5991,2510)
	(6005,2507)(6019,2502)(6033,2497)
	(6046,2490)(6058,2483)(6070,2475)
	(6081,2467)(6092,2458)(6101,2448)
	(6110,2437)(6120,2425)(6128,2411)
	(6137,2395)(6145,2379)(6152,2361)
	(6158,2342)(6164,2323)(6168,2304)
	(6172,2285)(6174,2266)(6175,2248)
	(6175,2231)(6174,2215)(6173,2200)
	(6170,2186)(6165,2172)(6160,2158)
	(6154,2144)(6147,2131)(6139,2119)
	(6129,2107)(6120,2096)(6110,2086)
	(6099,2077)(6089,2068)(6078,2061)
	(6068,2055)(6058,2050)(6045,2045)
	(6033,2040)(6020,2036)(6006,2033)
	(5992,2031)(5977,2030)(5963,2029)
	(5948,2028)(5934,2029)(5921,2030)
	(5908,2031)(5895,2032)(5882,2034)
	(5869,2037)(5855,2039)(5839,2043)
	(5822,2047)(5802,2052)(5781,2057)
	(5762,2062)(5730,2070)
\path(5853.693,2070.000)(5730.000,2070.000)(5839.141,2011.791)
\put(0,2677){\makebox(0,0)[lb]{\smash{{{\SetFigFont{10}{12.0}{\rmdefault}{\mddefault}{\updefault}$\P{n-1}{0}$::}}}}}
\put(2760,3502){\makebox(0,0)[lb]{\smash{{{\SetFigFont{10}{12.0}{\rmdefault}{\mddefault}{\updefault}$\ltrue$}}}}}
\put(2100,2752){\makebox(0,0)[lb]{\smash{{{\SetFigFont{10}{12.0}{\rmdefault}{\mddefault}{\updefault}$\ltrue$}}}}}
\put(3885,2977){\makebox(0,0)[lb]{\smash{{{\SetFigFont{10}{12.0}{\rmdefault}{\mddefault}{\updefault}$\ltrue$}}}}}
\put(3885,2227){\makebox(0,0)[lb]{\smash{{{\SetFigFont{10}{12.0}{\rmdefault}{\mddefault}{\updefault}$\ltrue$}}}}}
\put(6285,3202){\makebox(0,0)[lb]{\smash{{{\SetFigFont{10}{12.0}{\rmdefault}{\mddefault}{\updefault}$\term_0$}}}}}
\put(5250,3097){\makebox(0,0)[lb]{\smash{{{\SetFigFont{10}{12.0}{\rmdefault}{\mddefault}{\updefault}$ab_{n-1}$}}}}}
\put(6285,2227){\makebox(0,0)[lb]{\smash{{{\SetFigFont{10}{12.0}{\rmdefault}{\mddefault}{\updefault}$\term_0$}}}}}
\put(0,802){\makebox(0,0)[lb]{\smash{{{\SetFigFont{10}{12.0}{\rmdefault}{\mddefault}{\updefault}$\P{0}{n-1}$::}}}}}
\put(4485,1327){\makebox(0,0)[lb]{\smash{{{\SetFigFont{10}{12.0}{\rmdefault}{\mddefault}{\updefault}$\term_{n-1}$}}}}}
\put(3450,1222){\makebox(0,0)[lb]{\smash{{{\SetFigFont{10}{12.0}{\rmdefault}{\mddefault}{\updefault}$ab_{0}$}}}}}
\put(4485,352){\makebox(0,0)[lb]{\smash{{{\SetFigFont{10}{12.0}{\rmdefault}{\mddefault}{\updefault}$\term_{n-1}$}}}}}
\put(1365,742){\makebox(0,0)[lb]{\smash{{{\SetFigFont{10}{12.0}{\rmdefault}{\mddefault}{\updefault}$sb_{0}$}}}}}
\put(3450,277){\makebox(0,0)[lb]{\smash{{{\SetFigFont{10}{12.0}{\rmdefault}{\mddefault}{\updefault}$cm_{0}$}}}}}
\put(1935,1102){\makebox(0,0)[lb]{\smash{{{\SetFigFont{10}{12.0}{\rmdefault}{\mddefault}{\updefault}$\ab_{n-1}$}}}}}
\put(1860,352){\makebox(0,0)[lb]{\smash{{{\SetFigFont{10}{12.0}{\rmdefault}{\mddefault}{\updefault}$\sb_{n-1} \lor \cm_{n-1}$}}}}}
\put(5250,2152){\makebox(0,0)[lb]{\smash{{{\SetFigFont{10}{12.0}{\rmdefault}{\mddefault}{\updefault}$cm_{n-1}$}}}}}
\end{picture}
}}
\ec
\rule{\textwidth}{1pt}
\caption{Pair program $\P{n-1}{0} \pl \P{0}{n-1}$.}
\label{fig:twophase-0}
\efg

\begin{figure}
\rule{\textwidth}{1pt}
\bc
\scalebox{0.8}{\setlength{\unitlength}{0.00083333in}
\begingroup\makeatletter\ifx\SetFigFont\undefined%
\gdef\SetFigFont#1#2#3#4#5{%
  \reset@font\fontsize{#1}{#2pt}%
  \fontfamily{#3}\fontseries{#4}\fontshape{#5}%
  \selectfont}%
\fi\endgroup%
{\renewcommand{\dashlinestretch}{30}
\begin{picture}(6946,3719)(0,-10)
\put(1245,787){\ellipse{600}{600}}
\drawline(1035,1027)(1035,1027)
\put(1080,727){\makebox(0,0)[lb]{\smash{{{\SetFigFont{10}{12.0}{\rmdefault}{\mddefault}{\updefault}$st_{1}$}}}}}
\put(5460,307){\ellipse{600}{600}}
\path(5685,525)(5686,525)(5690,528)
	(5699,534)(5715,544)(5736,556)
	(5759,570)(5783,584)(5806,597)
	(5827,608)(5847,617)(5864,625)
	(5880,630)(5895,634)(5910,637)
	(5923,639)(5936,640)(5949,640)
	(5963,640)(5977,638)(5991,635)
	(6005,632)(6019,627)(6033,622)
	(6046,615)(6058,608)(6070,600)
	(6081,592)(6092,583)(6101,573)
	(6110,562)(6120,550)(6128,536)
	(6137,520)(6145,504)(6152,486)
	(6158,467)(6164,448)(6168,429)
	(6172,410)(6174,391)(6175,373)
	(6175,356)(6174,340)(6173,325)
	(6170,311)(6165,297)(6160,283)
	(6154,269)(6147,256)(6139,244)
	(6129,232)(6120,221)(6110,211)
	(6099,202)(6089,193)(6078,186)
	(6068,180)(6058,175)(6045,170)
	(6033,165)(6020,161)(6006,158)
	(5992,156)(5977,155)(5963,154)
	(5948,153)(5934,154)(5921,155)
	(5908,156)(5895,157)(5882,159)
	(5869,162)(5855,164)(5839,168)
	(5822,172)(5802,177)(5781,182)
	(5762,187)(5730,195)
\path(5853.693,195.000)(5730.000,195.000)(5839.141,136.791)
\put(5250,277){\makebox(0,0)[lb]{\smash{{{\SetFigFont{10}{12.0}{\rmdefault}{\mddefault}{\updefault}$cm_{1}$}}}}}
\put(3360,802){\ellipse{600}{600}}
\put(3165,742){\makebox(0,0)[lb]{\smash{{{\SetFigFont{10}{12.0}{\rmdefault}{\mddefault}{\updefault}$sb_{1}$}}}}}
\put(5460,1282){\ellipse{600}{600}}
\put(3660,3382){\ellipse{600}{600}}
\put(1560,2902){\ellipse{600}{600}}
\put(3660,2407){\ellipse{600}{600}}
\drawline(1035,3127)(1035,3127)
\path(1545,787)(3085,787)
\path(2950.000,749.500)(3085.000,787.000)(2950.000,824.500)
\path(3660,907)(5165,1242)
\path(5041.373,1176.064)(5165.000,1242.000)(5025.077,1249.272)
\path(3645,697)(5165,347)
\path(5025.028,340.749)(5165.000,347.000)(5041.857,413.837)
\path(840.000,772.000)(960.000,802.000)(840.000,832.000)
\path(960,802)(660,802)
\drawline(5160,1327)(5160,1327)
\path(1860,3007)(3365,3342)
\path(3241.373,3276.064)(3365.000,3342.000)(3225.077,3349.272)
\path(1845,2797)(3365,2447)
\path(3225.028,2440.749)(3365.000,2447.000)(3241.857,2513.837)
\drawline(3360,3427)(3360,3427)
\path(1140.000,2872.000)(1260.000,2902.000)(1140.000,2932.000)
\path(1260,2902)(960,2902)
\path(1485,952)(1487,953)(1491,956)
	(1499,961)(1511,969)(1528,980)
	(1550,994)(1576,1010)(1605,1029)
	(1638,1050)(1673,1072)(1710,1094)
	(1747,1117)(1783,1140)(1820,1161)
	(1855,1182)(1888,1202)(1920,1221)
	(1951,1238)(1980,1254)(2009,1270)
	(2035,1284)(2061,1297)(2087,1309)
	(2112,1321)(2136,1332)(2161,1342)
	(2185,1352)(2210,1362)(2235,1371)
	(2260,1380)(2286,1389)(2313,1398)
	(2340,1406)(2368,1415)(2397,1423)
	(2426,1431)(2457,1438)(2487,1446)
	(2519,1453)(2551,1460)(2583,1467)
	(2615,1473)(2648,1479)(2681,1485)
	(2714,1490)(2747,1495)(2780,1500)
	(2812,1504)(2844,1508)(2876,1512)
	(2908,1516)(2940,1519)(2972,1522)
	(3003,1524)(3035,1527)(3063,1529)
	(3092,1531)(3120,1533)(3150,1535)
	(3180,1536)(3211,1538)(3242,1539)
	(3275,1540)(3308,1541)(3341,1542)
	(3375,1543)(3410,1544)(3445,1544)
	(3481,1545)(3517,1545)(3553,1545)
	(3589,1545)(3626,1545)(3662,1544)
	(3698,1544)(3734,1543)(3769,1542)
	(3804,1541)(3839,1540)(3873,1539)
	(3907,1538)(3940,1536)(3973,1535)
	(4005,1533)(4037,1531)(4069,1529)
	(4100,1527)(4131,1525)(4163,1522)
	(4194,1520)(4226,1517)(4259,1514)
	(4293,1511)(4327,1508)(4363,1504)
	(4400,1500)(4439,1496)(4480,1491)
	(4522,1486)(4566,1481)(4612,1475)
	(4660,1469)(4709,1463)(4759,1457)
	(4809,1450)(4859,1444)(4908,1437)
	(4955,1431)(4999,1426)(5038,1420)
	(5074,1416)(5103,1412)(5128,1408)
	(5146,1406)(5175,1402)
\path(5052.026,1388.678)(5175.000,1402.000)(5060.225,1448.115)
\path(5685,1477)(5686,1477)(5690,1480)
	(5699,1486)(5715,1496)(5736,1508)
	(5759,1522)(5783,1536)(5806,1549)
	(5827,1560)(5847,1569)(5864,1577)
	(5880,1582)(5895,1586)(5910,1589)
	(5923,1591)(5936,1592)(5949,1592)
	(5963,1592)(5977,1590)(5991,1587)
	(6005,1584)(6019,1579)(6033,1574)
	(6046,1567)(6058,1560)(6070,1552)
	(6081,1544)(6092,1535)(6101,1525)
	(6110,1514)(6120,1502)(6128,1488)
	(6137,1472)(6145,1456)(6152,1438)
	(6158,1419)(6164,1400)(6168,1381)
	(6172,1362)(6174,1343)(6175,1325)
	(6175,1308)(6174,1292)(6173,1277)
	(6170,1263)(6165,1249)(6160,1235)
	(6154,1221)(6147,1208)(6139,1196)
	(6129,1184)(6120,1173)(6110,1163)
	(6099,1154)(6089,1145)(6078,1138)
	(6068,1132)(6058,1127)(6045,1122)
	(6033,1117)(6020,1113)(6006,1110)
	(5992,1108)(5977,1107)(5963,1106)
	(5948,1105)(5934,1106)(5921,1107)
	(5908,1108)(5895,1109)(5882,1111)
	(5869,1114)(5855,1116)(5839,1120)
	(5822,1124)(5802,1129)(5781,1134)
	(5762,1139)(5730,1147)
\path(5853.693,1147.000)(5730.000,1147.000)(5839.141,1088.791)
\path(3885,3577)(3886,3577)(3890,3580)
	(3899,3586)(3915,3596)(3936,3608)
	(3959,3622)(3983,3636)(4006,3649)
	(4027,3660)(4047,3669)(4064,3677)
	(4080,3682)(4095,3686)(4110,3689)
	(4123,3691)(4136,3692)(4149,3692)
	(4163,3692)(4177,3690)(4191,3687)
	(4205,3684)(4219,3679)(4233,3674)
	(4246,3667)(4258,3660)(4270,3652)
	(4281,3644)(4292,3635)(4301,3625)
	(4310,3614)(4320,3602)(4328,3588)
	(4337,3572)(4345,3556)(4352,3538)
	(4358,3519)(4364,3500)(4368,3481)
	(4372,3462)(4374,3443)(4375,3425)
	(4375,3408)(4374,3392)(4373,3377)
	(4370,3363)(4365,3349)(4360,3335)
	(4354,3321)(4347,3308)(4339,3296)
	(4329,3284)(4320,3273)(4310,3263)
	(4299,3254)(4289,3245)(4278,3238)
	(4268,3232)(4258,3227)(4245,3222)
	(4233,3217)(4220,3213)(4206,3210)
	(4192,3208)(4177,3207)(4163,3206)
	(4148,3205)(4134,3206)(4121,3207)
	(4108,3208)(4095,3209)(4082,3211)
	(4069,3214)(4055,3216)(4039,3220)
	(4022,3224)(4002,3229)(3981,3234)
	(3962,3239)(3930,3247)
\path(4053.693,3247.000)(3930.000,3247.000)(4039.141,3188.791)
\path(3885,2625)(3886,2625)(3890,2628)
	(3899,2634)(3915,2644)(3936,2656)
	(3959,2670)(3983,2684)(4006,2697)
	(4027,2708)(4047,2717)(4064,2725)
	(4080,2730)(4095,2734)(4110,2737)
	(4123,2739)(4136,2740)(4149,2740)
	(4163,2740)(4177,2738)(4191,2735)
	(4205,2732)(4219,2727)(4233,2722)
	(4246,2715)(4258,2708)(4270,2700)
	(4281,2692)(4292,2683)(4301,2673)
	(4310,2662)(4320,2650)(4328,2636)
	(4337,2620)(4345,2604)(4352,2586)
	(4358,2567)(4364,2548)(4368,2529)
	(4372,2510)(4374,2491)(4375,2473)
	(4375,2456)(4374,2440)(4373,2425)
	(4370,2411)(4365,2397)(4360,2383)
	(4354,2369)(4347,2356)(4339,2344)
	(4329,2332)(4320,2321)(4310,2311)
	(4299,2302)(4289,2293)(4278,2286)
	(4268,2280)(4258,2275)(4245,2270)
	(4233,2265)(4220,2261)(4206,2258)
	(4192,2256)(4177,2255)(4163,2254)
	(4148,2253)(4134,2254)(4121,2255)
	(4108,2256)(4095,2257)(4082,2259)
	(4069,2262)(4055,2264)(4039,2268)
	(4022,2272)(4002,2277)(3981,2282)
	(3962,2287)(3930,2295)
\path(4053.693,2295.000)(3930.000,2295.000)(4039.141,2236.791)
\put(0,2902){\makebox(0,0)[lb]{\smash{{{\SetFigFont{10}{12.0}{\rmdefault}{\mddefault}{\updefault}$\P{0}{1}$::}}}}}
\put(0,802){\makebox(0,0)[lb]{\smash{{{\SetFigFont{10}{12.0}{\rmdefault}{\mddefault}{\updefault}$\P{1}{0}$::}}}}}
\put(2760,1627){\makebox(0,0)[lb]{\smash{{{\SetFigFont{10}{12.0}{\rmdefault}{\mddefault}{\updefault}$\ltrue$}}}}}
\put(2100,877){\makebox(0,0)[lb]{\smash{{{\SetFigFont{10}{12.0}{\rmdefault}{\mddefault}{\updefault}$\sb_0$}}}}}
\put(3885,1102){\makebox(0,0)[lb]{\smash{{{\SetFigFont{10}{12.0}{\rmdefault}{\mddefault}{\updefault}$\ab_0$}}}}}
\put(3885,352){\makebox(0,0)[lb]{\smash{{{\SetFigFont{10}{12.0}{\rmdefault}{\mddefault}{\updefault}$\cm_0$}}}}}
\put(6285,1327){\makebox(0,0)[lb]{\smash{{{\SetFigFont{10}{12.0}{\rmdefault}{\mddefault}{\updefault}$\term_0$}}}}}
\put(5250,1222){\makebox(0,0)[lb]{\smash{{{\SetFigFont{10}{12.0}{\rmdefault}{\mddefault}{\updefault}$ab_{1}$}}}}}
\put(6285,352){\makebox(0,0)[lb]{\smash{{{\SetFigFont{10}{12.0}{\rmdefault}{\mddefault}{\updefault}$\term_0$}}}}}
\put(4485,3427){\makebox(0,0)[lb]{\smash{{{\SetFigFont{10}{12.0}{\rmdefault}{\mddefault}{\updefault}$\term_{1}$}}}}}
\put(3450,3322){\makebox(0,0)[lb]{\smash{{{\SetFigFont{10}{12.0}{\rmdefault}{\mddefault}{\updefault}$ab_{0}$}}}}}
\put(4485,2452){\makebox(0,0)[lb]{\smash{{{\SetFigFont{10}{12.0}{\rmdefault}{\mddefault}{\updefault}$\term_{1}$}}}}}
\put(1365,2842){\makebox(0,0)[lb]{\smash{{{\SetFigFont{10}{12.0}{\rmdefault}{\mddefault}{\updefault}$sb_{0}$}}}}}
\put(3450,2377){\makebox(0,0)[lb]{\smash{{{\SetFigFont{10}{12.0}{\rmdefault}{\mddefault}{\updefault}$cm_{0}$}}}}}
\put(2085,3202){\makebox(0,0)[lb]{\smash{{{\SetFigFont{10}{12.0}{\rmdefault}{\mddefault}{\updefault}$\ltrue$}}}}}
\put(2085,2452){\makebox(0,0)[lb]{\smash{{{\SetFigFont{10}{12.0}{\rmdefault}{\mddefault}{\updefault}$\ltrue$}}}}}
\end{picture}
}}
\ec
\rule{\textwidth}{1pt}
\caption{Pair program $\P{0}{1} \pl \P{1}{0}$.}
\label{fig:twophase-1}
\end{figure}

\begin{figure}
\rule{\textwidth}{1pt}\\[0.1in]
\hspace*{0.5in}
\scalebox{0.8}{\setlength{\unitlength}{0.00083333in}
\begingroup\makeatletter\ifx\SetFigFont\undefined%
\gdef\SetFigFont#1#2#3#4#5{%
  \reset@font\fontsize{#1}{#2pt}%
  \fontfamily{#3}\fontseries{#4}\fontshape{#5}%
  \selectfont}%
\fi\endgroup%
{\renewcommand{\dashlinestretch}{30}
\begin{picture}(8238,3869)(0,-10)
\put(1860,2452){\makebox(0,0)[lb]{\smash{{{\SetFigFont{10}{12.0}{\rmdefault}{\mddefault}{\updefault}$(\sb_{n-1} \lor \cm_{n-1}) \gc$}}}}}
\put(1860,2257){\makebox(0,0)[lb]{\smash{{{\SetFigFont{10}{12.0}{\rmdefault}{\mddefault}{\updefault}$\ltrue$}}}}}
\put(5460,1282){\ellipse{600}{600}}
\put(1245,787){\ellipse{600}{600}}
\put(3360,802){\ellipse{600}{600}}
\put(5460,307){\ellipse{600}{600}}
\put(3660,3532){\ellipse{600}{600}}
\put(1560,3052){\ellipse{600}{600}}
\put(3660,2557){\ellipse{600}{600}}
\drawline(1035,1177)(1035,1177)
\path(1545,787)(3085,787)
\path(2950.000,749.500)(3085.000,787.000)(2950.000,824.500)
\path(3660,907)(5165,1242)
\path(5041.373,1176.064)(5165.000,1242.000)(5025.077,1249.272)
\path(3645,697)(5165,347)
\path(5025.028,340.749)(5165.000,347.000)(5041.857,413.837)
\path(840.000,772.000)(960.000,802.000)(840.000,832.000)
\path(960,802)(660,802)
\drawline(5160,1327)(5160,1327)
\drawline(1035,1027)(1035,1027)
\path(1860,3157)(3365,3492)
\path(3241.373,3426.064)(3365.000,3492.000)(3225.077,3499.272)
\path(1845,2947)(3365,2597)
\path(3225.028,2590.749)(3365.000,2597.000)(3241.857,2663.837)
\drawline(3360,3577)(3360,3577)
\path(1140.000,3022.000)(1260.000,3052.000)(1140.000,3082.000)
\path(1260,3052)(960,3052)
\path(1485,952)(1487,953)(1491,956)
	(1499,961)(1511,969)(1528,980)
	(1550,994)(1576,1010)(1605,1029)
	(1638,1050)(1673,1072)(1710,1094)
	(1747,1117)(1783,1140)(1820,1161)
	(1855,1182)(1888,1202)(1920,1221)
	(1951,1238)(1980,1254)(2009,1270)
	(2035,1284)(2061,1297)(2087,1309)
	(2112,1321)(2136,1332)(2161,1342)
	(2185,1352)(2210,1362)(2235,1371)
	(2260,1380)(2286,1389)(2313,1398)
	(2340,1406)(2368,1415)(2397,1423)
	(2426,1431)(2457,1438)(2487,1446)
	(2519,1453)(2551,1460)(2583,1467)
	(2615,1473)(2648,1479)(2681,1485)
	(2714,1490)(2747,1495)(2780,1500)
	(2812,1504)(2844,1508)(2876,1512)
	(2908,1516)(2940,1519)(2972,1522)
	(3003,1524)(3035,1527)(3063,1529)
	(3092,1531)(3120,1533)(3150,1535)
	(3180,1536)(3211,1538)(3242,1539)
	(3275,1540)(3308,1541)(3341,1542)
	(3375,1543)(3410,1544)(3445,1544)
	(3481,1545)(3517,1545)(3553,1545)
	(3589,1545)(3626,1545)(3662,1544)
	(3698,1544)(3734,1543)(3769,1542)
	(3804,1541)(3839,1540)(3873,1539)
	(3907,1538)(3940,1536)(3973,1535)
	(4005,1533)(4037,1531)(4069,1529)
	(4100,1527)(4131,1525)(4163,1522)
	(4194,1520)(4226,1517)(4259,1514)
	(4293,1511)(4327,1508)(4363,1504)
	(4400,1500)(4439,1496)(4480,1491)
	(4522,1486)(4566,1481)(4612,1475)
	(4660,1469)(4709,1463)(4759,1457)
	(4809,1450)(4859,1444)(4908,1437)
	(4955,1431)(4999,1426)(5038,1420)
	(5074,1416)(5103,1412)(5128,1408)
	(5146,1406)(5175,1402)
\path(5052.026,1388.678)(5175.000,1402.000)(5060.225,1448.115)
\path(5685,1477)(5686,1477)(5690,1480)
	(5699,1486)(5715,1496)(5736,1508)
	(5759,1522)(5783,1536)(5806,1549)
	(5827,1560)(5847,1569)(5864,1577)
	(5880,1582)(5895,1586)(5910,1589)
	(5923,1591)(5936,1592)(5949,1592)
	(5963,1592)(5977,1590)(5991,1587)
	(6005,1584)(6019,1579)(6033,1574)
	(6046,1567)(6058,1560)(6070,1552)
	(6081,1544)(6092,1535)(6101,1525)
	(6110,1514)(6120,1502)(6128,1488)
	(6137,1472)(6145,1456)(6152,1438)
	(6158,1419)(6164,1400)(6168,1381)
	(6172,1362)(6174,1343)(6175,1325)
	(6175,1308)(6174,1292)(6173,1277)
	(6170,1263)(6165,1249)(6160,1235)
	(6154,1221)(6147,1208)(6139,1196)
	(6129,1184)(6120,1173)(6110,1163)
	(6099,1154)(6089,1145)(6078,1138)
	(6068,1132)(6058,1127)(6045,1122)
	(6033,1117)(6020,1113)(6006,1110)
	(5992,1108)(5977,1107)(5963,1106)
	(5948,1105)(5934,1106)(5921,1107)
	(5908,1108)(5895,1109)(5882,1111)
	(5869,1114)(5855,1116)(5839,1120)
	(5822,1124)(5802,1129)(5781,1134)
	(5762,1139)(5730,1147)
\path(5853.693,1147.000)(5730.000,1147.000)(5839.141,1088.791)
\path(5685,525)(5686,525)(5690,528)
	(5699,534)(5715,544)(5736,556)
	(5759,570)(5783,584)(5806,597)
	(5827,608)(5847,617)(5864,625)
	(5880,630)(5895,634)(5910,637)
	(5923,639)(5936,640)(5949,640)
	(5963,640)(5977,638)(5991,635)
	(6005,632)(6019,627)(6033,622)
	(6046,615)(6058,608)(6070,600)
	(6081,592)(6092,583)(6101,573)
	(6110,562)(6120,550)(6128,536)
	(6137,520)(6145,504)(6152,486)
	(6158,467)(6164,448)(6168,429)
	(6172,410)(6174,391)(6175,373)
	(6175,356)(6174,340)(6173,325)
	(6170,311)(6165,297)(6160,283)
	(6154,269)(6147,256)(6139,244)
	(6129,232)(6120,221)(6110,211)
	(6099,202)(6089,193)(6078,186)
	(6068,180)(6058,175)(6045,170)
	(6033,165)(6020,161)(6006,158)
	(5992,156)(5977,155)(5963,154)
	(5948,153)(5934,154)(5921,155)
	(5908,156)(5895,157)(5882,159)
	(5869,162)(5855,164)(5839,168)
	(5822,172)(5802,177)(5781,182)
	(5762,187)(5730,195)
\path(5853.693,195.000)(5730.000,195.000)(5839.141,136.791)
\path(3885,3727)(3886,3727)(3890,3730)
	(3899,3736)(3915,3746)(3936,3758)
	(3959,3772)(3983,3786)(4006,3799)
	(4027,3810)(4047,3819)(4064,3827)
	(4080,3832)(4095,3836)(4110,3839)
	(4123,3841)(4136,3842)(4149,3842)
	(4163,3842)(4177,3840)(4191,3837)
	(4205,3834)(4219,3829)(4233,3824)
	(4246,3817)(4258,3810)(4270,3802)
	(4281,3794)(4292,3785)(4301,3775)
	(4310,3764)(4320,3752)(4328,3738)
	(4337,3722)(4345,3706)(4352,3688)
	(4358,3669)(4364,3650)(4368,3631)
	(4372,3612)(4374,3593)(4375,3575)
	(4375,3558)(4374,3542)(4373,3527)
	(4370,3513)(4365,3499)(4360,3485)
	(4354,3471)(4347,3458)(4339,3446)
	(4329,3434)(4320,3423)(4310,3413)
	(4299,3404)(4289,3395)(4278,3388)
	(4268,3382)(4258,3377)(4245,3372)
	(4233,3367)(4220,3363)(4206,3360)
	(4192,3358)(4177,3357)(4163,3356)
	(4148,3355)(4134,3356)(4121,3357)
	(4108,3358)(4095,3359)(4082,3361)
	(4069,3364)(4055,3366)(4039,3370)
	(4022,3374)(4002,3379)(3981,3384)
	(3962,3389)(3930,3397)
\path(4053.693,3397.000)(3930.000,3397.000)(4039.141,3338.791)
\path(3885,2775)(3886,2775)(3890,2778)
	(3899,2784)(3915,2794)(3936,2806)
	(3959,2820)(3983,2834)(4006,2847)
	(4027,2858)(4047,2867)(4064,2875)
	(4080,2880)(4095,2884)(4110,2887)
	(4123,2889)(4136,2890)(4149,2890)
	(4163,2890)(4177,2888)(4191,2885)
	(4205,2882)(4219,2877)(4233,2872)
	(4246,2865)(4258,2858)(4270,2850)
	(4281,2842)(4292,2833)(4301,2823)
	(4310,2812)(4320,2800)(4328,2786)
	(4337,2770)(4345,2754)(4352,2736)
	(4358,2717)(4364,2698)(4368,2679)
	(4372,2660)(4374,2641)(4375,2623)
	(4375,2606)(4374,2590)(4373,2575)
	(4370,2561)(4365,2547)(4360,2533)
	(4354,2519)(4347,2506)(4339,2494)
	(4329,2482)(4320,2471)(4310,2461)
	(4299,2452)(4289,2443)(4278,2436)
	(4268,2430)(4258,2425)(4245,2420)
	(4233,2415)(4220,2411)(4206,2408)
	(4192,2406)(4177,2405)(4163,2404)
	(4148,2403)(4134,2404)(4121,2405)
	(4108,2406)(4095,2407)(4082,2409)
	(4069,2412)(4055,2414)(4039,2418)
	(4022,2422)(4002,2427)(3981,2432)
	(3962,2437)(3930,2445)
\path(4053.693,2445.000)(3930.000,2445.000)(4039.141,2386.791)
\put(0,802){\makebox(0,0)[lb]{\smash{{{\SetFigFont{10}{12.0}{\rmdefault}{\mddefault}{\updefault}$\P{i}{I}$::}}}}}
\put(2760,1627){\makebox(0,0)[lb]{\smash{{{\SetFigFont{10}{12.0}{\rmdefault}{\mddefault}{\updefault}$\ltrue \gc \ltrue$}}}}}
\put(6285,1327){\makebox(0,0)[lb]{\smash{{{\SetFigFont{10}{12.0}{\rmdefault}{\mddefault}{\updefault}$\term_{i-1} \gc \term_{i+1}$}}}}}
\put(5250,1222){\makebox(0,0)[lb]{\smash{{{\SetFigFont{10}{12.0}{\rmdefault}{\mddefault}{\updefault}$ab_{i}$}}}}}
\put(6285,352){\makebox(0,0)[lb]{\smash{{{\SetFigFont{10}{12.0}{\rmdefault}{\mddefault}{\updefault}$\term_{i-1} \gc \term_{i+1}$}}}}}
\put(1080,727){\makebox(0,0)[lb]{\smash{{{\SetFigFont{10}{12.0}{\rmdefault}{\mddefault}{\updefault}$st_{i}$}}}}}
\put(3165,742){\makebox(0,0)[lb]{\smash{{{\SetFigFont{10}{12.0}{\rmdefault}{\mddefault}{\updefault}$sb_{i}$}}}}}
\put(5250,277){\makebox(0,0)[lb]{\smash{{{\SetFigFont{10}{12.0}{\rmdefault}{\mddefault}{\updefault}$cm_{i}$}}}}}
\put(0,3052){\makebox(0,0)[lb]{\smash{{{\SetFigFont{10}{12.0}{\rmdefault}{\mddefault}{\updefault}$\P{0}{I}$::}}}}}
\put(3450,3472){\makebox(0,0)[lb]{\smash{{{\SetFigFont{10}{12.0}{\rmdefault}{\mddefault}{\updefault}$ab_{0}$}}}}}
\put(4485,2602){\makebox(0,0)[lb]{\smash{{{\SetFigFont{10}{12.0}{\rmdefault}{\mddefault}{\updefault}$\term_{n-1} \gc \term_1$}}}}}
\put(1365,2992){\makebox(0,0)[lb]{\smash{{{\SetFigFont{10}{12.0}{\rmdefault}{\mddefault}{\updefault}$sb_{0}$}}}}}
\put(3450,2527){\makebox(0,0)[lb]{\smash{{{\SetFigFont{10}{12.0}{\rmdefault}{\mddefault}{\updefault}$cm_{0}$}}}}}
\put(4485,3577){\makebox(0,0)[lb]{\smash{{{\SetFigFont{10}{12.0}{\rmdefault}{\mddefault}{\updefault}$\term_{n-1} \gc \term_1$}}}}}
\put(3585,352){\makebox(0,0)[lb]{\smash{{{\SetFigFont{10}{12.0}{\rmdefault}{\mddefault}{\updefault}$\cm_{i-1} \gc \ltrue$}}}}}
\put(3585,1102){\makebox(0,0)[lb]{\smash{{{\SetFigFont{10}{12.0}{\rmdefault}{\mddefault}{\updefault}$\ab_{i-1} \gc \ltrue$}}}}}
\put(1860,3427){\makebox(0,0)[lb]{\smash{{{\SetFigFont{10}{12.0}{\rmdefault}{\mddefault}{\updefault}$\ab_{n-1} \gc \ltrue$}}}}}
\put(1875,877){\makebox(0,0)[lb]{\smash{{{\SetFigFont{10}{12.0}{\rmdefault}{\mddefault}{\updefault}$\sb_{i-1} \gc \ltrue$}}}}}
\end{picture}
}}
\rule{\textwidth}{1pt}
\caption{The synthesized two phase commit protocol
	 $P^I = \P{0}{I} \pl (\,\pl_{1 \leq i < n} \P{i}{I})$.}
\label{fig:twophase-ring}
\end{figure}
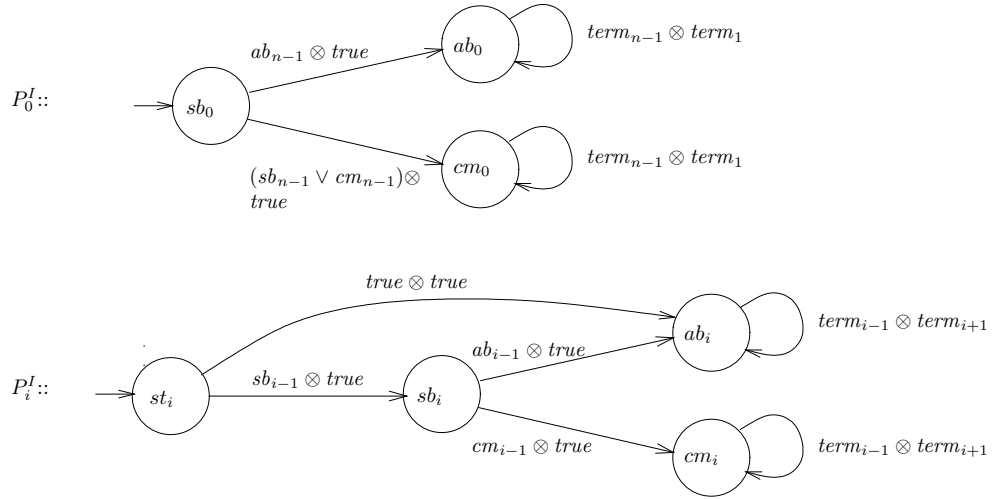

\begin{figure}
\bc
\scalebox{0.8}{\setlength{\unitlength}{0.00083333in}
\begingroup\makeatletter\ifx\SetFigFont\undefined%
\gdef\SetFigFont#1#2#3#4#5{%
  \reset@font\fontsize{#1}{#2pt}%
  \fontfamily{#3}\fontseries{#4}\fontshape{#5}%
  \selectfont}%
\fi\endgroup%
{\renewcommand{\dashlinestretch}{30}
\begin{picture}(8399,7422)(0,-10)
\path(4725,6462)(5925,6462)(5925,6012)
	(4725,6012)(4725,6462)
\path(225,2862)(1425,2862)(1425,2412)
	(225,2412)(225,2862)
\path(4725,2862)(5925,2862)(5925,2412)
	(4725,2412)(4725,2862)
\path(5025,6012)(4425,4662)
\path(4446.322,4783.842)(4425.000,4662.000)(4501.151,4759.473)
\path(4438,4212)(5038,2862)
\path(4961.849,2959.473)(5038.000,2862.000)(5016.678,2983.842)
\path(3525,1662)(4725,1662)(4725,1212)
	(3525,1212)(3525,1662)
\path(3525,462)(4725,462)(4725,12)
	(3525,12)(3525,462)
\drawline(3825,1662)(3825,1662)
\path(5025,2412)(4425,1662)
\path(4476.537,1774.445)(4425.000,1662.000)(4523.389,1736.963)
\drawline(4725,462)(4725,462)
\path(1425,2412)(3525,162)
\path(3421.190,229.257)(3525.000,162.000)(3465.053,270.196)
\path(5295.000,6582.000)(5325.000,6462.000)(5355.000,6582.000)
\path(5325,6462)(5325,7062)
\drawline(750,2862)(750,2862)
\path(4725,6012)(750,2862)
\path(825.417,2960.042)(750.000,2862.000)(862.682,2913.017)
\path(4725,4212)(7125,2862)
\path(7005.703,2894.684)(7125.000,2862.000)(7035.119,2946.979)
\path(7125,2412)(4725,1662)
\path(4830.589,1726.427)(4725.000,1662.000)(4848.486,1669.159)
\path(5925,2412)(7350,1662)
\path(7229.837,1691.342)(7350.000,1662.000)(7257.782,1744.437)
\path(7125,1662)(8325,1662)(8325,1212)
	(7125,1212)(7125,1662)
\path(7125,2862)(8325,2862)(8325,2412)
	(7125,2412)(7125,2862)
\path(3525,4662)(4725,4662)(4725,4212)
	(3525,4212)(3525,4662)
\path(3525,4212)(1425,2862)
\path(1509.719,2952.126)(1425.000,2862.000)(1542.164,2901.656)
\put(4425,5037){\makebox(0,0)[lb]{\smash{{{\SetFigFont{12}{14.4}{\rmdefault}{\mddefault}{\updefault}$n-1$}}}}}
\put(5025,3087){\makebox(0,0)[lb]{\smash{{{\SetFigFont{12}{14.4}{\rmdefault}{\mddefault}{\updefault}0}}}}}
\put(3300,162){\makebox(0,0)[lb]{\smash{{{\SetFigFont{12}{14.4}{\rmdefault}{\mddefault}{\updefault}0}}}}}
\put(0,7287){\makebox(0,0)[lb]{\smash{{{\SetFigFont{10}{12.0}{\rmdefault}{\mddefault}{\updefault}Two phase commit $\P{0}{n-1} \pl \P{n-1}{0}$}}}}}
\put(4875,6197){\makebox(0,0)[lb]{\smash{{{\SetFigFont{12}{14.4}{\rmdefault}{\mddefault}{\updefault}$st_{n-1}$}}}}}
\put(5400,6197){\makebox(0,0)[lb]{\smash{{{\SetFigFont{12}{14.4}{\rmdefault}{\mddefault}{\updefault}$sb_0$}}}}}
\put(4875,2597){\makebox(0,0)[lb]{\smash{{{\SetFigFont{12}{14.4}{\rmdefault}{\mddefault}{\updefault}$sb_{n-1}$}}}}}
\put(5400,2597){\makebox(0,0)[lb]{\smash{{{\SetFigFont{12}{14.4}{\rmdefault}{\mddefault}{\updefault}$cm_0$}}}}}
\put(3600,1397){\makebox(0,0)[lb]{\smash{{{\SetFigFont{12}{14.4}{\rmdefault}{\mddefault}{\updefault}$cm_{n-1}$}}}}}
\put(4200,1397){\makebox(0,0)[lb]{\smash{{{\SetFigFont{12}{14.4}{\rmdefault}{\mddefault}{\updefault}$cm_0$}}}}}
\put(3675,197){\makebox(0,0)[lb]{\smash{{{\SetFigFont{12}{14.4}{\rmdefault}{\mddefault}{\updefault}$ab_{n-1}$}}}}}
\put(4200,197){\makebox(0,0)[lb]{\smash{{{\SetFigFont{12}{14.4}{\rmdefault}{\mddefault}{\updefault}$ab_0$}}}}}
\put(900,2597){\makebox(0,0)[lb]{\smash{{{\SetFigFont{12}{14.4}{\rmdefault}{\mddefault}{\updefault}$sb_0$}}}}}
\put(4050,1812){\makebox(0,0)[lb]{\smash{{{\SetFigFont{12}{14.4}{\rmdefault}{\mddefault}{\updefault}$n-1$}}}}}
\put(5100,1587){\makebox(0,0)[lb]{\smash{{{\SetFigFont{12}{14.4}{\rmdefault}{\mddefault}{\updefault}$0$}}}}}
\put(7200,1397){\makebox(0,0)[lb]{\smash{{{\SetFigFont{12}{14.4}{\rmdefault}{\mddefault}{\updefault}$ab_{n-1}$}}}}}
\put(7800,1397){\makebox(0,0)[lb]{\smash{{{\SetFigFont{12}{14.4}{\rmdefault}{\mddefault}{\updefault}$cm_0$}}}}}
\put(7200,2597){\makebox(0,0)[lb]{\smash{{{\SetFigFont{12}{14.4}{\rmdefault}{\mddefault}{\updefault}$cm_{n-1}$}}}}}
\put(3600,4397){\makebox(0,0)[lb]{\smash{{{\SetFigFont{12}{14.4}{\rmdefault}{\mddefault}{\updefault}$sb_{n-1}$}}}}}
\put(4200,4397){\makebox(0,0)[lb]{\smash{{{\SetFigFont{12}{14.4}{\rmdefault}{\mddefault}{\updefault}$sb_0$}}}}}
\put(1800,2937){\makebox(0,0)[lb]{\smash{{{\SetFigFont{12}{14.4}{\rmdefault}{\mddefault}{\updefault}$n-1$}}}}}
\put(375,2597){\makebox(0,0)[lb]{\smash{{{\SetFigFont{12}{14.4}{\rmdefault}{\mddefault}{\updefault}$ab_{n-1}$}}}}}
\put(600,3087){\makebox(0,0)[lb]{\smash{{{\SetFigFont{12}{14.4}{\rmdefault}{\mddefault}{\updefault}$n-1$}}}}}
\put(7800,2597){\makebox(0,0)[lb]{\smash{{{\SetFigFont{12}{14.4}{\rmdefault}{\mddefault}{\updefault}$sb_0$}}}}}
\put(6825,3087){\makebox(0,0)[lb]{\smash{{{\SetFigFont{12}{14.4}{\rmdefault}{\mddefault}{\updefault}$n-1$}}}}}
\put(7050,1887){\makebox(0,0)[lb]{\smash{{{\SetFigFont{12}{14.4}{\rmdefault}{\mddefault}{\updefault}$n-1$}}}}}
\end{picture}
}}
\ec
\caption{Global state transition diagram of the pair-program
$\P{n-1}{0} \pl \P{0}{n-1}$.}
\label{fig:twophase-model-0}
\end{figure}

\begin{figure}
\bc
\scalebox{0.8}{\input{figs/twophase-model-1.eepic}}
\ec
\caption{Global state transition diagram of the pair-program
         $\P{0}{1} \pl \P{1}{0}$.}
\label{fig:twophase-model-1}
\end{figure}

\clearpage
\section{Synthesis of Dynamic Concurrent Programs}
\label{sec:dynamic:method}

\subsection{Dynamic Specifications}

A \intr{dynamic specification} consists of:
\begin{nlst1}

\item A ``universal'' set $\UPairs$ of \intr{pair-specifications}.
      A pair-specification has the form 
      $\tpl{\set{i,j},\spec_{ij}}$, where $i,j \in \Pids$, $i \ne j$,
      and $\spec_{ij} \in \ACTLmij$ specifies the interaction of
      processes $i$ and $j$. $\UPairs$ can be infinite. 

\item A finite set $\IPairs \sub \UPairs$, which gives the 
      pair-specifications which are \intr{in force}, that is, must be
      satisfied, initially.

\item A mapping $create: 2^\UPairs \mapsto 2^\UPairs$ which
   determines which new pair-specifications
   (in $\UPairs$) can be added to those that are in-force.
   If $\Pairs$ is the set of pair-specifications that are in-force
   and $\pspecij \in create(\Pairs)$, then
   $\Pairs \un \set{\pspecij}$ is a possible next value for the set
   of pair-specifications in-force.

\end{nlst1}

\noindent
We show in the sequel that the synthesized dynamic program satisfies
the dynamic specification in that every pair-specification is
satisfied from the time it comes into force. We make these notions
precise below.

\subsection{Overview of the Synthesis Method: Dynamic Addition of Pair-programs}
\label{sec:synth-overview}

Our synthesis method produces a dynamic concurrent program $\SP$.
$\SP$ consists of the \intr{conjunctive overlay} of a dynamically
increasing set of \intr{pair-programs}.
A {pair-program} is a static concurrent program consisting of
exactly two processes. $\SYij$ denotes a pair-program with
processes $i$ and $j$, and initial state set $S_{ij}^0$.
We use $\P{i}{j}$ for the synchronization skeleton of process $i$
within this pair-program, with the superscript $j$ indicating the
other process.
$\SHij$ denotes the shared variables in $\SYij$.
{The shared variable sets of different
pair-programs are disjoint: $\SH_{ij} \ints \SH_{i'j'} = \emptyset$ if
$\{i,j\} \neq \{i',j'\}$.}
The component processes of a pair-program (e.g., $\P{i}{j}$) are called
\intr{pair-processes}.
Define $\lP{i}{}$ to be the synchronization skeleton of $\P{i}{}$
with all the arc labels removed.

\bd[Conjunctive overlay, $\P{i}{j} \gc \P{i}{k}$]
Let $\P{i}{j}$ and $\P{i}{k}$ be pair-processes for $i$ such that
$\lP{i}{j} = \lP{i}{k}$. Then, \\
\halfind $\P{i}{j} \gc \P{i}{k}$ contains an arc from
     $s_i$ to $t_i$ with label
     $(\gdi{\l \in [1:n_j]} \CB{i}{j}{\l} \ar \CA{i}{j}{\l}) \gc
      (\gdi{\l \in [1:n_k]} \CB{i}{k}{\l} \ar \CA{i}{k}{\l})$\\
iff \\
\halfind $\P{i}{j}$ contains an arc from $s_i$ to $t_i$
     with label $\gdi{\l \in [1:n_j]} \CB{i}{j}{\l} \ar \CA{i}{j}{\l}$
and\\
\halfind $\P{i}{k}$ contains an arc from $s_i$ to $t_i$
     with label $\gdi{\l \in [1:n_k]} \CB{i}{k}{\l} \ar \CA{i}{k}{\l}$. 
\ed
Note that the $\gc$ operator is overloaded, and applies to both pair-processes
and to guarded commands. When applied to guarded commands, $\gc$ denotes the
``conjunction'' of guarded commands, so an arc with label
$(\gdi{\l \in [1:n_j]} \CB{i}{j}{\l} \ar \CA{i}{j}{\l}) \gc
      (\gdi{\l \in [1:n_k]} \CB{i}{k}{\l} \ar \CA{i}{k}{\l})$
can only be executed in a state in which 
$\CB{i}{j}{\l}$ holds for some $\l \in [n_j]$ and 
$\CB{i}{k}{\l}$ holds for some $k \in [n_j]$. Execution then involves the
parallel execution of the corresponding $\CA{i}{j}{\l}$ and $\CA{i}{k}{\l}$.
See \cite{AE98} for a full discussion of $\gd$ and $\gc$.
Conjunctive overlay viewed as a binary operation on both guarded commands and
pair-processes is commutative and associative, since the operands of $\gc$ are
treated identically.  Thus, we define and use the $n$-ary version of $\gc$
in the usual manner.

Given a dynamic concurrent program $\SP$, a new pair-program $\SYij$
can be dynamically added at run-time as follows. If $\SP$ already
contains $P_i$, then $P_i$ is
modified by taking the conjunctive overlay with $\P{i}{j}$, i.e.,
$P_i := P_i \gc \P{i}{j}$.
If $\SP$ does not contain $P_i$, then $P_i$ is
dynamically created and added as a new process,
 and is given the synchronization skeleton of
$\P{i}{j}$, i.e., $P_i := \P{i}{j}$.  Likewise for $P_j$.  We say that
$\SYij$ is \intr{active} once it has been added.
The ``synchronization skeleton code'' of the dynamic program thus changes at run
time, as pair-programs are added. Since each $P_i$ built up by successive
conjunctive overlays of pair-processes, the $n$-ary version of the $\gc$
operator can always be applied, provided that $\graph{P_i} = \graph{\P{i}{j}}$.
To assure this, we assume, in the sequel,
\vspace{-0.5ex}
\begin{quote}
For active pair-programs $\SYij$ and $\SYik$: $\graph{\P{i}{j}} = \graph{\P{i}{k}}$.
\end{quote}
\vspace{-0.5ex}
We emphasize that different pair-programs can have different
functionality, since the guarded commands which label the arcs of
$\P{i}{j}$ and $\P{i}{k}$ can be different.

Pair-programs are added only when a new pair-specification comes
into force, and is the means of satisfying the new pair-specification.
Thus, the transitions of $\SP$ are of two kinds: 
(1) \intr{normal} transitions,
which are atomic transitions (as described in Section~\ref{sec:model})
arising from execution of 
the conjunctive overlay of all active pair-programs,
and
(2) \intr{create} transitions, which correspond to making a new
pair-specification $\pspecij$ in-force, according to the $create$
mapping.  To satisfy $\pspecij$, we dynamically create a new
pair-program $\SYij$ such that $\SYij \sat \spec_{i,j}$, and
incorporate it into the existing dynamic program by performing a
{conjunctive overlay} with the currently active pair-programs.

\subsection{Technical Definitions}

If $\Pairs \sub \UPairs$, then define 
$\pairs{\I} = \{ \set{i,j} ~|~ \ex \spec_{ij} : 
	\tpl{\set{i,j},\spec_{ij}} \in \I \}$, and
$\procs{\I} = \{ i ~|~ \ex j : \set{i,j} \in \pairs{\I}\}$, and
$\Pairs(i) = \{j ~|~ \set{i,j} \in \pairs{I} \}$.
Processes $i$ and $j$ are \emph{neighbors} when
$\set{i,j} \in \pairs{\I}$.
If $\I \ne \emptyset$, then  $\I(i) \ne \emptyset$ for all $i \in \procs{I}$, by definition.
Thus, every process always has at least one neighbor.

An \intr{$i$-state} is a local state of  $\P{i}{j}$.
An $ij$-state is a global state of $\SYij$, i.e., (by Section~\ref{sec:model}) 
a tuple $(s_i, s_j, v_{ij}^1,\ldots,v_{ij}^m)$ where $s_i, s_j$ are
$i$-states, $j$-states, respectively, and
$v_{ij}^1,\ldots,v_{ij}^m$ give the values of all the variables in
$\SH_{ij}$.
When $i$ and $j$ are unspecified, we refer to an $ij$-state as a pair-state.

A \intr{configuration} is a tuple $\tpl{\I,\Pr,\Ps}$, where 
$\I \sub \UPairs$, 
$\Pr$ is a set of pair-programs $\SYij$, one for each 
	$\set{i,j} \in \pairs{\I}$, and 
$\Ps$ is a mapping from each $\set{i,j} \in \pairs{\I}$ to an
$ij$-state.
We refer to the components of $s$ as $s.\I$, $s.\Pr$, $s.\Ps$.
We write $\procs{s}$ for $\procs{s.\I}$, and 
$\pairs{s}$ for $\pairs{s.\I}$.
A \intr{consistent configuration} satisfies the constraint that 
all pair-states assign the same local state to
all common processes, i.e.,
for all $\set{i,j}, \set{i,k} \in \pairs{s}$, if
  $\Ps(\set{i,j}) = (s_i, s_j, v_{ij}^1,\ldots,v_{ij}^m)$ and
  $\Ps(\set{i,k}) = (s'_i, s_k, v_{ik}^1,\ldots,v_{ik}^m)$, then
  $s_i = s'_i$.
We assume henceforth that configurations are consistent, and our
definitions will respect this constraint.

For configuration $s$, $i \in \procs{s}$,  and atomic proposition $p_i \in \AP_i$, 
we define $s(p_i) = \Ps(\set{i,j})(p_i)$, where 
$\set{i,j} \in \pairs{s}$. By the above definitions and constraints, a $j$ such that 
$\set{i,j} \in \pairs{s}$ always exists when $i \in \procs{s}$,
and the value for $s(p_i)$ so defined is unique.

The state-to-formula operator
$\stof{s_i}$ converts an $i$-state $s_i$ into a
propositional formula:
$\stof{s_i} = ({\bigwedge}_{s_i(p_i) = true} p_i) ~\land~$
            $({\bigwedge}_{s_i(p_i) = false} \neg p_i)$,
where $p_i$ ranges over the members of $\AP_i$.
$\stof{s_i}$ characterizes $s_i$ in that $s_i \sat
\stof{s_i}$, and $s'_i \not\sat \stof{s_i}$ for all $s'_i \neq s_i$.
$\stof{s_{ij}}$ is defined similarly (but note that the variables in
$\SH_{ij}$ must be accounted for).

We define the \intr{state projection operator} $\up $, which is an overloaded
binary infix operator with several variants, depending on the type of
the operands.
For projection of $ij$-states onto a single process:
if $s_{ij} = (s_i, s_j, v_{ij}^1,\ldots,v_{ij}^m)$, then
   $s_{ij} \up i = s_i$.
For projection of $ij$-states onto the shared variables in $\SH_{ij}$:
if $s_{ij} = (s_i, s_j, v_{ij}^1,\ldots,v_{ij}^m)$, then
   $s_{ij} \up \SH_{ij} = (v_{ij}^1,\ldots,v_{ij}^m)$.
For projection of a configuration $s = \tpl{\I,\Pr,\Ps}$ onto a single process:
if $i \in \procs{s}$, then 
   $s \up i = \Ps(\set{i,j}) \up i$, where $\set{i,j} \in \pairs{s}$.
This is unique because configurations are consistent.
For projection of $s$ onto a pair-program:
if $\set{i,j} \in \pairs{s}$, then $s \up ij = \Ps(\set{i,j})$.
If $\set{i,j} \not\in \pairs{s}$, then $s \up ij$ is undefined.
If $J$ is a set of pairs such that $J \sub \pairs{s}$, then we define
the projection of $s$ onto $J$:
$s \pj J$ is the restriction of $s.\Ps$ to $J$.

\subsection{The Synthesis Method}

Given a dynamic specification, we synthesize a program $\SP$ as follows:
\begin{nlst1}

\item Initially, $\SP$ consists of the conjunctive overlay of the
pair-programs corresponding to the pair-specifications in $\IPairs$.

\item When a pair-specification $\tpl{\set{i,j},\spec_{ij}}$ is added,
as permitted by the $create$ mapping, synthesize a \intr{pair-program}
$\SYij$ using $\spec_{ij}$ as the specification, and add it to $\SP$ as
discussed in Section~\ref{sec:synth-overview} above.

\end{nlst1}
To synthesize pair-programs, any 
synthesis method which produces static concurrent programs in the
synchronization skeleton notation can be used, e.g.,
\cite{AAE98,AE01,EC82}.

Since the create transitions affect the actual code of $\SP$, we define them first.
The create transitions are determined by the intended meaning of the
\emph{create} rule, together with the constraint that creating a new
pair-program does not change the current state of existing
pair-programs.

\bd[Create transitions]
\label{def:create-trans}
Let $s,t$ be configurations. Then 
$(s, \create, t)$ is a \intrdef{create transition} iff
there exists $\set{i,j} \not\in \pairs{s}$ such that
\bn

\item $\tpl{\set{i,j},\spec_{ij}} \in create(s)$, i.e., 
      the rule for adding new pair-specifications allows the pair-specification
      $\tpl{\set{i,j},\spec_{ij}}$ to be added in global-state $s$.

\item $t.\I = s.\I  \un \tpl{\set{i,j},\spec_{ij}}$, and\\
      $t.\Pr = s.\Pr  \un \set{\SYij}$, where 
     $\SYij \sat \spec_{ij}$.

\item \label{def:gstd:create:consistent}
      $t \up ij$ is a reachable state of $\SYij$, and
      if $i \in \procs{s}$ then $t \up i = s \up i$, and
      if $j \in \procs{s}$ then $t \up j = s \up j$

\item for all $\{k,\l\} \in \pairs{s}: s.\Ps(\{k,\l\}) = t.\Ps(\{k,\l\})$

\en
\ed
Instead of a process index, we use a constant label $\create$ to indicate
a create transition.

Our synthesis method is given by the following.
\bd[Pairwise synthesis]
\label{def:dyn-pair-syn}
In configuration $s$, the synthesized program
$\SP$ is $\parallel_{i \in \procs{s}} P_i$, where 
$P_i = \gci{j \in s.\I(i)} \P{i}{j}$.

The set of initial configurations $S_0$ of $\SP$ consists of all 
$s$ such that 
(1) $s.\I = \IPairs$, 
(2) $s.\Pr$ contains exactly one pair-program $\SYij$ for each
    $\pspecij \in \IPairs$,
(3) $\SYij \sat \spec_{ij}$, and
(4) $s.\Ps(\{i,j\}) \in S_{ij}^0$ for all $\{i,j\} \in \pairs{s}$.
\ed
Another way to characterize process $P_i$ of $\SP$ is that 
$(s_i, \gci{j \in s.\I(i)} \gdi{\l \in [1:n_j]} \CB{i}{j}{\l} \ar \CA{i}{j}{\l}, t_i)$
is an arc in $P_{i}$
iff
$\forall j \in s.\I(i):$ 
$(s_i, \gdi{\l \in [1:n_j]} \CB{i}{j}{\l} \ar \CA{i}{j}{\l}, t_i)$
is an arc in $\P{i}{j}$.
Definition~\ref{def:dyn-pair-syn} gives the initial configurations $S_0$
of $\SP$, and the code of $\SP$ as a function of the $s.\I$ and
$s.\Pr$ components of the current configuration $s$. The code of $\SP$
does not depend on the the $s.\Ps$ component of $s$, which gives the
values of the atomic propositions and shared variables, i.e., the
state.  Definition~\ref{def:create-trans} shows how $s.\I$ and
$s.\Pr$ are changed by create transitions.  We assume that the
$S_{ij}^0$ are such that $S_0 \ne \emptyset$, i.e., there exist
consistent configurations that project onto a state in each
$S_{ij}^0$.

Since a configuration of $\SP$ determines both the state and the code
of all processes, the normal transitions that can be executed in a
configuration are determined intrinsically by that configuration,
Definition~\ref{def:dyn-pair-syn}, and the semantics of
synchronization skeletons, as follows.

\bd[Normal transitions]
\label{def:normal-trans}
Let $s,t$ be configurations and $i \in \procs{s}$. Then 
$(s, i, t)$ is a \intrdef{normal transition} iff
\bn

\item
there exist local states $s \pj i$, $t \pj i$ of $P_i$ such that,
for all $\set{i,j} \in \pairs{s}$, there exists an arc
        $(s \up i,
          \gdi{\l \in [n_j]} \CB{i}{j}{\l} \ar \CA{i}{j}{\l}, 
          t \up i)$
in $\P{i}{j}$ such that\\
        $~~\ex m \in [n_j]:$
             $s \up ij(\CB{i}{j}{m}) = true$ and\\
\hspace*{4em}  $<(s \pj ij) \pj \SH_{ij}> \CA{i}{j}{m} <(t \pj ij) \pj \SH_{ij}>$

\item for all $j$ in $\procs{s} - \{i\}$: $s \up j = t \up j$, and

\item for all $\set{j,k}$ in $\pairs{s}$, $i \not\in \set{j,k}$:
                      $s \up {jk} = t \up {jk}$.

\item $s.\I = t.\I$ and $s.\Pr = t.\Pr$

   \en
\ed
Thus, $P_i$ can execute a transition from global state $s$ to global state $t$
only if, for every $\set{i,j} \in \pairs{s}$, $\P{i}{j}$ can execute a
transition from $s \pj ij$ to $t \up ij$.
Also, $P_i$ reads the local state of its neighbors,
and reads/writes variables that are shared pairwise, i.e., between
$P_i$ and exactly one neighbor. Thus $\SP$ enjoys a \intr{spatial 
locality} property, which is useful when implementing $\SP$ in
atomic read/write memory.

$< (s \pj {ij}) \up \SH_{ij} > A < (t \pj {ij}) \up \SH_{ij} >$
is Hoare triple notation \cite{Ho69} for total correctness,
which in this case means that execution of $A$ always
terminates,\footnote{Termination is obvious, since the right-hand side
of $A$ is a list of constants.}
and, when the shared variables in $\SH_{ij}$ have the values
assigned by $s \pj ij$, leaves these variables with the values assigned by
$t \pj {ij}$.
$s \up ij(\CB{i}{j}{m}) = true$
states that the value of guard $\CB{i}{j}{m}$ in
state $s_{ij}$ is $true$.

The semantics of the synthesized program $\SP$ is given by its global
state transition diagram (GSTD), which is obtained by starting with the
initial configurations, and taking the closure under all the normal
and create transitions.

\bd[Global-state transition diagram of $\SP$]
\label{def:dynamic:gstd}
The semantics of $\SP$ is given by the
structure $\MP = (S_0, S, R_n, R_c)$ where
\bn

\item $S_0$ is the set of initial configurations of $\SP$, and
consists of all the
configurations $s_0$ such that $s_0 = \tpl{\IPairs,\Pr,\Ps}$,
$\Pr = \{ \SYij ~|~ \set{i,j} \in \pairs{\IPairs} \}$, and
$\Ps(\set{i,j}) \in S^0_{ij}$, i.e., 
the pair-specifications in $\IPairs$ are initially active, and
all pair-programs are in one of their start states.

\item $S$ is the set of all configurations such that 
	(1) $S_0 \sub S$ and
        (2) if $s \in S$ and there is a normal or create transition
	from $s$ to $t$, then $t \in S$.

\item $R_n \subseteq S \times \Pids \times S$ is a
transition relation consisting of the normal transitions of $\SP$, 
as given by Definition~\ref{def:normal-trans}.

\item $R_c \subseteq S \times \create \times S$ is a
transition relation consisting of the create transitions of $\SP$,
as given by Definition~\ref{def:create-trans}.
\en
\ed
It is clear that $R_c$ and $R_n$ are disjoint.

The creation of a pair-program is modeled in the above definition as
a single transition. At a lower level of abstraction, this creation is
realized by a protocol which synchronizes the ``activation'' of
$\SYij$ with the current computation of $P_i$ and $P_j$, if they are
already present. We give details in the full paper.

Let $M_{ij} = (S^0_{ij}, S_{ij}, R_{ij}, V_{ij})$ be the GSTD of
$\SYij$ as defined in Section~\ref{sec:ACTLS}.
$M_{ij}$ gives the semantics of $\SYij$ \emph{executing in isolation}.

\subsection{The Creation Protocol}
\label{sec:dynamic:creation}

When a new pair-program $\SYij$  is to be added, it must be
synchronized with $P_i$ and $P_j$, if these are already present, so
that the (pair-consistency) requirement is not violated.

\smpage{
\begin{tabbing}
aaaa\=aaaa\=aaaa\=aaaa\=\kill
    \>$\CREATE(\SYij)$\\[1ex]
1.  \>\IF $P_i$ is alive, \THEN send $P_i$ a request to halt execution;\\
2.  \>\IF $P_j$ is alive, \THEN send $P_j$ a request to halt execution;\\
3.  \>Wait for the necessary acknowledgments from $P_i$, $P_j$;\\
4.  \>Select a reachable state $s_{ij}$ of $M_{ij}$ such that 
	$s_{ij} \pj i = s_i$ if $P_i$ is alive, and  
	$s_{ij} \pj j = s_j$ if $P_j$ is alive.\\
    \>(We require that the creation rule imposes sufficient constraints on
       pair-program creation so\\
    \>that this is guaranteed to hold).\\
5.  \>Set the current state of $\SYij$ to $s_{ij}$\\
6.  \>Send $P_i$, $P_j$ permission to resume execution
\end{tabbing}
}

\section{Soundness of the Method for Dynamic Programs}
\label{sec:dynamic:soundness}

Let $\pi$ be a computation path of $\SP$. 
Let $J \sub \Pids \times \Pids$ be such that 
$J \sub \pairs{s.\I}$ for all $s$ along $\SP$.
Then, the
\intr{path-projection} of $\pi$ onto $J$, denoted $\pi \up
J$, is obtained as follows. 
Start with the first configuration $s$ along $\pi$ such that $\pairs{s} \ints J
\ne \emptyset$. (If no such configuration exists, then $\pi \pj J$ is
the empty sequence.)
Replace every configuration $t$ that occurs after $s$ along $\pi$ by $t
\up J$, and then remove all transitions $t \la{i} t'$  along $\pi$ such that 
$P_i$ is not a process in some pair in $J$, 
coalescing the source and target states of all such
transitions,  which must be the same, since they do not refer to $P_i$.
Define $M_J$ to be the $\MP$ for the case when $\IPairs = J$, and no
create transitions occur, i.e., the set of active pairs is always $J$.

Let $M_{ij} = (S^0_{ij}, S_{ij}, R_{ij})$ 
be the global state transition diagram of $\SYij$, as given by
Definition~\ref{def:pair-structure}. 
$S^0_{ij}$, $S_{ij}$ are the set of initial states, set of all states, respectively,
of $M_{ij}$.
$R_{ij} \subseteq S_{ij} \times \{i,j\} \times S_{ij}$
is the sets of transitions of $M_{ij}$.
$M_{ij}$ and $\MP$
can be interpreted as $\ACTL$ structures. $M_{ij}$
gives the semantics of $\SYij$ \emph{executing in isolation},
and $\MP$ gives the semantics of $\SP$.  Our main
soundness result below (the large model theorem) relates the $\ACTL$ formulae that hold in
$\MP$ to those that hold in $M_{ij}$.
We characterize transitions in $\MP$ as compositions
of transitions in all the relevant $M_{ij}$:

\bl[Transition mapping] %
\label{lem:dynamic:trans-map}
For all configurations $s, t \in S$ and $i \in \procs{s}$:\\
$s \la{i} t \in R_n ~\mathrm{iff}$\\
\ind \ind \ind $\fa j \in s.\I(i) \d
                  s \up ij \la{i} t \up ij \in R_{ij}$  and\\
\ind \ind \ind $\fa \set{j,k} \in \pairs{s}, i \not\in \set{j,k} \d
                  s \up {jk} = t \up {jk}$.
\el
\bpr
In configuration $s$, the constraints on a transition by $P_i$ are given by
exactly the pair-programs of which $P_i$ is a member, i.e., those 
$(i,j) \in \pairs{s}$. If all such pairs permit a transition 
($\fa j \in s.\I(i) \d s \up ij \la{i} t \up ij \in R_{ij}$), 
and if all pair-programs in which $P_i$ is not a member do not execute a
transition
($\fa \set{j,k} \in \pairs{s}, i \not\in \set{j,k} \d s \up {jk} = t \up {jk}$),
then $P_i$ can indeed execute the transition $s \la{i} t$, according to the
semantics of $\MP$. The other direction follows by similar reasoning. 
The technical formulation of this argument follows exactly the same lines as the
proof of  Lemma 6.4.1 in \cite{AE98}.
\epr

\bco[Transition mapping]
\label{cor:dynamic:trans-map}
For all configurations $s, t \in S$, $J \sub \pairs{s}$, and $i \in \procs{J}$,
if  $s \la{i} t \in R_n$, then $s \up J \la{i} t \up J \in R_J$.
\eco

\bl[Path mapping] %
\label{lem:dynamic:path-map}
If $\pi$ is a path in $M$, and let $J \sub \Pids \times \Pids$ be such
that $J \sub \pairs{s}$ for every configuration $s$ along $\pi$.
Then $\pi \up J$ is a path in $M_J$.
\el
\bpr
The proof carries over from \cite{AE98} with the straightforward modifications to
deal with $\create$ transitions.
\epr

In particular, when $J = \{ (i,j)\}$, Lemma~\ref{lem:dynamic:path-map} forms
the basis for our soundness proof, since it relates computations of
the synthesized program $\SP$ to computations of the pair-programs.

\subsection{Deadlock-Freedom}
\label{sec:dynamic:deadlock}

In our dynamic model, the definition of wait-for-graph is essentially the same 
as the static case (Definition~\ref{def:static:wait-for-graph}), except that 
the set of process nodes are also a function of the current configuration.

\bd[Wait-for-graph  $W(s)$]
\label{def:dynamic:wait-for-graph}
Let $s$ be an arbitrary configuration. The {\em wait-for-graph} $W(s)$ of $s$
is a directed bipartite graph, where
\begin{nlst1}

\item the nodes of $W(s)$ are
   \begin{nlst2}
   \item the processes $\{ P_{i} ~|~ i \in \procs{s} \}$, and
   \item the arcs
             $\{ a_i ~|~ i \in \procs{s} \mbox{~and~} a_i \in P_{i}
                           \mbox{~and~} s \up i = a_i^I.start \}$
   \end{nlst2}

\item there is an edge from $\PP{i}{I}$ to every node of the form
      $a_i$ in $W(s)$, and

\item there is an edge from $a_i$ to $\PP{j}{I}$ in $W(s)$ if and only if
                  $\set{i,j} \in \pairs{s}$ and $a_i \in W(s)$ and
                  $s \up ij(a_i.guard_j) = \lfalse$.
\end{nlst1}
\ed
Recall that $a_i.guard_j$ is the conjunct of the guard of arc $a_i$
which references the state shared by $P_i$ and $P_j$ (in effect,
$\AP_j$ and $\SH_{ij}$).
As before, we characterize a deadlock as the occurrence in the wait-for-graph of
a \intr{supercycle}:
\bd[Supercycle]
\label{def:dynamic:supercycle}
$SC$ is a \intrdef{supercycle} in $W(s)$ if and only if:
\begin{nlst1}
   \item $SC$ is nonempty,
   \item if $\PP{i}{I} \in SC$ then for all $a_i$ such that
$a_i \in W(s)$, $\PP{i}{I} \lra a_i \in SC$, and
   \item if $a_i \in  SC$ then there exists $\PP{j}{I}$ such that
$a_i \lra \PP{j}{I} \in W(s)$ and $a_i \lra \PP{j}{I} \in SC$.
\end{nlst1}
\ed
Note that this definition implies that $SC$ is a subgraph of $W(s)$.

To extend the wait-for-graph condition
(Section~\ref{sec:static:wait-for-cond})  to the dynamic model, we
need to take the create transitions ($R_c$) into account.
Thus, we modify the wait-for-graph condition as follows.
In addition to the static Wait-For-Graph Condition of
Definition~\ref{def:static:wait-for-cond}, we require that a newly added
pair-machine have at least one of its processes initially enabled.

\bd[Dynamic wait-for-graph condition]
\label{def:dynamic:wait-for-cond}
Let $k \in \Pids$, and let $t_k$ be an arbitrary local state of $\hat{P}_k$, and
let $n$ be the number of outgoing arcs of $t_k$ in $\hat{P}_k$. 
Let $s,t$ be arbitrary configurations such that either 

\bn

\item \label{def:dynamic:wait-for-cond:normal}
$(s, k, t) \in R_n$, 
$\pairs{s} = \pairs{t} = \{ \{j,k\}, \{k,\l_1\},\ldots, \{k,\l_n\} \}$, 
$k \not\in \{j, \l_1,\ldots,\l_n\}$, and $t \pj k = t_k$, or

\item \label{def:dynamic:wait-for-cond:create}
$(s, \create, t) \in R_c$, 
$\pairs{s} = \{ \{k,\l_1\},\ldots, \{k,\l_n\} \}$, 
$\pairs{t} = \{ \{j,k\}, \{k,\l_1\},\ldots, \{k,\l_n\} \}$, 
$k \not\in \{j, \l_1,\ldots,\l_n\}$, and $t \pj k = t_k$.

\en
Then,\medskip\\ 
\ind 	$\fa a_j \d (a_j \lra \PP{k}{} \not\in W(t))$
	~or~
        $\ex a_k \in W(t) \d
               (\fa \l \in \{\l_1,\ldots,\l_n\} \d
                      a_k \lra \PP{\l}{} \not\in W(t))$.

\ed

\bt[Dynamic supercycle-free wait-for-graph]
\label{thm:dynamic:supercycle-free-waitfor}
If the wait-for-graph condition holds, and $W(s_0)$ is
supercycle-free for every initial configuration $s_0 \in S_0$, then for
every reachable configuration $t$ of $\MP$, $W(t)$ is supercycle-free.
\et
\bpr
Similar to the proof of Theorem~\ref{thm:static:supercycle-free-waitfor}
with straightforward adaptations to deal with the create transitions 
(assumption~\ref{def:dynamic:wait-for-cond:create} of 
Definition~\ref{def:dynamic:wait-for-cond}).
\epr

\subsubsection{Establishing Deadlock-freedom}

We show that the absence of supercycles in the wait-for-graph of a
configuration implies that there is at least one enabled move in that configuration.
The proofs are very similar to the static case, and are omitted.

\bp[Supercycle \cite{AE98}]
\label{prop:dynamic:supercycle}
If $W(s)$ is supercycle-free, then
     some move $a_i$ has no outgoing edges in $W(s)$.
\ep

\bt[Deadlock freedom]
\label{thm:dynamic:dead-free}
If, for every reachable configuration $s$ of $\MP$, $W(s)$ is supercycle-free, then
$\MP, S_0 \sat \AG \EX \ltrue$.
\et

\subsection{Liveness}
\label{sec:dynamic:liveness}

To assure liveness properties of the synthesized program $\SP$, we 
assume a form of weak fairness. Let $\CL(f)$ be the set of all
subformulae of $f$, including $f$ itself. Let $ex_i$ be an assertion
that is true along a transition in a structure iff that transition
results from executing process $i$. Let $en_i$ hold in a configuration $s$
iff $\PP{i}{I}$ has some arc that is enabled in $s$.
Let $\normal$ be an  assertion
that is true along all transitions of $\MP$ that are drawn from $R_n$.
Let $\pi$ be a fullpath of $\MP$. Define
$\states{\pi} = \{ s ~|~ \mbox{$s$ occurs along $\pi$} \}$.
Define $\procs{\pi} = \UN_{s \in \states{\pi}} \procs{s}$, and
       $\pairs{\pi} = \UN_{s \in \states{\pi}} \pairs{s}$.

\bd[Weak blocking fairness $\f_b$]
\label{def:dynamic:weak-blocking-fairness}

\hspace*{\fill}
  \mbox{$~~{\f_b(\pi)} ~\df~ 
          \AND_{i \in \procs{\pi}} \ea(\blk_i \land en_i) \imp \io ex_i$}
\hspace*{\fill}
\ed
Weak blocking fairness requires that a process that is continuously
enabled and in a sometimes-blocking state is eventually executed.

\bd $($\textup{\textbf{Weak eventuality fairness,}} $\f_\l$$)$\\
\label{def:dynamic:weak-eventuality-fairness}
\hspace*{\fill}
\mbox{${\f_\l}(\pi) ~\df~ \AND_{(i,j) \in \pairs{\pi}}
         (\ea en_i \lor \ea en_j) \land \ea \pnd_{ij} \imp \io (ex_i \lor ex_j)$.}
\hspace*{\fill}
\ed
Weak eventuality fairness requires that if an eventuality is
continuously pending, and one of $\PP{i}{I}$ or $\PP{j}{I}$ is continuously
enabled, then eventually one of them will be executed.

\bd[Creation fairness $\f_c$] 
\label{def:dynamic:creation-fairness}
${\f_c} ~\df~ \io \normal$.
\ed
A fullpath $\pi$ satisfies creation fairness iff it contains an
infinite number of normal transitions.

A fullpath $\pi$ is \emph{fair} iff 
$\pi \sat_L \f_b(\pi) \land \f_\l(\pi) \land \f_c$,
where $\sat_L$ is the satisfaction
relation of propositional linear-time temporal logic \cite{Em90,MW84}.
Our overall fairness notion $\f$ is thus the conjunction of weak
blocking fairness, weak eventuality fairness, and creation fairness:
$\f \df \f_b \land \f_\l \land \f_c$.

Let $\aen_j \df	\fa a_j^i \in \P{j}{i} \d (\stof{a_j^i.start} \imp  a_j^i.guard)$,
i.e., $\aen_j$
holds iff every arc of $\P{j}{i}$
whose start state is a component of the current $ij$-state $s_{ij}$ is
also enabled in $s$.
We say that $P_k$ \emph{blocks} $P_i$ in configuration $s$ iff, in
$W(s)$, there is a path from $P_i$ to $P_k$. 
Define $\Wt{ij}{s}$ to be the set of all $k$ such that 
there is a path in $W(s)$ from at least one of $P_i$ or $P_j$ to $P_k$.
Thus, $\Wt{ij}{s}$ is the set of processes that block the pair-program
$\SYij$ from executing some arc of  $\P{i}{j}$ or $\P{j}{i}$.

\bd[Liveness condition for dynamic programs]
\label{def:dynamic:liveness-cond}
The liveness condition is the conjunction of the following:
\bn 

\item \label{def:dynamic:liveness-cond:pair-machines}
Let $s$ be an arbitrary reachable configuration.
Then, for every $\{i,j\} \in \pairs{s}$:\linebreak
	$M_{ij}, S_{ij}^0 \sat \AG\A(\G ex_i \imp \ea \aen_j)$

\item \label{def:dynamic:liveness-cond:creation}
Let $s$ be an arbitrary reachable configuration.
For every $\set{i,j} \in \pairs{s}$ such that 
$s \sat \pnd_{ij}$, the following must hold.
There exists a finite $W \sub \Pids$  such that 
for all $t$ reachable from $s$ along paths in which 
$\pnd_{ij}$ holds in all configurations, 
$\Wt{ij}{t} \sub W$.

\en
\ed
The first condition above is a ``local one,'' i.e., it is evaluated on
pair-programs in isolation. It requires that, for every pair-program
$\SYij$, when executing in isolation, that if $\Pij$ can execute
continuously along some path, then there exists a suffix of that path
along which $\Pij$ does not block any arc of $\Pji$.  The second
condition is ``global,'' it requires that a process is not forever
delayed because new processes which block it are constantly being
added.

Given the liveness condition and the absence of deadlocks and
the use of $\f$-fair scheduling, we can show that one of $\PP{i}{I}$
or $\PP{j}{I}$ is guaranteed to be executed from any configuration 
whose $ij$-projection has a pending eventuality.
Let $\satf$ be the satisfaction relation of $\CTLS$ when the path
quantifiers $\A$ and $\E$ are restricted to fair fullpaths ($\A$: for
all fair fullpaths, $\E$: for some fair fullpath) \cite{EL87}.

\bl[Progress for dynamic programs]
\label{lem:dynamic:progress}
Let $s$ be an arbitrary reachable configuration and
$\set{i,j} \in \pairs{s}$. If
\bn
   \item \label{ass:dynamic:liveness} the liveness condition holds, and
   \item \label{ass:dynamic:supercycle} for every reachable configuration $u$, $W(u)$ is
supercycle-free, and
   \item \label{ass:dynamic:pending} $M_{ij}, s \up ij \sat \neg h_{ij} \land \AF h_{ij}$
		             for some $h_{ij} \in \CL(\spec_{ij})$, then
\en
\hspace*{\fill}
	$\MP, s \sat_{\f} \AF(ex_i \lor ex_j)$.
\hspace*{\fill}
\el
\bpr
By assumption~\ref{ass:dynamic:supercycle} and Theorem~\ref{thm:dynamic:dead-free}, 
$\MP, S_0 \sat \AG \EX \ltrue$. Hence every fullpath in $\MP$ is infinite.
Let $\pi$ be an arbitrary $\f$-fair fullpath starting in $s$.
If $\MP,\pi \sat \F(ex_i \lor ex_j)$, then we are done. Hence we assume
\bleqn{(*)}
	$\pi \sat \G(\neg ex_i \land \neg ex_j)$
\eleqn
in the remainder of the proof.
Let $t$ be an arbitrary configuration along $\pi$. By 
clause~\ref{def:dynamic:liveness-cond:creation} of the liveness
condition for dynamic programs
(Definition~\ref{def:dynamic:liveness-cond}),
$\Wt{ij}{t} \sub W$ for some finite $W \sub \Pids$. 
Hence, these exists a configuration $v$ along $\pi$ such that, for all
subsequent configurations $w$ along $\pi$, $\Wt{ij}{w} \sub \Wt{ij}{v}$, i.e.,
after  $v$, the set of processes that block $\SYij$ does not increase.
Now consider the static concurrent program 
$P_J$ with interconnection relation 
$J = \{ \set{k,l} ~|~ 
	\set{k,l} \in \pairs{v} \mbox{ and } \set{k,l} \sub \Wt{ij}{v} \}$
and initial state
set $\{v \pj J\}$. By applying Lemma~\ref{lem:static:progress} to $P_J$, we
conclude that $M_J, v \pj J \sat_{\f} \AF(ex_i \lor ex_j)$.
Now let $\rho_J = \pi^v \pj J$, where $\pi^v$ is the infinite suffix of
$\pi$ starting in $v$. We now establish
\bleqn{(**)}
	$\rho_J$ is an infinite path in $M_J$
\eleqn
given the assumption that (*) holds.
From (*) and weak eventuality fairness 
(Definition~\ref{def:dynamic:weak-eventuality-fairness}),
we see that $\Wt{ij}{t}$ is nonempty for every configuration $t$ along $\pi$,
since otherwise one of $P_i$, $P_j$ would be executed.
By definition, there is no path in $W(t)$ from a process in $\Wt{ij}{t}$
to a process outside $\Wt{ij}{t}$. Hence, by 
assumption~\ref{ass:dynamic:supercycle} and
Proposition~\ref{prop:dynamic:supercycle}, there exists some 
$P_k \in \Wt{ij}{t}$ such that $P_k$ has an enabled move in configuration $t$.
Since this holds for all configurations $t$ along $\pi$, we conclude by 
Weak blocking fairness 
(Definition~\ref{def:dynamic:weak-blocking-fairness}), that 
infinitely often along $\pi$, some process in $\Wt{ij}{v}$ is
executed.
Hence, by Definition~\ref{def:pproj} and the definition of $J$, $\rho_J$ is infinite.

From Lemma~\ref{lem:dynamic:path-map} $\rho_J$ is a path in $M_J$. 
Hence,  $\rho_J$ is a fullpath in $M_J$. 
By Definition~\ref{def:pproj}, the first state of $\rho_J$ is $v \pj J$.
Hence, by  $M_J, v \pj J \sat_{\f} \AF(ex_i \lor ex_j)$, we have
$\rho_J \sat \F(ex_i \lor ex_j)$. 
From $\rho_J = \pi^v \pj J$ and Definition~\ref{def:pproj}, we conclude 
$\pi^v \sat \F(ex_i \lor ex_j)$. Hence, 
$\pi \sat \F(ex_i \lor ex_j)$, contrary to assumption.
\epr

\subsection{The Large Model Theorem for Dynamic Programs}
\label{sec:dynamic:large-model}

The large model theorem establishes the soundness of our synthesis
method.
The large-model theorem states that any subformula of $\spec_{ij}$
which holds in the $ij$-projection of a configuration $s$ also holds
in $s$ itself. That is, correctness properties satisfied by a
pair-program executing in isolation also hold in the synthesized
program $\SP$.

\bt[Large model]
\label{thm:dynamic:large-model}
Let $i,j \in \Pids$ and let $s$ be an arbitrary
reachable configuration in $\MP$ such that $\pspecij \in s.\I$, 
where $\spec_{ij}$ is an $\ACTLmij$ formula.
If
\bn

\item the liveness condition  for dynamic programs holds,

\item $W(u)$ is supercycle-free for every reachable configuration $u$ in $\MP$, and

\item $M_{ij}, s \up {ij} \sat f_{ij}$ for some $f_{ij} \in \CL(\spec_{ij})$,
\en
then\\
\hspace*{\fill}
	$\MP, s \satf f_{ij}$.
\hspace*{\fill}
\et
\bpr
The theorem follows from Theorem~\ref{thm:dynamic:dead-free} and
Lemma~\ref{lem:dynamic:progress} in essentially the same way that
Theorem~\ref{thm:static:large-model} follows from
Theorem~\ref{thm:static:dead-free} and
Lemma~\ref{lem:static:progress}, i.e., the static case.
The proof is 
very similar, since the statements (but not the proofs) of 
Theorem~\ref{thm:dynamic:dead-free} and Lemma~\ref{lem:dynamic:progress} 
are identical to those of 
Theorem~\ref{thm:static:dead-free} and Lemma~\ref{lem:static:progress}.
The only difference in the proof is
in dealing with create transitions. This is straightforward, since 
$\SYij$ is created with its current state set to one of its reachable states,
and so the same projection relationships hold between $\MP$ and 
$M_{ij}$ in the dynamic case as between $M_I$ and $M_{ij}$ in the static case,
in particular, Lemma~\ref{lem:dynamic:path-map} provides the
exact dynamic analogue for Lemma~\ref{lem:static:gen-pmap}, and is 
the only projection result used in establishing the large model theorem.
The only difference is that in
the dynamic case the projection starts from the point
that $\SYij$ is created. Since we do not require computation paths to start from
an initial state, this does not pose a problem. We note that the only 
result for static programs that involves reachability is
Corollary~\ref{cor:static:reach}, and this is only used to establish
deadlock-freedom for the static case. For the dynamic case, deadlock freedom is 
guaranteed by the dynamic wait-for-graph condition 
(Definition~\ref{def:dynamic:wait-for-cond}), which contains an explicit clause
(clause~\ref{def:dynamic:wait-for-cond:create}) to deal with creation.
\epr

We note the important case of $f_{ij} = \AG g_{ij}$, i.e., $f_{ij}$ expresses a 
\intr{global} property, since $g_{ij}$ holds in all configurations reachable
from $s$.

\section{Implementation in Atomic read/write Shared Memory}
\label{sec:atomic-implementation}

We now show how the synthesized program can be implemented in atomic
read/write memory. 
To break down the atomicity of an arc in the synthesized program, we
require that, in all pair-programs, all guards are \emph{temporarily
stable}, \cite{Ka86}, that is, once the guard holds, it continues to
hold until some arc is executed, not necessarily the arc corresponding
to the guard.

We generalize this discussion as follows.
Let $(s_i,  \gdi{\l \in \oneton} B_\l \ar A_\l, t_i)$ be an arc of
$\P{i}{j}$ in pair-program $\SYij$. 
We require
\bleqn{(TSTAB)}
$M_{ij}, S_{ij}^0 \sat 
\AND_{\l \in \oneton} 
     \AG( (\stof{s_i} \land \CB{i}{j}{\l}) \imp 
          \A[(\stof{s_i} \land \CB{i}{j}{\l}) \,\Uw\, \neg s_i] 
        )$.
\eleqn
Now consider, in process $P_i$ of $\SP$, the arc in 
$P_{i}$ $s_i$ to $t_i$. By Definition~\ref{def:dyn-pair-syn}, this arc has the
label
$\gci{j \in s.\I(i)} \gdi{\l \in [1:n_j]} \CB{i}{j}{\l} \ar \CA{i}{j}{\l}$
in configuration $s$. Since $P_i$ will be explicitly involved in any creation
step which adds a pair of which $P_i$ is a member, we assume that this label 
does not change, for the time being. Now, $P_i$ can evaluate each of 
the $\CB{i}{j}{\l}$ sequentially, rather than simultaneously, 
since once true, each $\CB{i}{j}{\l}$ will remain true until 
$P_i$ executes either the above arc or some other arc.
Once $P_i$ has observed that 
${j \in s.\I(i)}$, there exists ${\l \in [1:n_j]}$ such that
$\CB{i}{j}{\l}$ holds, then $P_i$ can execute the arc.
The condition (TSTAB) can be checked in polynomial time by the
model-checking algorithm of \cite{CES86}.

Execution of the arc will also involve the simultaneous execution of the assignments 
$\CA{i}{j}{\l}$. To break this multiple assignment down into atomic read and
write operations, 
we use efficient solutions to the dining/drinking philosophers problem
\cite{SP88,CM88} to guarantee mutual exclusion of neighboring
processes. Once a process has excluded all its neighbors (i.e., it
``has all the forks''), it can then perform the multiple assignment sequentially.
The following subsection gives details of this implementation.

As an alternative to using dining/drinking philosophers,
if we have available hardware operations such as
compare-and-swap. or load-linked/store conditional, then we can use the
constructions of \cite{M97,M00}. These algorithms permit the efficient, 
wait-free implementation of the multiple assignments.
\subsection{Implementation using underlying dining/drinking philosophers algorithm}
\label{sec:impl:dphils}

The problem is to implement every move
$a_i = \gci{j \in s.\I(i)} \gdi{\l \in [1:n_j]} \CB{i}{j}{\l} \ar \CA{i}{j}{\l}$
of every process $P_i$, in configuration $s$. The implementation consists of the
following three procedures.
The first, $\POLL(P_i,a_i)$ repeatedly polls all the guards of the move $a_i$,
until a guard $\CB{i}{j}{\l}$ for each neighbor $P_j$ of $P_i$ is found which is
true. When this occurs, the move $a_i$ can be executed.

\smpage{
\begin{tabbing}
aaaa\=aaaa\=aaaa\=aaaa\=\kill
    \>$\POLL(P_i,a_i)$\\[1ex]
1.  \>$X[a_i] :=  s.\I(i)$;\\
2.  \>\REPEAT\\
    \>  \>poll all the $\CB{i}{j}{\l}$ for $j \in X$, $\l \in [1:n_j]$;\\
    \>  \>$\FOR$ every $j$ such that $\CB{i}{j}{\l}$ polled $\ltrue$ for some $\l$\\
    \>  \>   \>$X := X - \{j\}$;\\
    \>  \>   \>$choice_i^j[a_i] := \l$ \\
    \>\UNTIL $X[a_i] = \emptyset$

\end{tabbing}
}

Now in a local state $s_i$, $P_i$ will usually have a choice of several moves.
The second procedure, $\CHOOSE(P_i, s_i)$,
repeatedly poll the guards of all such moves, until one
is found all of whose guards are true. This move can then be executed by $P_i$.
The actual execution is carried out by the $\EXECUTE(P_i, s_i)$ procedure. 
$\EXECUTE(P_i, s_i)$ first invokes $\CHOOSE(P_i, s_i)$ to determine which move to
execute.
It then obtains the exclusive access to all the shared variables that execution
of $a_i$ updates, and exclusive access to the atomic propositions of $P_i$.
Once all necessary locks are obtained, the move chosen move $a_i$ can be
executed in an ``atomic'' manner.

\smpage{
\begin{tabbing}
aaaa\=aaaa\=aaaa\=aaaa\=\kill
    \>$\CHOOSE(P_i, s_i)$\\[1ex]
1.  \>Let $a_i^1 \ldots a_i^k$ be all the moves of $P_i$ with start state $s_i$;\\
2.  \>Invoke $\POLL(P_i,a_i^1) \ldots \POLL(P_i,a_i^k)$ simultaneously, and in an 
      ``interleaved'' manner, i.e., interleave\\
    \>the executions of   $\POLL(P_i,a_i^1) \ldots \POLL(P_i,a_i^k)$;\\
3.  \>Let $a_i^c = 
(s_i,\gci{j \in s.\I(i)} \gdi{\l \in [1:n_j]} \CB{i}{j}{\l} \ar \CA{i}{j}{\l},t_i)$
 be the first move for which $X[a_i^c] = \emptyset$ becomes true\\
4.  \>$\RETURN(a_i^c, choice_i^j)$
\end{tabbing}
}

\smpage{
\begin{tabbing}
aaaa\=aaaa\=aaaa\=aaaa\=\kill
    \>$\EXECUTE(P_i, s_i)$\\[1ex]
1.  \>Invoke $\CHOOSE(P_i, s_i)$ and let $a_i^c, choice_i^j$ be the returned values;\\
2.  \>\FORALL ${j \in s.\I(i)}$ \DO\\
    \>  \>obtain a lock on all variables in $\CA{i}{j}{\l}$, $\l =
          choice_i^j[a_i^c]$, e.g., by using a drinking philosophers algorithm;\\
    \>  \>obtain a lock on the atomic propositions of $P_i$ (i.e., those in $\AP_i$)\\
3.  \>\FORALL ${j \in s.\I(i)}$ \DO\\
    \>  \>execute $\CA{i}{j}{\l}$, $\l = choice_i^j[a_i^c]$;\\
    \>  \>change the local state of $P_i$ to $t_i$\\
4.  \>\FORALL ${j \in s.\I(i)}$ \DO\\
5.  \>  \>release all locks\\
\end{tabbing}
}
\newcommand{\chvar}{6\xspace}
\newcommand{\chprop}{7\xspace}
\newcommand{\chst}{3\xspace}

The overall implementation is given by the procedure $\MAIN(P_i)$, which
implements the process $P_i$. $\MAIN(P_i)$ repeatedly invokes 
$\EXECUTE(P_i,s_i)$, where $s_i$ is the current local state of $P_i$.
The low-level concurrent program $P_r$ is then given by the concurrent
composition of $\MAIN(P_i)$ for every process $P_i$ that has been created so far.
Let $M_r$ be the global state transition diagram of $P_r$.
$M_r$ can be formally defined in a similar manner to $\MP$
(Definition~\ref{def:dynamic:gstd}). 

\smpage{
\begin{tabbing}
aaaa\=aaaa\=aaaa\=aaaa\=\kill
    \>$\MAIN(P_i)$\\[1ex]
1.  \>Let $s_i$ be an initial local state of $P_i$;\\
2.  \>\REPEAT forever\\
    \>  \>invoke $\EXECUTE(P_i,s_i)$;\\
    \>  \>update $s_i$ to be the resulting local state of $P_i$;\\
3.  \>  \>participate in any outstanding $\CREATE$ protocols, if a
          request to suspend execution has\\
    \>  \>been received
\end{tabbing}
}

Note that $P_i$ participates in executions of $\CREATE$ only when it
is not executing normal transitions. This prevents the interleaving of
the low atomicity implementations of normal and create transitions.
Thus, in particular, during the low atomicity execution of a single 
normal transition, the value of $s.\I(i)$, i.e., the set of neighbors
of $P_i$, does not change. This is essential to the correctness of the
implementation.

\subsection{Soundness of the Implementation in Atomic read/write shared memory}
\label{sec:atomic-soundness}

We show that $M_r$ satisfies the same $\ACTLm_{ij}$ formulae as $\MP$.
Roughly, we can consider $M_r$ to consist of a ``stretched out'' version of
$\MP$, in which each transition of $\MP$ is replaced by a sequence of transitions,
together with all of the possible interleavings that result from this 
refinement of the transitions in $\MP$. Due to our use of locking, this
refinement does not generate any configurations that are unreachable in $\MP$.
Likewise, paths in $M_r$ have ``corresponding'' paths in $\MP$.
Hence, so correctness is preserved.

Let $s, u$ be configurations of $\MP$, $M_r$ respectively. Then define 
$s \sim u$ iff $\fa p \in \AP: s(p) = u(p)$ and $(\fa x \in \SH: s(x) = u(x))$.
Let $\pi, \rho$ be fullpaths of $\MP$, $M_r$ respectively. Then define 
$\pi \sim \rho$ iff 
$\pi$ can be written as a sequence of finite bocks of configurations $\pi_1,\pi_2,\ldots$,
$\rho$ can be written as a sequence of finite bocks of configurations $\rho_1,\rho_2,\ldots$,
and for all $i \ge 0$, for every $s$ in $\pi_i$ and every $u$ in $\rho_i$, 
$s \sim u$.

\bl
\label{lem:low-atomicity:dynamic:path-corr}
Let $s, u$ be configurations of $\MP$, $M_r$ respectively such that $s \sim u$.
Then, for every fullpath $\rho$ of $M_r$ starting in $u$, there exists
a fullpath $\pi$ of $\MP$ starting in $s$ such that $\pi \sim \rho$.
\el
\bpr
We assume that line \chst of $\EXECUTE(P_i,s_i)$ is executed
atomically. This is reasonable, since exclusive access locks to all
the shared variables and atomic propositions modified by 
line \chst of $\EXECUTE(P_i,s_i)$ are obtained first. We do not assume 
the atomic execution of any other part of the implementation algorithm.

Given $\rho$, consider the subsequence
of the transitions of $\rho$ given by the transitions that correspond
to the execution of line \chst of $\EXECUTE(P_i,s_i)$. These are the
only transitions of $\rho$ which change the shared variables and
atomic propositions, and
so affect the truth of $\sim$.
From the construction of the implementation algorithm, 
we can show that there exists a fullpath $\pi$ of $\MP$ starting in
$s$ which executes the same sequence of changes to the shared
variables and atomic propositions. It follows that $\pi \sim \rho$.
\epr

\bt
\label{lem:low-atomicity:dynamic:reduction}
Let $s,u$ be configurations of $\MP$, $M_r$ respectively such that $s \sim u$. 
Let $\pi,\rho$ be fullpaths of $\MP$, $M_r$ respectively such that $\pi \sim \rho$. 
Let $f$ be any formula of $\ACTLSm$.
Then,\\
\ind If $\MP, s \satf f$, then $M_r, u \satf f$.\\
\ind If $\MP, \pi \satf f$, then $M_r, \rho \satf f$.
\et
\bpr
The proof is by induction on the structure of $f$, i.e., by induction on the
number of times rules S2, S3, and P1--3 of the definition of $\ACTLS$ syntax are
applied to generate $f$. Rule S1 of that definition gives the base case.

\textit{Base case}: $f$ is one of $\ltrue$, $\lfalse$, $p$, $\neg p$ for some atomic
           proposition $p$.
Since $s$ and $u$ agree on all atomic propositions, $M_r, u \sat f$ follows
immediately from $\MP,s \sat f$.

\textit{Induction step}: There are several cases.

\case{1}{S2 is applied, and $f$ is $g \lor h$, a state formula} 
Hence $\MP,s \sat g \lor h$. 
By $\ACTLS$ semantics, $\MP,s \sat g$ or $\MP,s \sat h$. 
By the induction hypothesis, $M_r,u \sat g$ or $M_r,u \sat h$. 
Hence, by $\ACTLS$ semantics, $M_r,u \sat g \lor h$.

\case{2}{S2 is applied, and $f$ is $g \land h$, a state formula} 
Hence $\MP,s \sat g \land h$. 
By $\ACTLS$ semantics, $\MP,s \sat g$ and $\MP,s \sat h$. 
By the induction hypothesis, $M_r,u \sat g$ and $M_r,u \sat h$. 
Hence, by $\ACTLS$ semantics, $M_r,u \sat g \land h$.

\case{3}{S3 is applied, and $f$ is $\A g$, a state formula}
Hence $g$ is a path formula. Assume $\MP,s \sat \A g$. 
Let $\rho$ be an arbitrary fullpath of $M_r$ starting in $u$.
By Lemma~\ref{lem:low-atomicity:dynamic:path-corr}, there exists a
fullpath $\pi$ of $\MP$ starting in $s$ such that $\pi \sim \rho$.
Since  $\MP,s \sat \A g$, we have $\MP, \pi \sat g$, by $\ACTLS$ semantics.
From $\pi \sim \rho$ and the induction hypothesis, we obtain $M_r, \rho \sat g$.
Since $\rho$ was chosen arbitrarily from the fullpaths starting in $u$,
we conclude $M_r, u \sat \A g$, by $\ACTLS$ semantics.

\case{4}{P1 is applied, and $f$ is $g$, where $f$ is a path formula and $g$ is a
state formula}.
Assume $\MP,\pi \sat f$. Hence, $\MP,s \sat g$, where $s$ is the first state of 
$\pi$. Let $u$ be the first state of $\rho$. Then, $s \sim u$, by the definition
of $\pi \sim \rho$. By the induction hypothesis,  $\MP,s \sat g$, and 
$s \sim u$, we obtain $M_r,u \sat g$. Hence, by $\ACTLS$ semantics, 
$M, \rho \sat f$.

\case{5}{P2 is applied, and $f$ is $g \lor h$, a path formula} 
Hence $\MP,\pi \sat g \lor h$. 
By $\ACTLS$ semantics, $\MP,\pi \sat g$ or $\MP,\pi \sat h$. 
By the induction hypothesis, $M_r,\rho \sat g$ or $M_r,\rho \sat h$. 
Hence, by $\ACTLS$ semantics, $M_r,\rho \sat g \lor h$.

\case{6}{P2 is applied, and $f$ is $g \land h$, a path formula} 
Hence $\MP,\pi \sat g \land h$. 
By $\ACTLS$ semantics, $\MP,\pi \sat g$ and $\MP,\pi \sat h$. 
By the induction hypothesis, $M_r,\rho \sat g$ and $M_r,\rho \sat h$. 
Hence, by $\ACTLS$ semantics, $M_r,\rho \sat g \land h$.

\case{7}{P3 is applied, and $f$ is $g \U h$, a path formula} 
Assume $\MP,\pi \sat f$. Hence, there exists $i \ge 1$ such that 
$\MP, \pi^{i'} \sat h$ and $(\fa i: 1 \le i < i': \MP, \pi^i \sat g)$.
Let $j'$ be the smallest natural number such that $\pi^{i'} \corr \rho^{j'}$.
By the induction hypothesis, $M_r, \rho^{j'} \sat h$.
Let $j$ be any natural number such that $1 \le j < j'$.
By the definition of $\pi \corr \rho$,
there exists some $i$ such that $1 \le i < i'$ and $\pi^i \corr \rho^j$.
Since $1 \le i < i'$, we have $\MP, \pi^i \sat g$.
Hence, by the induction hypothesis, $M_r, \rho^j \sat g$.
We have thus shown 
$M_r, \rho^{j'} \sat h$ and $(\fa j: 1 \le j < j': M_r, \rho^j \sat g)$.
By $\ACTLS$ semantics, $M_r, \rho \sat g \U h$.

\case{8}{P3 is applied, and $f$ is $g \Uw h$, a path formula} 
Assume $\MP,\pi \sat f$. Hence, by $\ACTLS$ semantics,
$\MP,\pi \sat g \U h$ or $\MP,\pi \sat \G g$.
$\MP,\pi \sat g \U h$ is just Case 6 above. 
$\MP,\pi \sat \G gh$ can also be treated with an argument analogous
to that of Case 6. Hence, we can establish 
$M_r,\rho \sat g \U h$ or $M_r,\rho \sat \G g$.
Thus, $M_r,\rho \sat g \Uw h$.
\epr

\bt[Large model theorem for low-atomicity implementation]
\label{lem:low-atomicity:dynamic:soundness}
Let $i,j \in \Pids$ and let $u$ be an arbitrary
reachable configuration in $M_r$ such that $\pspecij \in u.\I$, 
where $\spec_{ij}$ is an $\ACTLmij$ formula.
If
\bn

\item the liveness condition for dynamic programs holds,

\item $W(v)$ is supercycle-free for every reachable configuration $v$ in $M_r$, and

\item $M_{ij}, s \up {ij} \sat f_{ij}$ for some $f_{ij} \in \CL(\spec_{ij})$,
\en
then\\
\hspace*{\fill}
	$M_r, u \satf f_{ij}$.
\hspace*{\fill}
\et
\bpr
Immediate from Theorem~\ref{thm:dynamic:large-model} and
Theorem~\ref{lem:low-atomicity:dynamic:reduction}. 
\epr

\section{Example---The Eventually Serializable Data Service}
\label{sec:example:esds}

The eventually-serializable data service (ESDS) of \cite{FGLLS99,LLSG92} is a
replicated, distributed data service that trades off immediate
consistency for improved efficiency. A shared data object is
replicated, and the response to an operation at a particular replica
may be out of date, i.e., not reflecting the effects of other
operations that have not yet been received by that replica. Thus,
operations may be reordered \emph{after} the response is issued.
Replicas communicate amongst each other the operations they receive,
so that eventually every operation ``stabilizes,'' i.e., its ordering
is fixed with respect to all other operations. Clients may require an operation
to be \intr{strict}, i.e., stable at the time of response (and so
it cannot be reordered after the response is issued). Clients may also
specify, in an operation $x$, a set $x.\prev$ of other
operations that should precede $x$ (client-specified constraints, $\CSC$).
We let $\Op$ be the (countable) set of all operations, 
$\R$ the set of all replicas (which may increase dynamically),
$\client{x}$ be the client issuing operation $x$,
$\replica{x}$  be the replica that handles operation $x$.
We use $x$ to index over operations, 
$c$ to index over clients, and $r,r'$ to index over replicas.
For each operation $x$, we define a client process $C_c^x$ and a 
replica process $R_r^x$, where $c= \client{x}$, $r = \replica{x}$.
Thus, a client consists of many processes, one for each operation it
issues. As the client issues operations, these processes are created
dynamically.
Likewise a replica consists of many processes, one for each operation it
processes. Thus, we can use dynamic process creation and
finite-state processes to model an infinite-state system, 
such as the one here, which in general handles an unbounded number
of operations with time.
The pair-specifications are as follows. The local structure
specification of a process  are implicitly conjoined with any pair-specification
referring to that process.
The atomic predicates have the following meaning for operation $x$.
$in$ is the initial state. 
$wt$ means that $x$ is submitted but not yet done.
$dn$ means that $x$ is done.
$st$ means that $x$ is table.
$snt$ means that the result of $x$ has been sent to the client.
We give pair-programs for a strict operation $x$. The
pair-programs for a non-strict operation are similar, except that the
transitions from $dn_r^x$ to $st_r^x$ to $[st_r^x ~ snt_r^x]$ can
also be performed in the reverse order (i.e., there is a branch from
the $dn_r^x$ state), since the result of $x$ can be sent before $x$ stabilizes.
For example, Figure~\ref{fig:notstrict} gives the 
pair-program $R_r^x \pl R_{r'}^{x}$ when $x$ is not strict.

\vspace{1.0ex}

\emph{\ul{Local structure of clients $C_c^x$}}
\begin{lst}
   \item[$in_c^x$:]
      $x$ is initially pending

   \item[$\AG( in_c^x \imp ( \AX_c wt_c^x \land \EX_c wt_c^x ) )$ $\land$
	 $\AG( wt_c^x \imp   \AX_c dn_c^x  )$ $\land$
	 $\AG( dn_c^x \imp ( \AX_c dn_c^x \land \EX_c dn_c^x ) )$:]
      $C_c^x$ moves from $in_c^x$ to $wt_c^x$ to $dn_c^x$, and
	 thereafter remains in $dn_c^x$, and
      $C_c^x$ can always move from $in_c^x$ to $wt_c^x$.

   \item[$\AG( ( in_c^x \equiv \neg (wt_c^x \lor  dn_c^x)) \land
               (wt_c^x \equiv \neg ( in_c^x \lor  dn_c^x)) \land
               ( dn_c^x \equiv \neg ( in_c^x  \lor wt_c^x))
         )$:]
      $C_c^x$ is always in exactly one of the states $in_c^x$ (initial
               state),
               $wt_c^x$ ($x$ has been submitted, and the client is
               waiting for a response), or $dn_c^x$ ($x$ is done).
\end{lst}

\vspace{1.0ex}

\emph{\ul{Local structure of replicas $R_r^x$}}
This is as shown in Figures~\ref{fig:submit}, \ref{fig:prev}, and \ref{fig:strict}.
We omit the temporal logic formulae to
save space. They are constructed in an analogous manner to those for
the clients

\vspace{1.0ex}

\emph{\ul{Client-replica interaction, $C_c^x \pl R_{r}^{x}$}},
$x \in \Op$, $c = \client{x}$, $r = \replica{x}$
\begin{lst}

   \item[$\AG(wt_r^x \imp wt_c^x)$:]
	$x$ is not received by its replica before it is submitted

   \item[$\AG(wt_c^x \imp \AF wt_r^x)$:]
	every submitted $x$ is eventually received by its replica

   \item[$\AG(wt_c^x \imp \AF dn_c^x)$:]
	every submitted $x$ is eventually performed

   \item[$\AG(dn_c^x \imp \AG dn_c^x)$:]
	once an operation $x$ is done, it remains done

\end{lst}

\bfg
\begin{center}
\scalebox{0.8}{\setlength{\unitlength}{0.00083333in}
\begingroup\makeatletter\ifx\SetFigFont\undefined%
\gdef\SetFigFont#1#2#3#4#5{%
  \reset@font\fontsize{#1}{#2pt}%
  \fontfamily{#3}\fontseries{#4}\fontshape{#5}%
  \selectfont}%
\fi\endgroup%
{\renewcommand{\dashlinestretch}{30}
\begin{picture}(8336,1597)(0,-10)
\put(8062.500,404.500){\arc{530.330}{3.9270}{8.6394}}
\path(7992.894,179.569)(7875.000,217.000)(7961.409,128.493)
\path(4275,367)(5250,367)
\path(5130.000,337.000)(5250.000,367.000)(5130.000,397.000)
\path(7230.040,336.841)(7350.000,367.000)(7229.960,396.841)
\path(7350,367)(5837,365)
\path(600,367)(525,367)(600,367)
	(675,367)(750,367)(825,367)(900,367)
\path(780.000,337.000)(900.000,367.000)(780.000,397.000)
\put(6142.500,1229.500){\arc{471.010}{4.0625}{8.5039}}
\path(6122.436,1024.410)(6000.000,1042.000)(6099.754,968.862)
\path(600,1267)(525,1267)(600,1267)
	(675,1267)(750,1267)(825,1267)(900,1267)
\path(780.000,1237.000)(900.000,1267.000)(780.000,1297.000)
\put(1185,352){\ellipse{600}{600}}
\put(1050,322){\makebox(0,0)[lb]{\smash{{{\SetFigFont{10}{12.0}{\rmdefault}{\mddefault}{\updefault}$in_r^x$}}}}}
\put(2265,352){\ellipse{600}{600}}
\put(3975,352){\ellipse{600}{600}}
\put(3840,322){\makebox(0,0)[lb]{\smash{{{\SetFigFont{10}{12.0}{\rmdefault}{\mddefault}{\updefault}$dn_r^x$}}}}}
\put(5550,307){\ellipse{600}{600}}
\put(5415,337){\makebox(0,0)[lb]{\smash{{{\SetFigFont{10}{12.0}{\rmdefault}{\mddefault}{\updefault}$st_r^x$}}}}}
\put(7650,367){\ellipse{600}{600}}
\path(1485,352)(1965,352)
\path(1845.000,322.000)(1965.000,352.000)(1845.000,382.000)
\path(3555.000,322.000)(3675.000,352.000)(3555.000,382.000)
\path(3675,352)(2565,352)
\put(1500,517){\makebox(0,0)[lb]{\smash{{{\SetFigFont{10}{12.0}{\rmdefault}{\mddefault}{\updefault}$wt_c^x$}}}}}
\put(2100,292){\makebox(0,0)[lb]{\smash{{{\SetFigFont{10}{12.0}{\rmdefault}{\mddefault}{\updefault}$wt_r^x$}}}}}
\put(0,292){\makebox(0,0)[lb]{\smash{{{\SetFigFont{17}{20.4}{\rmdefault}{\mddefault}{\updefault}$R_r^x$}}}}}
\put(7500,442){\makebox(0,0)[lb]{\smash{{{\SetFigFont{10}{12.0}{\rmdefault}{\mddefault}{\updefault}$snt_r^x$}}}}}
\put(7500,442){\makebox(0,0)[lb]{\smash{{{\SetFigFont{10}{12.0}{\rmdefault}{\mddefault}{\updefault}$snt_r^x$}}}}}
\put(7500,217){\makebox(0,0)[lb]{\smash{{{\SetFigFont{10}{12.0}{\rmdefault}{\mddefault}{\updefault}$st_r^x$}}}}}
\put(1185,1252){\ellipse{600}{600}}
\put(3165,1252){\ellipse{600}{600}}
\put(5775,1252){\ellipse{600}{600}}
\path(1485,1252)(2865,1252)
\path(2745.000,1222.000)(2865.000,1252.000)(2745.000,1282.000)
\path(5355.000,1222.000)(5475.000,1252.000)(5355.000,1282.000)
\path(5475,1252)(3465,1252)
\put(1050,1222){\makebox(0,0)[lb]{\smash{{{\SetFigFont{10}{12.0}{\rmdefault}{\mddefault}{\updefault}$in_c^x$}}}}}
\put(3045,1222){\makebox(0,0)[lb]{\smash{{{\SetFigFont{10}{12.0}{\rmdefault}{\mddefault}{\updefault}$wt_c^x$}}}}}
\put(5640,1222){\makebox(0,0)[lb]{\smash{{{\SetFigFont{10}{12.0}{\rmdefault}{\mddefault}{\updefault}$dn_c^x$}}}}}
\put(3675,1417){\makebox(0,0)[lb]{\smash{{{\SetFigFont{10}{12.0}{\rmdefault}{\mddefault}{\updefault}$dn_r^x \lor st_r^x$}}}}}
\put(0,742){\makebox(0,0)[lb]{\smash{{{\SetFigFont{17}{20.4}{\rmdefault}{\mddefault}{\updefault}$||$}}}}}
\put(0,1192){\makebox(0,0)[lb]{\smash{{{\SetFigFont{17}{20.4}{\rmdefault}{\mddefault}{\updefault}$C_c^x$}}}}}
\end{picture}
}
}
\end{center}
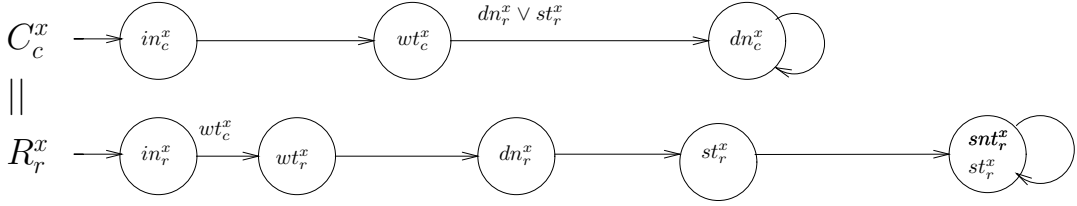
\caption{Client-replica interaction: pair-program $C_c^x \pl R_{r}^{x}$,
         $r = \replica{x}$.}
\label{fig:submit}
\efg

\emph{\ul{$\CSC$ constraints, pair-machine $R_r^x \pl R_{r'}^{x'}$}},
$x \in \Op$, $x' \in x.\prev$, $r = \replica{x}$, $r' = \replica{x'}$
\begin{lst}

   \item[$\AG(dn_r^x \imp dn_{r'}^{x'})$:]
	every operation in $x.\prev$ is performed before $x$ is

   \item[$\AG(dn_r^x \imp \AG dn_r^x) \land \AG(dn_{r'}^{x'} \imp \AG dn_{r'}^{x'})$:]
	once an operation is done, it remains done

\end{lst}

\bfg
\begin{center}
\scalebox{0.8}{\setlength{\unitlength}{0.00083333in}
\begingroup\makeatletter\ifx\SetFigFont\undefined%
\gdef\SetFigFont#1#2#3#4#5{%
  \reset@font\fontsize{#1}{#2pt}%
  \fontfamily{#3}\fontseries{#4}\fontshape{#5}%
  \selectfont}%
\fi\endgroup%
{\renewcommand{\dashlinestretch}{30}
\begin{picture}(8499,1597)(0,-10)
\put(8062.500,404.500){\arc{530.330}{3.9270}{8.6394}}
\path(7992.894,179.569)(7875.000,217.000)(7961.409,128.493)
\path(4275,367)(5250,367)
\path(5130.000,337.000)(5250.000,367.000)(5130.000,397.000)
\path(7230.040,336.841)(7350.000,367.000)(7229.960,396.841)
\path(7350,367)(5837,365)
\path(600,367)(525,367)(600,367)
	(675,367)(750,367)(825,367)(900,367)
\path(780.000,337.000)(900.000,367.000)(780.000,397.000)
\put(8062.500,1304.500){\arc{530.330}{3.9270}{8.6394}}
\path(7992.894,1079.569)(7875.000,1117.000)(7961.409,1028.493)
\path(4275,1267)(5250,1267)
\path(5130.000,1237.000)(5250.000,1267.000)(5130.000,1297.000)
\path(7230.040,1236.841)(7350.000,1267.000)(7229.960,1296.841)
\path(7350,1267)(5837,1265)
\path(600,1267)(525,1267)(600,1267)
	(675,1267)(750,1267)(825,1267)(900,1267)
\path(780.000,1237.000)(900.000,1267.000)(780.000,1297.000)
\put(2625,1417){\makebox(0,0)[lb]{\smash{{{\SetFigFont{10}{12.0}{\rmdefault}{\mddefault}{\updefault}$dn_{r'}^{x'} \lor st_{r'}^{x'}$}}}}}
\put(1185,1252){\ellipse{600}{600}}
\put(1050,1222){\makebox(0,0)[lb]{\smash{{{\SetFigFont{10}{12.0}{\rmdefault}{\mddefault}{\updefault}$in_r^x$}}}}}
\put(2265,1252){\ellipse{600}{600}}
\put(3975,1252){\ellipse{600}{600}}
\put(3840,1222){\makebox(0,0)[lb]{\smash{{{\SetFigFont{10}{12.0}{\rmdefault}{\mddefault}{\updefault}$dn_r^x$}}}}}
\put(5550,1207){\ellipse{600}{600}}
\put(5415,1237){\makebox(0,0)[lb]{\smash{{{\SetFigFont{10}{12.0}{\rmdefault}{\mddefault}{\updefault}$st_r^x$}}}}}
\put(2265,352){\ellipse{600}{600}}
\put(7650,367){\ellipse{600}{600}}
\put(3975,352){\ellipse{600}{600}}
\put(5550,307){\ellipse{600}{600}}
\put(1185,352){\ellipse{600}{600}}
\path(1485,352)(1965,352)
\path(1845.000,322.000)(1965.000,352.000)(1845.000,382.000)
\path(3555.000,322.000)(3675.000,352.000)(3555.000,382.000)
\path(3675,352)(2565,352)
\put(2100,292){\makebox(0,0)[lb]{\smash{{{\SetFigFont{10}{12.0}{\rmdefault}{\mddefault}{\updefault}$wt_{r'}^{x'}$}}}}}
\put(0,292){\makebox(0,0)[lb]{\smash{{{\SetFigFont{17}{20.4}{\rmdefault}{\mddefault}{\updefault}$R_{r'}^{x'}$}}}}}
\put(7500,442){\makebox(0,0)[lb]{\smash{{{\SetFigFont{10}{12.0}{\rmdefault}{\mddefault}{\updefault}$snt_r^x$}}}}}
\put(7500,442){\makebox(0,0)[lb]{\smash{{{\SetFigFont{10}{12.0}{\rmdefault}{\mddefault}{\updefault}$snt_{r'}^{x'}$}}}}}
\put(7500,217){\makebox(0,0)[lb]{\smash{{{\SetFigFont{10}{12.0}{\rmdefault}{\mddefault}{\updefault}$st_{r'}^{x'}$}}}}}
\put(3840,322){\makebox(0,0)[lb]{\smash{{{\SetFigFont{10}{12.0}{\rmdefault}{\mddefault}{\updefault}$dn_{r'}^{x'}$}}}}}
\put(5415,337){\makebox(0,0)[lb]{\smash{{{\SetFigFont{10}{12.0}{\rmdefault}{\mddefault}{\updefault}$st_{r'}^{x'}$}}}}}
\put(1050,322){\makebox(0,0)[lb]{\smash{{{\SetFigFont{10}{12.0}{\rmdefault}{\mddefault}{\updefault}$in_{r'}^{x'}$}}}}}
\put(7650,1267){\ellipse{600}{600}}
\path(1485,1252)(1965,1252)
\path(1845.000,1222.000)(1965.000,1252.000)(1845.000,1282.000)
\path(3555.000,1222.000)(3675.000,1252.000)(3555.000,1282.000)
\path(3675,1252)(2565,1252)
\put(2100,1192){\makebox(0,0)[lb]{\smash{{{\SetFigFont{10}{12.0}{\rmdefault}{\mddefault}{\updefault}$wt_r^x$}}}}}
\put(0,1192){\makebox(0,0)[lb]{\smash{{{\SetFigFont{17}{20.4}{\rmdefault}{\mddefault}{\updefault}$R_r^x$}}}}}
\put(7500,1342){\makebox(0,0)[lb]{\smash{{{\SetFigFont{10}{12.0}{\rmdefault}{\mddefault}{\updefault}$snt_r^x$}}}}}
\put(7500,1342){\makebox(0,0)[lb]{\smash{{{\SetFigFont{10}{12.0}{\rmdefault}{\mddefault}{\updefault}$snt_r^x$}}}}}
\put(7500,1117){\makebox(0,0)[lb]{\smash{{{\SetFigFont{10}{12.0}{\rmdefault}{\mddefault}{\updefault}$st_r^x$}}}}}
\put(0,742){\makebox(0,0)[lb]{\smash{{{\SetFigFont{17}{20.4}{\rmdefault}{\mddefault}{\updefault}$||$}}}}}
\end{picture}
}
}
\end{center}
\caption{$\CSC$ constraints: pair-program $R_r^x \pl R_{r'}^{x'}$, $r = \replica{x}$,
         $x' \in x.\prev$, $r' = \replica{x'}$.}
\label{fig:prev}
\efg

\emph{\ul{Strictness constraints, pair-machine $R_r^x \pl R_{r'}^{x}$}},
$x \in \Op$, $x.strict$, $r = \replica{x}$,  $r' \in \R - \{\replica{x}\}$
\begin{lst}

   \item[$\AG(snt_r^x \imp \AND_i st_i^x)$:]
	a strict operation is not performed until it is stable at all replicas

   \item[$\AG(snt_r^x \imp \AG snt_r^x) \land \AG(st_{r}^{x} \imp \AG st_{r}^{x})$:]
	once operation results are sent, they remain sent, and once an
	operation is stable, it remains stable

\end{lst}

\vspace{1.0ex}

\emph{\ul{Eventual stabilization, $R_r^x \pl R_{r'}^{x}$}},
$x \in \Op$, $r = \replica{x}$, $r' \in \R - \{\replica{x}\}$
\begin{lst}

   \item[$\AG(wt_r^x \imp \AND_i \AF st_i^x)$:]
	every submitted operation eventually stabilizes

\end{lst}

\vspace{1.0ex}

\bfg
\begin{center}
\scalebox{0.8}{\setlength{\unitlength}{0.00083333in}
\begingroup\makeatletter\ifx\SetFigFont\undefined%
\gdef\SetFigFont#1#2#3#4#5{%
  \reset@font\fontsize{#1}{#2pt}%
  \fontfamily{#3}\fontseries{#4}\fontshape{#5}%
  \selectfont}%
\fi\endgroup%
{\renewcommand{\dashlinestretch}{30}
\begin{picture}(8336,2002)(0,-10)
\put(8062.500,1484.500){\arc{530.330}{3.9270}{8.6394}}
\path(7992.894,1259.569)(7875.000,1297.000)(7961.409,1208.493)
\put(6892.500,284.500){\arc{471.010}{4.0625}{8.5039}}
\path(6872.436,79.410)(6750.000,97.000)(6849.754,23.862)
\path(4275,1447)(5250,1447)
\path(5130.000,1417.000)(5250.000,1447.000)(5130.000,1477.000)
\path(7230.040,1416.841)(7350.000,1447.000)(7229.960,1476.841)
\path(7350,1447)(5837,1445)
\path(600,1447)(525,1447)(600,1447)
	(675,1447)(750,1447)(825,1447)(900,1447)
\path(780.000,1417.000)(900.000,1447.000)(780.000,1477.000)
\path(600,322)(525,322)(600,322)
	(675,322)(750,322)(825,322)(900,322)
\path(780.000,292.000)(900.000,322.000)(780.000,352.000)
\put(4350,1597){\makebox(0,0)[lb]{\smash{{{\SetFigFont{10}{12.0}{\rmdefault}{\mddefault}{\updefault}$dn_{r'}^x$}}}}}
\put(5400,472){\makebox(0,0)[lb]{\smash{{{\SetFigFont{10}{12.0}{\rmdefault}{\mddefault}{\updefault}$st_r^x$}}}}}
\put(5925,1822){\makebox(0,0)[lb]{\smash{{{\SetFigFont{10}{12.0}{\rmdefault}{\mddefault}{\updefault}$[v := val(x,lb_r)]$}}}}}
\put(1185,1432){\ellipse{600}{600}}
\put(1050,1402){\makebox(0,0)[lb]{\smash{{{\SetFigFont{10}{12.0}{\rmdefault}{\mddefault}{\updefault}$in_r^x$}}}}}
\put(2265,1432){\ellipse{600}{600}}
\put(3975,1432){\ellipse{600}{600}}
\put(3840,1402){\makebox(0,0)[lb]{\smash{{{\SetFigFont{10}{12.0}{\rmdefault}{\mddefault}{\updefault}$dn_r^x$}}}}}
\put(5550,1387){\ellipse{600}{600}}
\put(5415,1417){\makebox(0,0)[lb]{\smash{{{\SetFigFont{10}{12.0}{\rmdefault}{\mddefault}{\updefault}$st_r^x$}}}}}
\put(7650,1447){\ellipse{600}{600}}
\put(1185,307){\ellipse{600}{600}}
\put(6525,307){\ellipse{600}{600}}
\put(4875,307){\ellipse{600}{600}}
\put(3165,307){\ellipse{600}{600}}
\path(1485,1432)(1965,1432)
\path(1845.000,1402.000)(1965.000,1432.000)(1845.000,1462.000)
\path(3555.000,1402.000)(3675.000,1432.000)(3555.000,1462.000)
\path(3675,1432)(2565,1432)
\path(1485,307)(2865,307)
\path(2745.000,277.000)(2865.000,307.000)(2745.000,337.000)
\path(4455.000,277.000)(4575.000,307.000)(4455.000,337.000)
\path(4575,307)(3465,307)
\path(6105.447,290.264)(6225.000,322.000)(6104.578,350.258)
\path(6225,322)(5190,307)
\put(2100,1372){\makebox(0,0)[lb]{\smash{{{\SetFigFont{10}{12.0}{\rmdefault}{\mddefault}{\updefault}$wt_r^x$}}}}}
\put(0,1372){\makebox(0,0)[lb]{\smash{{{\SetFigFont{17}{20.4}{\rmdefault}{\mddefault}{\updefault}$R_r^x$}}}}}
\put(7500,1522){\makebox(0,0)[lb]{\smash{{{\SetFigFont{10}{12.0}{\rmdefault}{\mddefault}{\updefault}$snt_r^x$}}}}}
\put(7500,1522){\makebox(0,0)[lb]{\smash{{{\SetFigFont{10}{12.0}{\rmdefault}{\mddefault}{\updefault}$snt_r^x$}}}}}
\put(7500,1297){\makebox(0,0)[lb]{\smash{{{\SetFigFont{10}{12.0}{\rmdefault}{\mddefault}{\updefault}$st_r^x$}}}}}
\put(0,847){\makebox(0,0)[lb]{\smash{{{\SetFigFont{17}{20.4}{\rmdefault}{\mddefault}{\updefault}$||$}}}}}
\put(1050,277){\makebox(0,0)[lb]{\smash{{{\SetFigFont{10}{12.0}{\rmdefault}{\mddefault}{\updefault}$in_{r'}^x$}}}}}
\put(6390,307){\makebox(0,0)[lb]{\smash{{{\SetFigFont{10}{12.0}{\rmdefault}{\mddefault}{\updefault}$st_{r'}^x$}}}}}
\put(4740,307){\makebox(0,0)[lb]{\smash{{{\SetFigFont{10}{12.0}{\rmdefault}{\mddefault}{\updefault}$dn_r^x$}}}}}
\put(3045,307){\makebox(0,0)[lb]{\smash{{{\SetFigFont{10}{12.0}{\rmdefault}{\mddefault}{\updefault}$wt_{r'}^x$}}}}}
\put(0,322){\makebox(0,0)[lb]{\smash{{{\SetFigFont{17}{20.4}{\rmdefault}{\mddefault}{\updefault}$R_{r'}^{x}$}}}}}
\put(2550,1822){\makebox(0,0)[lb]{\smash{{{\SetFigFont{10}{12.0}{\rmdefault}{\mddefault}{\updefault}$[lb_r(x) := next(lb_r)]$}}}}}
\put(3450,472){\makebox(0,0)[lb]{\smash{{{\SetFigFont{10}{12.0}{\rmdefault}{\mddefault}{\updefault}$dn_r^x \ar skip$}}}}}
\put(3450,697){\makebox(0,0)[lb]{\smash{{{\SetFigFont{10}{12.0}{\rmdefault}{\mddefault}{\updefault}$[lb_{r'}(x) := lb_r(x)]$}}}}}
\end{picture}
}
}
\end{center}
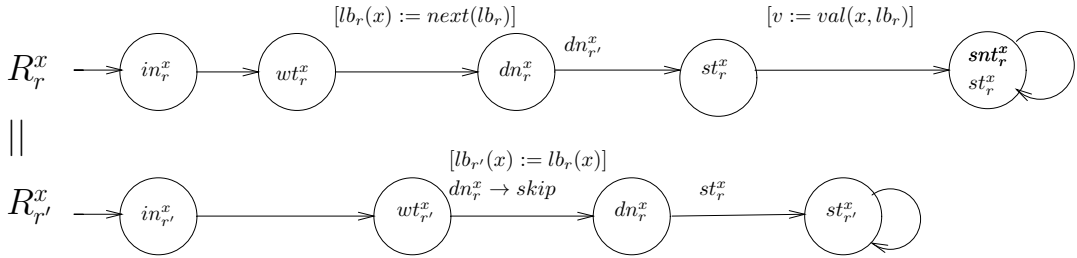
\caption{The Pair-program $R_r^x \pl R_{r'}^{x}$, when $x$ is strict, 
         $r = \replica{x}$, $r' \in \R - \{\replica{x}\}$.}
\label{fig:strict}
\efg

\emph{\ul{Rule for Dynamic process creation}}
At any point, a client $C_c$ can create the pair-programs required
for the processing of a new operation $x$, for which $\client{x} =
C_c$.
These pair-programs are $C_c^x \pl R_{r}^{x}$ where $r = \replica{x}$,
$R_r^x \pl R_{r'}^{x'}$ where $x' \in x.\prev$, $r' = \replica{x'}$, and
$R_r^x \pl R_{i}^{x}$ $r = \replica{x}$, $i \in \R$.
It is permissible for $\replica{x}$ to be a ``new'' replica, i.e., one
that currently does not occur in any pair-program. Thus, the set of
``current replicas'' can be expanded at run-time. This
is done implicitly when the first operation which is processed by that
replica is instantiated.
Likewise, a ``new'' client can submit an
operation for the first time. Thus, clients can also be created
dynamically.

For each pair-specification, we synthesize a pair-program satisfying
it, e.g., using the method of \cite{EC82}.
Figures~\ref{fig:submit}, \ref{fig:prev}, and \ref{fig:strict}
show the resulting pair-programs.
We then apply
Definition~\ref{def:dyn-pair-syn} to synthesize the ESDS program with
a dynamic number of clients and replicas,
shown in
Figure~\ref{fig:esds}. The ESDS program, and the
pair-program  $R_r^x \pl R_{r'}^{x}$ of Figure~\ref{fig:strict} both
manipulate some ``underlying'' data, i.e., data which is updated,
but not referenced in any guard, and so does not affect
control-flow. This data consists of a labeling
function $lb_r$ which assigns to each operation $x$ at replica $r$ a
label, drawn from a well-ordered set. The assignment 
$lb_r(x) := next(lb_r)$ takes the smallest label not yet allocated by
$lb_r$ and assigns it to $lb_r(x)$. 
The labels encode ordering information for the
operations. The assignment $v := val(x,lb_r)$ computes a value $v$ for
operation $x$, using the ordering given by $lb_r$:
operations with a smaller label are ordered before operations with a
larger label. In the figures, these assignments to underlying data are
shown within $[..]$ brackets, alongside the arc-labels obtained by
pairwise synthesis. They are not used when verifying
correctness properties; the ordering constraints given by the
$x.\prev$ sets are sufficient to verify that the client-specified
constraints are obeyed.
Finally, we add self-loops to the final local state of every process
for technical reasons related to establishing deadlock-freedom.

Correctness of the ESDS program follows
immediately from Theorem~\ref{thm:dynamic:large-model}, since the
conjunction of the pair-specifications gives us the desired
correctness properties (formulae of the forms $\AG( p_i \imp \AX_i q_i)$,
$\AG( p_i \imp \EX_i q_i)$ are not in $\ACTLmij$, but were shown
to be preserved in \cite{AE98}, and the proof given there
still applies).

\bfg
\begin{center}
\scalebox{0.8}{\setlength{\unitlength}{0.00083333in}
\begingroup\makeatletter\ifx\SetFigFont\undefined%
\gdef\SetFigFont#1#2#3#4#5{%
  \reset@font\fontsize{#1}{#2pt}%
  \fontfamily{#3}\fontseries{#4}\fontshape{#5}%
  \selectfont}%
\fi\endgroup%
{\renewcommand{\dashlinestretch}{30}
\begin{picture}(8336,3434)(0,-10)
\put(8062.500,2534.500){\arc{530.330}{3.9270}{8.6394}}
\path(7992.894,2309.569)(7875.000,2347.000)(7961.409,2258.493)
\put(6892.500,284.500){\arc{471.010}{4.0625}{8.5039}}
\path(6872.436,79.410)(6750.000,97.000)(6849.754,23.862)
\path(4275,2497)(5400,3172)
\path(5312.536,3084.536)(5400.000,3172.000)(5281.666,3135.985)
\path(600,2497)(525,2497)(600,2497)
	(675,2497)(750,2497)(825,2497)(900,2497)
\path(780.000,2467.000)(900.000,2497.000)(780.000,2527.000)
\path(4275,2497)(5400,1822)
\path(5281.666,1858.015)(5400.000,1822.000)(5312.536,1909.464)
\path(7255.873,2416.750)(7350.000,2497.000)(7229.182,2470.487)
\path(7350,2497)(5987,1820)
\path(7229.120,2523.229)(7350.000,2497.000)(7255.684,2577.028)
\path(7350,2497)(5987,3170)
\path(600,322)(525,322)(600,322)
	(675,322)(750,322)(825,322)(900,322)
\path(780.000,292.000)(900.000,322.000)(780.000,352.000)
\put(6525,1972){\makebox(0,0)[lb]{\smash{{{\SetFigFont{10}{12.0}{\rmdefault}{\mddefault}{\updefault}$[v := val(x,lb_r)]$}}}}}
\put(6525,2947){\makebox(0,0)[lb]{\smash{{{\SetFigFont{10}{12.0}{\rmdefault}{\mddefault}{\updefault}$dn_{r'}^x$}}}}}
\put(5400,472){\makebox(0,0)[lb]{\smash{{{\SetFigFont{10}{12.0}{\rmdefault}{\mddefault}{\updefault}$st_r^x$}}}}}
\put(4500,1972){\makebox(0,0)[lb]{\smash{{{\SetFigFont{10}{12.0}{\rmdefault}{\mddefault}{\updefault}$dn_{r'}^x$}}}}}
\put(4050,3022){\makebox(0,0)[lb]{\smash{{{\SetFigFont{10}{12.0}{\rmdefault}{\mddefault}{\updefault}$[v := val(x,lb_r)]$}}}}}
\put(1185,2482){\ellipse{600}{600}}
\put(1050,2452){\makebox(0,0)[lb]{\smash{{{\SetFigFont{10}{12.0}{\rmdefault}{\mddefault}{\updefault}$in_r^x$}}}}}
\put(2265,2482){\ellipse{600}{600}}
\put(3975,2482){\ellipse{600}{600}}
\put(3840,2452){\makebox(0,0)[lb]{\smash{{{\SetFigFont{10}{12.0}{\rmdefault}{\mddefault}{\updefault}$dn_r^x$}}}}}
\put(5700,3112){\ellipse{600}{600}}
\put(5565,3142){\makebox(0,0)[lb]{\smash{{{\SetFigFont{10}{12.0}{\rmdefault}{\mddefault}{\updefault}$st_r^x$}}}}}
\put(5700,1762){\ellipse{600}{600}}
\put(5565,1792){\makebox(0,0)[lb]{\smash{{{\SetFigFont{10}{12.0}{\rmdefault}{\mddefault}{\updefault}$st_r^x$}}}}}
\put(7650,2497){\ellipse{600}{600}}
\put(1185,307){\ellipse{600}{600}}
\put(6525,307){\ellipse{600}{600}}
\put(4875,307){\ellipse{600}{600}}
\put(3165,307){\ellipse{600}{600}}
\path(1485,2482)(1965,2482)
\path(1845.000,2452.000)(1965.000,2482.000)(1845.000,2512.000)
\path(3555.000,2452.000)(3675.000,2482.000)(3555.000,2512.000)
\path(3675,2482)(2565,2482)
\path(1485,307)(2865,307)
\path(2745.000,277.000)(2865.000,307.000)(2745.000,337.000)
\path(4455.000,277.000)(4575.000,307.000)(4455.000,337.000)
\path(4575,307)(3465,307)
\path(6105.447,290.264)(6225.000,322.000)(6104.578,350.258)
\path(6225,322)(5190,307)
\put(2100,2422){\makebox(0,0)[lb]{\smash{{{\SetFigFont{10}{12.0}{\rmdefault}{\mddefault}{\updefault}$wt_r^x$}}}}}
\put(0,2422){\makebox(0,0)[lb]{\smash{{{\SetFigFont{17}{20.4}{\rmdefault}{\mddefault}{\updefault}$R_r^x$}}}}}
\put(7500,2572){\makebox(0,0)[lb]{\smash{{{\SetFigFont{10}{12.0}{\rmdefault}{\mddefault}{\updefault}$snt_r^x$}}}}}
\put(7500,2572){\makebox(0,0)[lb]{\smash{{{\SetFigFont{10}{12.0}{\rmdefault}{\mddefault}{\updefault}$snt_r^x$}}}}}
\put(7500,2347){\makebox(0,0)[lb]{\smash{{{\SetFigFont{10}{12.0}{\rmdefault}{\mddefault}{\updefault}$st_r^x$}}}}}
\put(2550,2872){\makebox(0,0)[lb]{\smash{{{\SetFigFont{10}{12.0}{\rmdefault}{\mddefault}{\updefault}$[lb_r(x) := next(lb_r)]$}}}}}
\put(1050,277){\makebox(0,0)[lb]{\smash{{{\SetFigFont{10}{12.0}{\rmdefault}{\mddefault}{\updefault}$in_{r'}^x$}}}}}
\put(6390,307){\makebox(0,0)[lb]{\smash{{{\SetFigFont{10}{12.0}{\rmdefault}{\mddefault}{\updefault}$st_{r'}^x$}}}}}
\put(4740,307){\makebox(0,0)[lb]{\smash{{{\SetFigFont{10}{12.0}{\rmdefault}{\mddefault}{\updefault}$dn_r^x$}}}}}
\put(3045,307){\makebox(0,0)[lb]{\smash{{{\SetFigFont{10}{12.0}{\rmdefault}{\mddefault}{\updefault}$wt_{r'}^x$}}}}}
\put(0,322){\makebox(0,0)[lb]{\smash{{{\SetFigFont{17}{20.4}{\rmdefault}{\mddefault}{\updefault}$R_{r'}^{x}$}}}}}
\put(3450,847){\makebox(0,0)[lb]{\smash{{{\SetFigFont{10}{12.0}{\rmdefault}{\mddefault}{\updefault}$[lb_{r'}(x) := lb_r(x)]$}}}}}
\put(3450,622){\makebox(0,0)[lb]{\smash{{{\SetFigFont{10}{12.0}{\rmdefault}{\mddefault}{\updefault}$dn_r^x \ar skip$}}}}}
\put(0,1372){\makebox(0,0)[lb]{\smash{{{\SetFigFont{17}{20.4}{\rmdefault}{\mddefault}{\updefault}$||$}}}}}
\end{picture}
}
}
\end{center}
\caption{The Pair-program $R_r^x \pl R_{r'}^{x}$, when $x$ is not strict}
\label{fig:notstrict}
\efg

\bfg
\begin{center}
\scalebox{0.8}{\setlength{\unitlength}{0.00083333in}
\begingroup\makeatletter\ifx\SetFigFont\undefined%
\gdef\SetFigFont#1#2#3#4#5{%
  \reset@font\fontsize{#1}{#2pt}%
  \fontfamily{#3}\fontseries{#4}\fontshape{#5}%
  \selectfont}%
\fi\endgroup%
{\renewcommand{\dashlinestretch}{30}
\begin{picture}(8336,2677)(0,-10)
\put(6892.500,284.500){\arc{471.010}{4.0625}{8.5039}}
\path(6872.436,79.410)(6750.000,97.000)(6849.754,23.862)
\put(6142.500,2309.500){\arc{471.010}{4.0625}{8.5039}}
\path(6122.436,2104.410)(6000.000,2122.000)(6099.754,2048.862)
\put(8062.500,1259.500){\arc{530.330}{3.9270}{8.6394}}
\path(7992.894,1034.569)(7875.000,1072.000)(7961.409,983.493)
\path(600,322)(525,322)(600,322)
	(675,322)(750,322)(825,322)(900,322)
\path(780.000,292.000)(900.000,322.000)(780.000,352.000)
\path(600,2347)(525,2347)(600,2347)
	(675,2347)(750,2347)(825,2347)(900,2347)
\path(780.000,2317.000)(900.000,2347.000)(780.000,2377.000)
\path(4275,1222)(5250,1222)
\path(5130.000,1192.000)(5250.000,1222.000)(5130.000,1252.000)
\path(7230.040,1191.841)(7350.000,1222.000)(7229.960,1251.841)
\path(7350,1222)(5837,1220)
\path(600,1222)(525,1222)(600,1222)
	(675,1222)(750,1222)(825,1222)(900,1222)
\path(780.000,1192.000)(900.000,1222.000)(780.000,1252.000)
\put(5400,472){\makebox(0,0)[lb]{\smash{{{\SetFigFont{10}{12.0}{\rmdefault}{\mddefault}{\updefault}$st_r^x$}}}}}
\put(4425,1372){\makebox(0,0)[lb]{\smash{{{\SetFigFont{10}{12.0}{\rmdefault}{\mddefault}{\updefault}$\gci{r' \in \R'} dn_{r'}^x$}}}}}
\put(5850,1597){\makebox(0,0)[lb]{\smash{{{\SetFigFont{10}{12.0}{\rmdefault}{\mddefault}{\updefault}$[v := val(x,lb_r)]$}}}}}
\put(2489,1412){\makebox(0,0)[lb]{\smash{{{\SetFigFont{10}{12.0}{\rmdefault}{\mddefault}{\updefault}$\gci{r' \in \R'} dn_{r'}^{x'} \lor st_{r'}^{x'}$}}}}}
\put(1185,1207){\ellipse{600}{600}}
\put(1050,1177){\makebox(0,0)[lb]{\smash{{{\SetFigFont{10}{12.0}{\rmdefault}{\mddefault}{\updefault}$in_r^x$}}}}}
\put(2265,1207){\ellipse{600}{600}}
\put(5550,1162){\ellipse{600}{600}}
\put(5415,1192){\makebox(0,0)[lb]{\smash{{{\SetFigFont{10}{12.0}{\rmdefault}{\mddefault}{\updefault}$st_r^x$}}}}}
\put(3975,1207){\ellipse{600}{600}}
\put(3840,1177){\makebox(0,0)[lb]{\smash{{{\SetFigFont{10}{12.0}{\rmdefault}{\mddefault}{\updefault}$dn_r^x$}}}}}
\put(1185,307){\ellipse{600}{600}}
\put(6525,307){\ellipse{600}{600}}
\put(4875,307){\ellipse{600}{600}}
\put(3165,307){\ellipse{600}{600}}
\put(1185,2332){\ellipse{600}{600}}
\put(3165,2332){\ellipse{600}{600}}
\put(5775,2332){\ellipse{600}{600}}
\put(7650,1222){\ellipse{600}{600}}
\path(1485,307)(2865,307)
\path(2745.000,277.000)(2865.000,307.000)(2745.000,337.000)
\path(4455.000,277.000)(4575.000,307.000)(4455.000,337.000)
\path(4575,307)(3465,307)
\path(6105.447,290.264)(6225.000,322.000)(6104.578,350.258)
\path(6225,322)(5190,307)
\path(1485,2332)(2865,2332)
\path(2745.000,2302.000)(2865.000,2332.000)(2745.000,2362.000)
\path(5355.000,2302.000)(5475.000,2332.000)(5355.000,2362.000)
\path(5475,2332)(3465,2332)
\path(1485,1207)(1965,1207)
\path(1845.000,1177.000)(1965.000,1207.000)(1845.000,1237.000)
\path(3555.000,1177.000)(3675.000,1207.000)(3555.000,1237.000)
\path(3675,1207)(2565,1207)
\put(0,1672){\makebox(0,0)[lb]{\smash{{{\SetFigFont{17}{20.4}{\rmdefault}{\mddefault}{\updefault}$||$}}}}}
\put(0,772){\makebox(0,0)[lb]{\smash{{{\SetFigFont{17}{20.4}{\rmdefault}{\mddefault}{\updefault}$||$}}}}}
\put(1050,277){\makebox(0,0)[lb]{\smash{{{\SetFigFont{10}{12.0}{\rmdefault}{\mddefault}{\updefault}$in_{r'}^x$}}}}}
\put(6390,307){\makebox(0,0)[lb]{\smash{{{\SetFigFont{10}{12.0}{\rmdefault}{\mddefault}{\updefault}$st_{r'}^x$}}}}}
\put(4740,307){\makebox(0,0)[lb]{\smash{{{\SetFigFont{10}{12.0}{\rmdefault}{\mddefault}{\updefault}$dn_r^x$}}}}}
\put(3045,307){\makebox(0,0)[lb]{\smash{{{\SetFigFont{10}{12.0}{\rmdefault}{\mddefault}{\updefault}$wt_{r'}^x$}}}}}
\put(0,322){\makebox(0,0)[lb]{\smash{{{\SetFigFont{17}{20.4}{\rmdefault}{\mddefault}{\updefault}$R_{r'}^{x}$}}}}}
\put(1050,2302){\makebox(0,0)[lb]{\smash{{{\SetFigFont{10}{12.0}{\rmdefault}{\mddefault}{\updefault}$in_c^x$}}}}}
\put(3045,2302){\makebox(0,0)[lb]{\smash{{{\SetFigFont{10}{12.0}{\rmdefault}{\mddefault}{\updefault}$wt_c^x$}}}}}
\put(5640,2302){\makebox(0,0)[lb]{\smash{{{\SetFigFont{10}{12.0}{\rmdefault}{\mddefault}{\updefault}$dn_c^x$}}}}}
\put(3675,2497){\makebox(0,0)[lb]{\smash{{{\SetFigFont{10}{12.0}{\rmdefault}{\mddefault}{\updefault}$dn_r^x$}}}}}
\put(0,2272){\makebox(0,0)[lb]{\smash{{{\SetFigFont{17}{20.4}{\rmdefault}{\mddefault}{\updefault}$C_c^x$}}}}}
\put(1500,1372){\makebox(0,0)[lb]{\smash{{{\SetFigFont{10}{12.0}{\rmdefault}{\mddefault}{\updefault}$wt_c^x$}}}}}
\put(2100,1147){\makebox(0,0)[lb]{\smash{{{\SetFigFont{10}{12.0}{\rmdefault}{\mddefault}{\updefault}$wt_r^x$}}}}}
\put(0,1147){\makebox(0,0)[lb]{\smash{{{\SetFigFont{17}{20.4}{\rmdefault}{\mddefault}{\updefault}$R_r^x$}}}}}
\put(7500,1297){\makebox(0,0)[lb]{\smash{{{\SetFigFont{10}{12.0}{\rmdefault}{\mddefault}{\updefault}$snt_r^x$}}}}}
\put(7500,1297){\makebox(0,0)[lb]{\smash{{{\SetFigFont{10}{12.0}{\rmdefault}{\mddefault}{\updefault}$snt_r^x$}}}}}
\put(7500,1072){\makebox(0,0)[lb]{\smash{{{\SetFigFont{10}{12.0}{\rmdefault}{\mddefault}{\updefault}$st_r^x$}}}}}
\put(3450,697){\makebox(0,0)[lb]{\smash{{{\SetFigFont{10}{12.0}{\rmdefault}{\mddefault}{\updefault}$[lb_{r'}(x) := lb_r(x)]$}}}}}
\put(3450,472){\makebox(0,0)[lb]{\smash{{{\SetFigFont{10}{12.0}{\rmdefault}{\mddefault}{\updefault}$dn_r^x \ar skip$}}}}}
\put(2499,1617){\makebox(0,0)[lb]{\smash{{{\SetFigFont{10}{12.0}{\rmdefault}{\mddefault}{\updefault}$[lb_r(x) := next(lb_r)]$}}}}}
\end{picture}
}
}
\end{center}
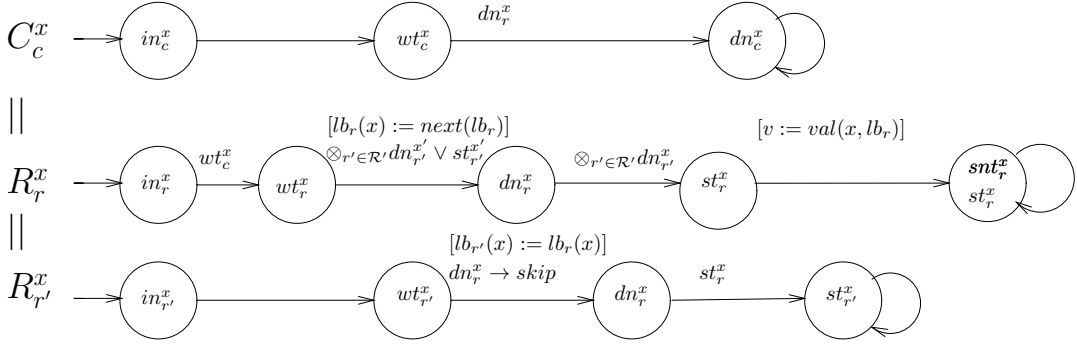
\caption{The Synthesized ESDS System. $c = \client{x}$, $r =
\replica{x}$. $x'$ ranges over $x.\prev$, 
and $r'$ ranges over $\R' = \R - \{\replica{x}\}$ in $\gci{r'}$. $R_{r'}^{x'}$ is
not shown since it is isomorphic to $R_r^x$.}
\label{fig:esds}
\efg

\section{Conclusions and Further Work}
\label{sec:conclusions}

We presented a synthesis method which deals with an arbitrary and
dynamically changing number of component processes without incurring
the exponential overhead due to state-explosion. Our method applies to
any process interconnection scheme, does not make any assumption of
similarity among the component processes, preserves all pairwise
correctness properties expressed as nexttime-free formulae of $\ACTL$,
and produces efficient low-grain atomicity programs which require only
operations commonly available in hardware.

Further work includes extending the method to a model of concurrent
computation which facilitates abstraction and refinement, via a notion
of external behavior, such as the model of \cite{AL01}, which also
handles dynamic process creation.  We also plan to deal with
fault-tolerance by incorporating the work of \cite{AAE98}, and to
investigate extending the method to other models of computation such
as real-time and probabilistic.

\bibliographystyle{alpha}
\bibliography{bibfiles/ABBREV,bibfiles/DIST,bibfiles/IOAUT,bibfiles/MODEL,bibfiles/SYNTH,bibfiles/LOGIC,bibfiles/MOBILE}

\end{document}